\documentclass[hyper,letterpaper,12pt]{article}
\usepackage{a4wide}
\usepackage{amsmath}
\usepackage[
      colorlinks=true,
      linkcolor=blue,
      urlcolor=blue,
      filecolor=black,
      citecolor=blue,
            ]{hyperref}
\usepackage{color}
\usepackage{graphicx}
\usepackage{amsfonts}
\usepackage{amssymb}
\usepackage{epstopdf}

\begin{document}
\title{\bf \Large Towards Complete Phase Diagrams of a Holographic P-wave Superconductor Model}

\author{\large
~Rong-Gen Cai$^1$\footnote{E-mail: cairg@itp.ac.cn}~,
~~Li Li$^1$\footnote{E-mail: liliphy@itp.ac.cn}~,
~~Li-Fang Li$^2$\footnote{E-mail: lilf@itp.ac.cn}~,
~~Run-Qiu Yang$^1$\footnote{E-mail: aqiu@itp.ac.cn}\\
\\
\small $^1$State Key Laboratory of Theoretical Physics,\\
\small Institute of Theoretical Physics, Chinese Academy of Sciences,\\
\small Beijing 100190,  China.\\
\small $^2$State Key Laboratory of Space Weather, \\
\small Center for Space Science and Applied Research, Chinese Academy of Sciences,\\
\small Beijing 100190, China.}
\date{\today}
\maketitle

\begin{abstract}
We study in detail the phase structure of a holographic p-wave
superconductor model in a five dimensional Einstein-Maxwell-complex
vector field theory with a negative cosmological constant. To
construct complete phase diagrams of the model, we consider both the
soliton and black hole backgrounds. In both two cases, there exist
second order, first order and zeroth order phase transitions, and
the so-called ``retrograde condensation" also happens. In
particular, in the soliton case with the mass of the vector field
being beyond a certain critical value, we find a series of phase
transitions happen such as
``insulator/superconductor/insulator/superconductor", as the
chemical potential continuously increases. We construct complete
phase diagrams in terms of temperature and chemical potential and
find some new phase boundaries.

\end{abstract}

\tableofcontents
\section{ Introduction}
The strong/weak duality characteristic of the anti-de
Sitter/conformal field theory  correspondence
(AdS/CFT)~\cite{Maldacena:1997re,Gubser:1998bc,Witten:1998qj}
enables us to study the properties of strong coupled systems by a
weak coupled AdS gravity with one higher dimension. There are a lot
of strong interacting phenomena in condensed matter physics, which
are thought to be a good place where the AdS/CFT correspondence is
applicable (see
refs.~\cite{Hartnoll:2009sz,Herzog:2009xv,McGreevy:2009xe,Horowitz:2010gk}
for reviews). In order to describe more general phenomena in this
framework,  recent efforts are devoted to generalizing the
correspondence to those systems with less symmetries (see
refs.~\cite{Horowitz:2012ky,Horowitz:2013jaa,Ling:2013aya,Donos:2011bh,Donos:2013wia,Rozali:2013ama,Cai:2013sua}
for example) and describing the far-from equilibrium problems
quantitatively (see
refs.~\cite{Murata:2010dx,Bhaseen:2012gg,Adams:2012pj,Garcia-Garcia:2013rha,Chesler:2013lia}
for example).

One of the most studied objects is the holographic superconductor
(superfluid).  In the simplest
model~\cite{Hartnoll:2008vx,Hartnoll:2008kx}, the onset of
superconductivity is characterized by the instability to form
complex scalar hair in the bulk black hole spacetime, which
corresponds to condensation of composite charged scalar operator
spontaneously breaking U(1) symmetry below a critical temperature in
the CFT side. This toy model describes a holographic s-wave
superconductor/conductor phase transition, i.e., a transition from
black hole with no scalar hair (normal phase/conductor) to the case
with scalar hair at low temperatures (superconducting phase). By
using AdS soliton as the background in this Einstein-Maxwell-complex
scalar theory, the authors of ref.~\cite{Nishioka:2009zj}
constructed a holographic model mimicking a s-wave
superconductor/insulator transition at zero temperature. The normal
insulating phase is described by a pure AdS soliton which exhibits a mass
gap and is used to describe a confining phase in a dual
theory~\cite{Witten:1998zw}. Adding a chemical potential $\mu$ to
the bulk, the AdS soliton will become unstable to developing scalar
hair for sufficiently large chemical potential, which describes a
superconducting phase. It was shown that as one changes the
temperature $T$ and the chemical potential $\mu$ there are as many
as four phases in this model, including the AdS soliton, the AdS
black hole and their superconducting phases. The complete phase
diagram in terms of $T$ and $\mu$ was constructed in
ref.~\cite{Horowitz:2010jq}.

Recently, the authors of ref.~\cite{Cai:2013pda} proposed a
holographic p-wave superconductor model  by introducing a complex
vector field $\rho_\mu$ charged under a Maxwell field $A_\mu$ in the
bulk, which is dual to a strong coupled system involving a charged
vector operator with a global U(1) symmetry. This setup meets the
minimal requirement to construct a holographic p-wave superconductor
model, since the condensate of the dual vector operator may
spontaneously break the U(1) symmetry as well as spatial rotation
symmetry. Actually, worked in the probe limit neglecting the back
reaction of the matter fields, the vector hair appears at low
temperatures and the condensed phase exhibits an infinite DC
conductivity and a gap in the AC conductivity. In this sense, this
Einstein-Maxwell-complex vector model can be regarded as a
holographic p-wave model.~\footnote{The first holographic p-wave
model was given in ref.~\cite{Gubser:2008wv} by introducing a SU(2)
Yang-Mills field into the bulk, where a gauge boson generated by one
SU(2) generator is dual to the vector order parameter. An
alternative holographic realization of p-wave superconductivity
comes from the condensation of a 2-form field in the
bulk~\cite{Aprile:2010ge}.} The probe approximation is only
justified in the limit of large $q$ with $q\rho_\mu$ and $qA_\mu$
fixed. In the paper~\cite{Cai:2013pda2}, the full back reaction was
taken into account in the four dimensional black hole background.
Depending on the charge $q$ and mass square $m^2$ of the complex
vector field, the model presents a rich phase structure. One can
find zeroth order, first order and second order phase transitions.
There is also a ``retrograde condensation" in which the hairy black
hole solution exists only for temperature above a critical value
and is thermodynamically subdominant. In particular, we can see in
this model the conductor/superconductor/conductor phase transitions
when the temperature is continuously lowered. In addition, the behavior
of entanglement entropy in this model was discussed in ref.~\cite{EE}.

The Einstein-Maxwell-complex vector model can be directly
generalized to the AdS soliton background to describe a holographic
p-wave superconductor/insulator phase transition. Actually, in the
probe approximation, the vector condensate and instability induced
by a magnetic field in a (4+1) dimensional AdS soliton background  was studied
in ref.~\cite{Cai:2013kaa}, where a vortex lattice structure can
form in the spatial directions perpendicular to the applied magnetic
field.\footnote{In the ref.\cite{Bu:2012mq}, a similar phenomenon that the magnetic field leads to the instability was also found in a SU(2) model in the AdS Schwarzschild background and hard wall cutoff model.}
Note that in ref.~\cite{Cai:2013kaa} the vector condensation
and AdS soliton instability can be totally induced by an applied
magnetic field because of the existence of the coupling between the
magnetic moment of the vector field and the background magnetic
field in this model.  Furthermore, it was found that this toy model
is a generalization of the SU(2) p-wave model~\cite{Gubser:2008wv}
in the sense that the vector field has a general mass and
gyromagnetic ratio.

As we have shown in the black hole case~\cite{Cai:2013pda2}, it is
worthwhile to consider the full back reaction of the matter fields
on the soliton geometry and to find all possible phase behaviors.
This is just one of the purposes of the present paper. On the other
hand, in order to construct complete phase diagrams of the model, we
will also study phase behaviors of the model in a (4+1) dimensional
black hole background by generalizing the study in
ref.~\cite{Cai:2013pda2}.  Since in this paper we do not turn on
magnetic field, the model is left with two parameters, i.e., the
charge $q$ and mass square $m^2$ of the complex vector field
$\rho_\mu$. More precisely, $m^2$ determines the scaling dimension
of the dual vector operator, while $q$ controls the strength of the
back reaction. Depending on $q$ and $m^2$, the model in the black
hole case exhibits all known phase structure reported in
ref.~\cite{Cai:2013pda2}. In the soliton geometry, we can find
second order transition, first order transition, zeroth order
transition as well as ``retrograde condensation" in the sense that
the hairy soliton solution appears only for chemical potential below
a critical value and has free energy larger than the pure AdS soliton.
Those phase behaviors can be comparable to those in the black hole
case. Nevertheless, there is an additional interesting behavior in
the soliton case.  In some region of model parameters, the
condensate will be absent in some range of chemical potential, where
the condensate would be expected to appear naively. This leads to a
series of phase transitions. In a typical example, as one increases the
chemical potential, we will encounter for the
insulator/superconductor/insulator/superconductor phase transitions.

Taking both the soliton geometry and black hole case into
account, we obtain four distinct solutions: AdS soliton, AdS
black hole and their corresponding hairy solutions. We construct the
full phase diagram as a function of chemical potential and
temperature for various $q$ and $m^2$. Due to the rich phase
behaviors of the model, the $T$-$\mu$ phase diagram is much more
complicated and interesting than the holographic s-wave
model~\cite{Horowitz:2010jq} and SU(2) p-wave
model~\cite{Akhavan:2010bf,Basu:2009vv}. We can find some new phase boundaries
that are never reported in the literature in the framework of
AdS/CFT correspondence.

The organization of this paper is as follows. In the next section,
we introduce the holographic model and give no-hair soliton and
black hole solutions. In section~\ref{Soliton}, we give our ansatz
for the hairy soliton solution and specify the boundary conditions
to be satisfied. Then the phase transition for various parameters
will be discussed in detail. In section~\ref{BHhair}, a parallel
discussion for the case of black hole geometry will be presented.
The complete phase diagrams in terms of temperature and chemical
potential are constructed in section~\ref{completepd}. The
conclusion and some discussions are included in
section~\ref{concdiss}. Some calculation details are shown in
appendixes.


\section{The Holographic Model and No-hair Solutions}
\label{sect:model}

Let us begin with a $(4+1)$ dimensional Einstein-Maxwell-complex vector field theory with a negative cosmological constant~\cite{Cai:2013pda}
\begin{equation}\label{action}
\begin{split}
S=\frac{1}{2\kappa^2}\int d^5 x
\sqrt{-\bar{g}}(\mathcal{R}+\frac{12}{L^2}+\mathcal{L}_m),\\
\mathcal{L}_m=-\frac{1}{4}F_{\mu\nu} F^{\mu \nu}-\frac{1}{2}\rho_{\mu\nu}^\dagger\rho^{\mu\nu}-m^2\rho_\mu^\dagger\rho^\mu+iq\gamma \rho_\mu\rho_\nu^\dagger F^{\mu\nu},
\end{split}
\end{equation}
where $L$ is the AdS radius which will be set to be unity and $\kappa^2\equiv 8\pi G $ is related to the gravitational constant in the bulk. $\bar{g}$ is the determinant of the bulk metric $g_{\mu\nu}$ and $\rho_\mu$ is the complex vector field with mass $m$ and charge $q$. We define $F_{\mu\nu}=\nabla_\mu A_\nu-\nabla_\nu A_\mu$ and $\rho_{\mu\nu}=D_\mu\rho_\nu-D_\nu\rho_\mu$ with the covariant derivative $D_\mu=\nabla_\mu-iqA_\mu$. The last non-minimal coupling term characterizes the magnetic moment of the vector field $\rho_\mu$, which is crucial in the presence of an applied  magnetic field~\cite{Cai:2013pda,Cai:2013kaa}. However, in the present study, since we only consider the case without external magnetic field, this term will not play any role. So this model is left with two parameters $m^2$ determining the scaling dimension of the dual operator and $q$ controlling the strength of the back reaction of matter fields on the background geometry.

The full equations of motion deduced from the action~\eqref{action} are the complex vector field equations
\begin{equation}\label{vector}
D^\nu\rho_{\nu\mu}-m^2\rho_\mu+iq\gamma\rho^\nu F_{\nu\mu}=0,
\end{equation}
Maxwell's equations
\begin{equation}\label{gauge}
\nabla^\nu F_{\nu\mu}=iq(\rho^\nu\rho_{\nu\mu}^\dagger-{\rho^\nu}^\dagger\rho_{\nu\mu})+iq\gamma\nabla^\nu(\rho_\nu\rho_\mu^\dagger-\rho_\nu^\dagger\rho_\mu),
\end{equation}
as well as the gravitational field equations
\begin{equation}\label{tensor}
\begin{split}
\mathcal{R}_{\mu\nu}-\frac{1}{2}\mathcal{R}g_{\mu\nu}&-\frac{6}{L^2}g_{\mu\nu}=\frac{1}{2}F_{\mu\lambda}{F_\nu}^\lambda+\frac{1}{2}\mathcal{L}_m g_{\mu\nu}\\
&+\frac{1}{2}\{[\rho_{\mu\lambda}^\dagger{\rho_\nu}^\lambda+m^2{\rho_\mu}^\dagger\rho_\nu-iq\gamma(\rho_\mu{\rho_\lambda}^\dagger-{\rho_\mu}^\dagger\rho_\lambda){F_\nu}^\lambda]+\mu\leftrightarrow\nu\}.
\end{split}
\end{equation}
Since $\rho_\mu$ is charged under the U(1) gauge field, according to AdS/CFT correspondence, its dual operator will carry the same charge under this symmetry and a vacuum expectation value of this operator will then trigger the U(1) symmetry breaking spontaneously. Thus, the condensate of the dual vector operator will break the U(1) symmetry as well as the spatial rotation symmetry since the condensate will pick out one direction as special. Therefore, viewing this vector field as an order parameter, the holographic model can be used to mimic a p-wave superconductor (superfluid) phase transition. The gravity background without vector hair ($\rho_\mu=0$)/with vector hair ($\rho_\mu\neq0$) is used to mimic the normal phase/superconducting phase in the dual system.

First, let us consider solutions with $\rho_{\mu}=0$ and $A_t =\phi(r)$. One solution with planar symmetry is the AdS Reissner-N\"{o}rdstrom (RN-AdS) black hole, which reads~\cite{Hartnoll:2009sz}
\begin{equation}\label{RNmetric}
\begin{split}
ds^2=-f(r)dt^2+\frac{dr^2}{f(r)}+r^2(dx^2+dy^2+dz^2),\\
f(r)=r^2\left[1-\left(1+\frac{\mu^2}{3r_h^2}\right)\left(\frac{r_h}{r}\right)^4+\frac{\mu^2}{3r_h^2}\left(\frac{r_h}{r}\right)^6\right], \quad \phi(r)=\mu\left[1-\left(\frac{r_h}{r}\right)^2\right],
\end{split}
\end{equation}
The horizon locates at $r_h$ and $\mu$ is the chemical potential of the dual field theory. The temperature $T$ of the black hole is
\begin{equation}\label{bhtemp}
T=\frac{r_h}{\pi}\left(1-\frac{\mu^2}{6r_h^2}\right)\;.
\end{equation}
In the following, we will assume the coordinate $z$ is compactified with period $\Gamma$, whose value will be given
below, while $x$ and $y$ are in $(-\infty, \infty)$.
There is another trivial planar solution, the so-called AdS soliton, which is given by
\begin{equation}\label{pure}
\begin{split}
ds^2=\frac{dr^2}{r^2g(r)}+r^2(-dt^2+dx^2+dy^2+g(r)d\eta^2),\\
g(r)=1-\frac{r_0^4}{r^4}, \quad \phi(r)=\mu,
\end{split}
\end{equation}
where $r_0$ is the tip of the soliton. To avoid the potential conical singularity at $r=r_0$, the spatial direction $\eta$ must make an identification with period
\begin{equation}
\Gamma=\frac{\pi}{r_0}.
\end{equation}
Thus, the AdS soliton is cigar shaped with the asymptotical geometry $R^{1,2}\times S^1$ near the AdS boundary $r\rightarrow\infty$. Because the spacetime exists only for $r>r_0$, the dual field theory is in a confined phase and has a mass gap, $E_g \sim r_0$. Since the time component of metric is regular at the tip, the soliton can be associated with any temperature in the Euclidean sector.

In the holographic setup, the RN-AdS  black hole~\eqref{RNmetric} corresponds to a conductor while the AdS soliton~\eqref{pure} describes an insulator~\cite{Nishioka:2009zj}. In order to obtain a superconducting phase in both soliton and black hole backgrounds, we need to find solutions with non-trivial $\rho_\mu$ in the bulk. That is what we will do in the following.

\section{AdS Soliton with Vector Hair}
\label{Soliton}
\subsection{Equations of motion and boundary conditions}
\label{sect:motion}
To construct homogeneous charged solutions with vector hair in the soliton background, we adopt the following ansatz
\begin{equation}\label{ansatz}
\begin{split}
ds^2=\frac{dr^2}{r^2g(r)}+r^2(-f(r)dt^2+h(r)dx^2+dy^2+g(r)e^{-\chi(r)}d\eta^2),\\
\rho_\nu dx^\nu=\rho_x(r)dx,\quad A_\nu dx^\nu=\phi(r)dt,
\end{split}
\end{equation}
where $g(r)$ vanishes at the tip $r=r_0$ of the soliton. Further, in order to obtain a smooth geometry at the tip $r_0$, $\eta$ should be made with an identification
\begin{equation}
\label{Gamma}
\eta\sim\eta+\Gamma,\qquad \Gamma=\frac{4\pi e^{\frac{\chi(r_0)}{2}}}{r_0^2 g'(r_0)}.
\end{equation}
This gives a dual picture of the boundary theory with a mass gap, which is reminiscent of an insulating phase.

One finds that the $r$ component of~\eqref{gauge} implies that the phase of $\rho_x$ must be a constant. Without loss of generality, we can take $\rho_x$ to be real. Then, the independent equations of motion in terms of the above ansatz are deduced as follows
\begin{equation}\label{eoms}
\begin{split}
\phi''-(\frac{f'}{2f}-\frac{g'}{g}-\frac{h'}{2h}+\frac{\chi'}{2}-\frac{3}{r})\phi'-\frac{2q^2\rho_x^2}{r^4gh}\phi=0,\\
\rho_x''+(\frac{f'}{2f}+\frac{g'}{g}-\frac{h'}{2h}-\frac{\chi'}{2}+\frac{3}{r})\rho_x'+\frac{q^2\phi^2}{r^4fg}\rho_x-\frac{m^2}{r^2g}\rho_x=0, \\
f''-(\frac{f'}{2f}-\frac{g'}{g}-\frac{h'}{2h}+\frac{\chi'}{2}-\frac{5}{r})f'-\frac{\phi'^2}{r^2}-\frac{2q^2\rho_x^2\phi^2}{r^6gh}=0,\\
\chi'-\frac{f'}{f}-\frac{2g'}{g}-\frac{h'}{h}+\frac{2{\rho_x'}^2}{3rh}-\frac{\phi'^2}{3rf}-\frac{2q^2\rho_x^2\phi^2}{3r^5fgh}+\frac{8}{rg}-\frac{8}{r}=0,\\
h''+(\frac{f'}{2f}+\frac{g'}{g}-\frac{h'}{2h}-\frac{\chi'}{2}+\frac{5}{r})h'+\frac{2{\rho_x'}^2}{r^2}-\frac{2q^2\rho_x^2\phi^2}{r^6fgh}+\frac{2m^2\rho_x^2}{r^4gh}=0,\\
(\frac{6}{r}-\frac{f'}{f}-\frac{h'}{h})\frac{g'}{g}+(\frac{f'}{f}+\frac{h'}{h})\chi'-\frac{f'h'}{fh}-\frac{2\rho_x'^2}{r^2h}+\frac{\phi'^2}{r^2f}+\frac{6q^2
\rho_x^2\phi^2}{r^6fgh}-\frac{2m^2\rho_x^2}{r^4gh}-\frac{24}{r^2g}+\frac{24}{r^2}=0,
\end{split}
\end{equation}
where the prime denotes the derivative with respect to $r$.

The full coupled equations of motion do not admit an analytical solution with non-trivial $\rho_x$. Therefore, we have to solve them numerically. We will use shooting method to solve
equations~\eqref{eoms}. In order to find the solutions for all the six functions $\mathcal{F}=\{\rho_x,\phi,f,g,h,\chi\}$ one must impose suitable boundary conditions at both conformal boundary $r\rightarrow\infty$ and the tip $r=r_0$.

In order to match the asymptotical AdS boundary, we have the general falloff near the boundary $r\rightarrow\infty$ as
\begin{equation} \label{boundary}
\begin{split}
&\phi=\mu-\frac{\rho}{r^2}+\ldots,\quad \rho_x=\frac{{\rho_x}_-}{r^{{\Delta}_-}}+\frac{{\rho_x}_{+}}{r^{{\Delta}_+}}+\ldots,\quad f=1+\frac{f_4}{r^4}+\ldots\\
&g=1+\frac{g_4}{r^4}+\ldots,\quad h=1+\frac{h_4}{r^4}+\ldots,\quad \chi=0+\frac{\chi_4}{r^4}+\ldots,
\end{split}
\end{equation}
where the dots stand for the higher order terms in the expansion of $1/r$ and ${\Delta}_\pm=1\pm\sqrt{1+m^2}$.~\footnote{The $m^2$ has a lower bound as $m^2=-1$ with ${\Delta}_+={\Delta}_-=1$. In that case, there exists a logarithmic term in the asymptotical expansion of $\rho_x$. One has to treat such a term as the source set to be zero to avoid the instability induced by this term~\cite{Horowitz:2008bn}. In this paper, however, we always consider the case with $m^2 >-1$.} In general, in the above expansion we must impose ${\rho_x}_-=0$, which  meets the requirement that the condensate appears spontaneously. According to the AdS/CFT dictionary, up to a normalization, the coefficients $\mu$, $\rho$, and ${\rho_x}_{+}$ are regarded as chemical potential, charge density and the $x$ component of the vacuum expectation of the vector operator $\hat{J^\mu}$ in the dual field theory, respectively.

We impose the regularity conditions at the tip $r=r_0$, which mean that all our functions have finite values and admit a series expansion in terms of $(r-r_0)$ as~\footnote{In some extremal cases, the functions do not admit such Taylor series expansion. In appendix~\ref{app1b}, it is shown that a type of solutions exists with a finite nonzero $\Gamma$, but is expressed in terms of fractional order of $(r-r_0)$. This type of solutions corresponds to the limit case of the discontinuous points in subsection~\ref{sect:superconducor1}}
\begin{equation}\label{series}
\mathcal{F}=\mathcal{F}(r_h)+\mathcal{F}'(r_h)(r-r_0)+\cdots.
\end{equation}
By plugging the expansion~\eqref{series} into~\eqref{eoms}, one can find that there are six independent parameters at the tip $\{r_0,\rho_x(r_0),\phi(r_0),f(r_0),h(r_0),\chi(r_0)\}$. However, there exist four useful scaling symmetries in the equations of motion, which read
\begin{equation} \label{scaling1}
\chi\rightarrow \chi+\lambda,\quad \eta\rightarrow e^{\lambda/2}\eta,
\end{equation}
\begin{equation} \label{scaling2}
\phi\rightarrow\lambda \phi,\quad t\rightarrow\lambda^{-1} t,\quad f\rightarrow\lambda^2 f,
\end{equation}
\begin{equation} \label{scaling3}
\rho_x\rightarrow\lambda \rho_x,\quad x\rightarrow\lambda^{-1} x,\quad h\rightarrow\lambda^2 h,
\end{equation}
and
\begin{equation} \label{scaling4}
r\rightarrow\lambda r,\quad (t,x,y,\eta)\rightarrow{\lambda^{-1}}(t,x,y,\eta),\quad(\phi,\rho_x)\rightarrow\lambda(\phi,\rho_x),
\end{equation}
where in each case $\lambda$ is a real positive constant.

By using above four scaling symmetries, we can first set $\{r_0=1,f(r_0)=1,h(r_0)=1,\chi(r_0)=0\}$ for performing numerics. After solving the coupled differential equations, one should use the first three symmetries again to satisfy the asymptotic conditions $f(\infty)=1$, $h(\infty)=1$ and $\chi(\infty)=0$.~\footnote{Here we assume the difference of $\chi(r)$ between at the tip and at the boundary is finite. However, it does not hold in the limit case of the discontinuous points discussed in subsection~\ref{sect:superconducor1}. The details will be discussed in appendix~\ref{app1b}} We choose $\phi(r_0)$ as the shooting parameter to match the source free condition, i.e., ${\rho_x}_-=0$. Finally, for fixed $m^2$ and $q$, we have a one-parameter family of solutions labeled by $\rho_x(r_0)$. After solving the set of equations, we can read off the condensate $\langle \hat{J^x}\rangle$, chemical potential $\mu$ and charge density $\rho$ from the corresponding coefficients in~\eqref{boundary}. It should be noticed that different solutions obtained in this way will have different periods $\Gamma$ for $\eta$ direction. We should use the last scaling symmetry to set all of the periods $\Gamma$ equal in order to obtain same boundary geometry. We shall fix $\Gamma$ to be $\pi$ in this paper.

\subsection{Free energy}
\label{sect:free}
In the soliton background, there are two kinds of solutions: one is the pure AdS soliton without the vector hair and the other is the hairy one. To determine which kind of solutions is thermodynamically favored, we must calculate the grand potential $\Omega$ of the system in the grand canonical ensemble. In gauge/gravity duality the grand potential $\Omega$ of the boundary thermal state is identified with temperature times the on-shell bulk action in Euclidean signature. One should also consider the Gibbons-Hawking boundary term for a well-defined Dirichlet variational principle and further a surface counterterm for removing divergence. Since we consider a stationary problem, the Euclidean action is related to the Minkowski one by a minus sign as
\begin{equation}\label{onshell}
-2\kappa^2 S_{Euclidean}=\int d^5x\sqrt{-\bar{g}}(\mathcal{R}+12+\mathcal{L}_m)+\int_{r\rightarrow\infty} d^4x\sqrt{-\bar{h}}(2\mathcal{K}-6),
\end{equation}
where $\bar{g}$ is the determinant of the bulk metric (one should not confuse it with the function $g(r)$ appearing in the metric ansatz), $\bar{h}$ is the determinant of the induced metric on the boundary, and $\mathcal{K}$ is the trace of the extrinsic curvature.~\footnote{In principle, we should also consider the surface counterterm for the charged vector field $\rho_\mu$, but one can easily see that this term makes no contribution under the source free condition, i.e., $\rho_{x-}=0$.} By using the equations of motion~\eqref{eoms} and the asymptotical expansions of matter and metric functions near the boundary, the grand potential $\Omega$ turns out to be
\begin{equation}\label{grand1}
\frac{2\kappa^2\Omega}{V_3}=g_4+4(f_4+h_4-\chi_4),
\end{equation}
where $V_3=\int dx dy d\eta$. There is a little subtle issue dealing with soliton case, since the soliton background has no horizon and the associated Hawking temperature vanishes. Nevertheless, one can introduce an arbitrary inverse temperature $1/T$ as the period of the Euclidean time coordinate. The integration over the Euclidean time in the Euclidean action just gives the factor $1/T$, which cancels the temperature factor in the Euclidean action and leads to a finite grand potential. In the case of $m^2>-1$, it is found that~$f_4+h_4-\chi_4=0$ (see appendix~\ref{app2}). Since we have scaled $\Gamma$ to be $\pi$, $g_4=-1$ for the pure AdS soliton solution, namely, in the normal insulating phase.

\subsection{Phase transition in AdS soliton backgrounds }
\label{sect:superconducor}
In what follows we will look for hairy soliton solutions numerically. Our strategy is to fix $m^2$ and for each $m^2$ we scan a wide range of $q$ which determines the strength of the back reaction of matter fields on the background. The precision of numerical calculation limits us to investigate the entire range of $q$. Nevertheless, our numerical calculation reveals that the system exhibits qualitatively different behavior depending on concrete value of $m^2$. More specially, there are two critical mass squares, $m_{c1}^2=0.218\pm0.001$ and $m_{c2}^2$. Although it is difficult to determine the exact value of $m_{c2}$ numerically, it is suggested to be zero by numerical analysis. Depending on $q$ and $m^2$, we can find second order, first order and zeroth order phase transitions. There also exists the ``retrograde condensation" in which the hairy soliton solution exists only for chemical potential below a critical value and is thermodynamically subdominant. What's more, a discontinuous condensed phase appears when $m^2>m^2_{c1}$ and $q$ is less than some critical value, which is a new kind of phase transition that does not exist in other models. We shall give  more details in the following. Since we would meet with more than one critical chemical potential at which a phase transition happens as we increase the chemical potential, for brevity, we shall denote the first critical chemical potential as $\mu_{c1}$, the second critical chemical potential as $\mu_{c2}$, and so on.

\subsubsection{Phase transition for $m^2>m_{c1}^2$}
\label{sect:superconducor1}
For the case $m^2>m_{c1}^2$, there exist three critical charges $q_1$, $q_2$ and $q_3$ with $q_1<q_2<q_3$. Their values depend on $m^2$ we choose. The three critical charges divide the parameter space of $q$ into four regions, shown in table~\ref{Taba}. The phase behavior changes qualitatively in each region. For the case $q\geq q_3$, the condensed phase will appear above $\mu_{c1}$ through a second order transition and becomes thermodynamically preferred. As we decrease $q$ to a value smaller than $q_3$, the order of transition from the normal (insulator) phase to the superconducting phase becomes first order. Much more interesting thing happens when we continue decreasing the value of $q$ past $q_2$. Starting from the small $\mu$ region, the system undergos a first order transition from the normal phase to the condensed phase at $\mu_{c1}$, and as $\mu$ increases to $\mu_{c2}$ there is a zeroth order transition back to the normal phase, then at the larger chemical potential $\mu_{c3}$, it comes back to the condensed phase again by a zeroth order transition. There is a  region $\mu_{c2}<\mu<\mu_{c3}$ in which hairy solutions do not exist, so the condensed phase should jump to the normal phase. As one can see clearly, there are three phase transitions as we increase the chemical potential. The later two zeroth order phase transitions have not been reported in any holographic model. For the case $q\leq q_1$, the system undergoes a zeroth order transition  from the normal phase to the condensed phase at $\mu_{c1}$. The phase transition and their orders are summarized in table~\ref{Taba}.

\begin{table}\scriptsize
  \centering
  \begin{tabular}{|c|c|c|c|c|c|c|c|}
    \hline
    \small{Charge} & \multicolumn{7}{|c|}{\small{Phase transition and its order,~~$m^2>m^2_{c1}$}}\\
    \hline
    $q\geq q_3$ & $\mu<\mu_{c1}$, S & $\mu_{c1}$, $2^{nd}$  & \multicolumn{5}{|c|}{$\mu>\mu_{c1}$, SC} \\
    \hline
    $q_2\leq q<q_3$&$\mu<\mu_{c1}$, S & $\mu_{c1}$, $1^{st}$& \multicolumn{5}{|c|}{$\mu>\mu_{c1}$, SC} \\
    \hline
    $q_1<q<q_2$&$\mu<\mu_{c1}$, S & $\mu_{c1}$,$1^{st}$ & $\mu_{c1}<\mu<\mu_{c2}$, SC & $\mu_{c2}$,$0^{th}$ & $\mu_{c2}<\mu<\mu_{c3}$, S & $\mu_{c3}$, $0^{th}$ & $\mu>\mu_{c3}$, SC \\
    \hline
    $q\leq q_1$ &\multicolumn{5}{|c|}{$\mu<\mu_{c1}$,S} & $\mu_{c1}$, $0^{th}$ & $\mu>\mu_{c1}$, SC\\
    \hline
  \end{tabular}
  \caption{The phase transition and its order with respect to the charge and chemical potential in the case of $m^2>m^2_{c1}$. In the table, S=normal phase/insulator with pure AdS soliton, SC=condensed phase/superconductor with hairy soliton solutions.  $0^{th},1^{st}$,and~$2^{nd}$ stand for the zeroth order, first order and second order phase transitions, respectively.}
\label{Taba}
\end{table}

As a typical example, we consider the case $m^2=5/4$. We find similar results for other values of $m^2$ in the region $m^2 >m_{c1}^2$. For $m^2=5/4$, the three critical values of charge are $q_3\simeq1.5600$, $q_2\simeq1.5345$, and $q_1\simeq1.5020$, respectively. The pure AdS soliton survives for arbitrary value of $q$. Nevertheless, we can find additional solutions with non-vanishing $\rho_x$, which  are thermodynamically preferred for sufficiently larger chemical potential. That is to say, for each value of $q$, there must be a phase transition from the normal phase to the condensed phase occurring at a certain chemical potential. From the perspective of dual system, it means that a charged vector operator obtains a non-zero vacuum expectation value $\langle\hat{J_x}\rangle\neq0$, which breaks the U(1) symmetry as well as the rotation symmetry in $x-y$ plane spontaneously.~\footnote{The breaking of the rotation symmetry in $x-y$ plane is due to the fact that $\langle\hat{J_x}\rangle$ chooses $x$ direction as special. However, this anisotropy does not display in the stress energy tensor of the dual field theory. In fact, the $xx$-component and $yy$-component of the stress energy tensor are equal. For more details one can see in appendix~\ref{app2}.}

\begin{figure}
  \includegraphics[width=0.5\textwidth]{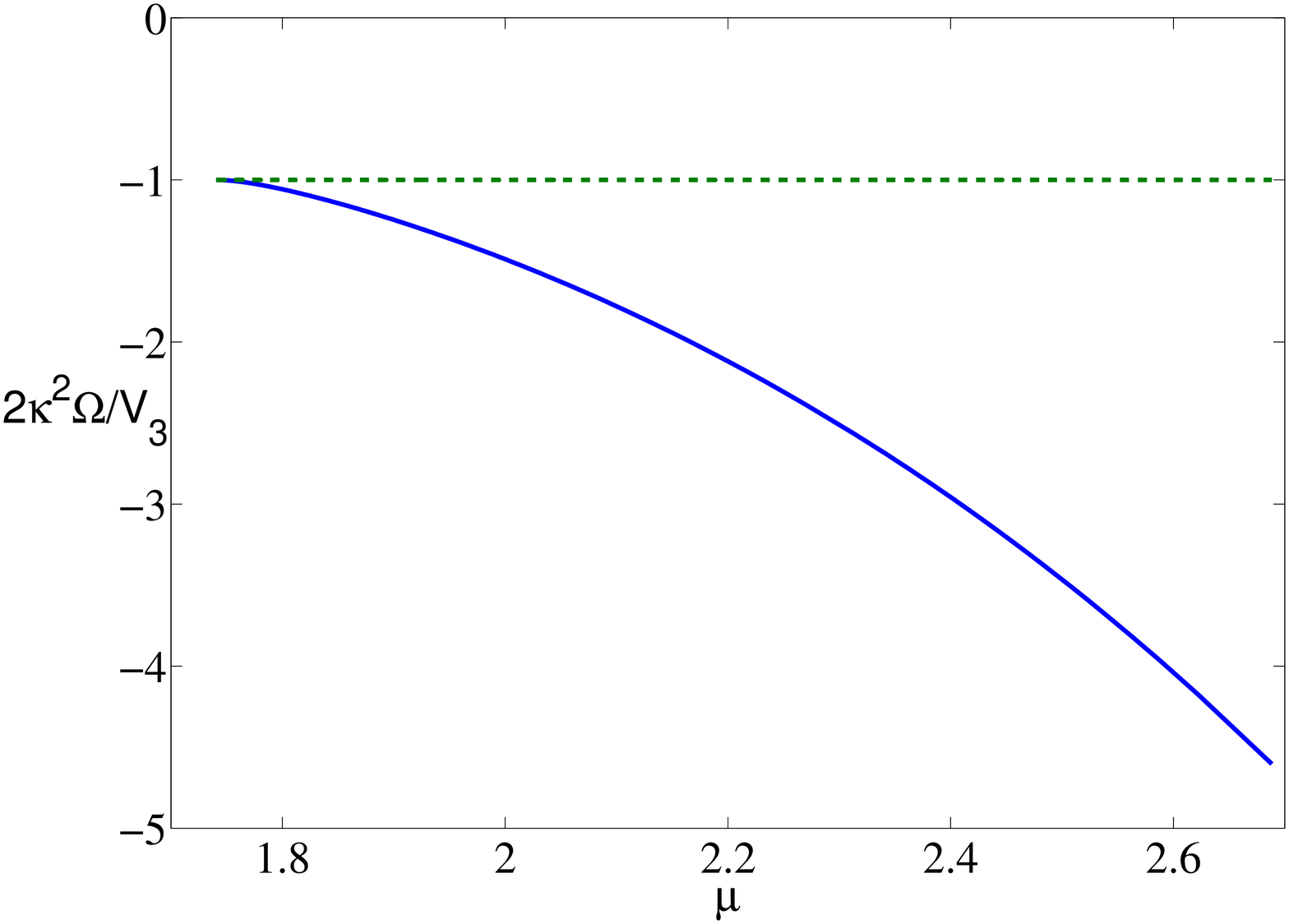}
  \includegraphics[width=0.5\textwidth]{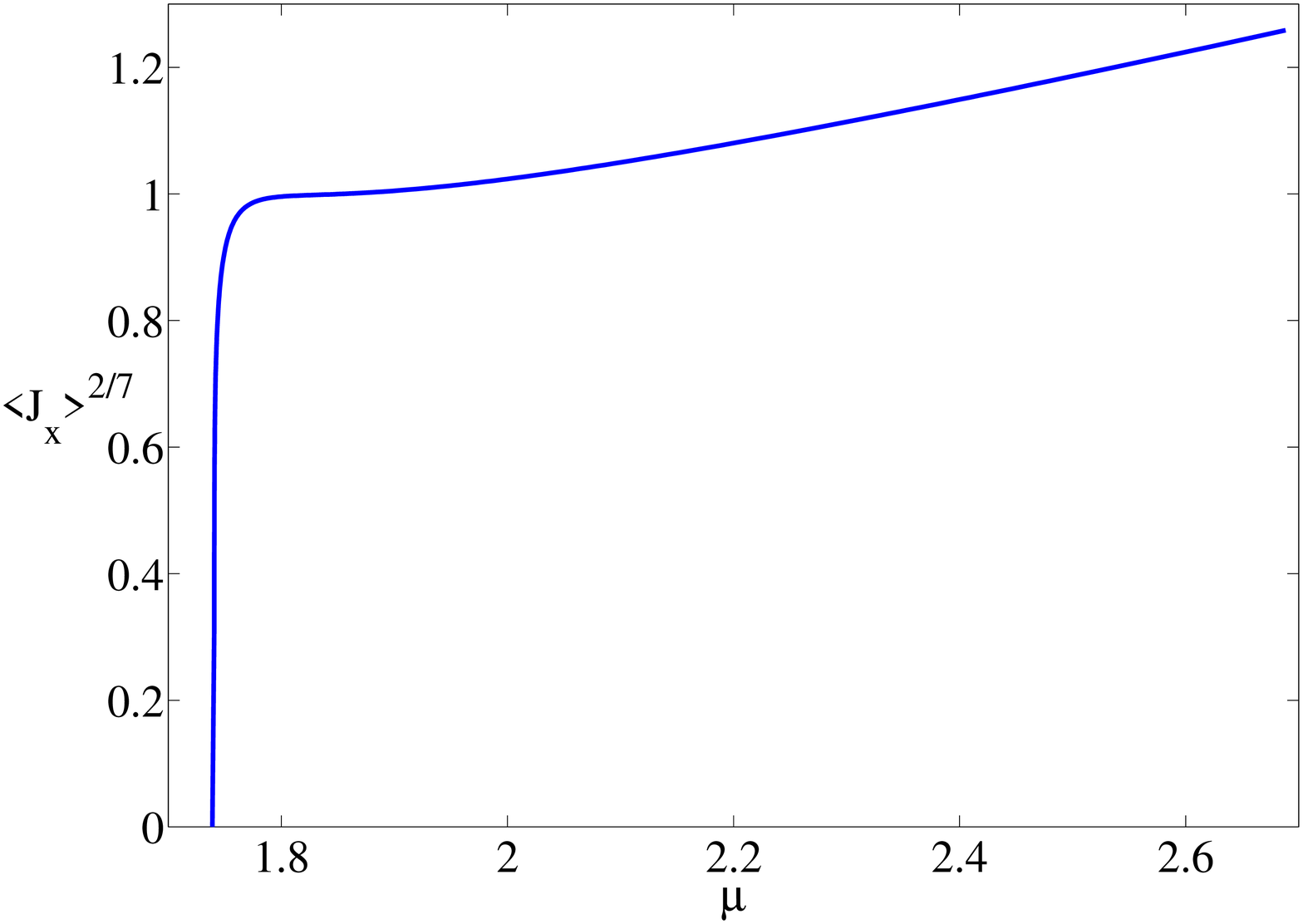}\\
  \caption{The grand potential and vacuum expectation value of vector operator with respect to the chemical potential in the case of $m^2=\frac54,~q=1.600$. In the left plot, the grand potentials of the pure soliton and hairy soliton are described by the green dashed line and blue solid line, respectively. The condensate begins at $\mu_{c1}\simeq1.7070$. The right plot shows the condensate of the vector operator $\langle \hat{J}_x\rangle\neq0$ when $\mu>\mu_{c1}$.}\label{m2qg4q1.6}
\end{figure}

For $q\geq q_3$, we take $q=1.600$ as a typical example. Figure~\ref{m2qg4q1.6} shows the grand potential and condensate as a function of chemical potential, from which one can see that $\langle\hat{J_x}\rangle$ arises continuously from zero at $\mu_{c1}$. The grand potential $\Omega$ in the left plot of figure~\ref{m2qg4q1.6} indicates that above $\mu_{c1}$ the configuration with non-vanishing vector ``hair" is indeed thermodynamically preferred to the normal phase. It is a second order phase transition with the critical behavior $\langle\hat{J_x}\rangle\sim(\mu/\mu_{c1}-1)^{1/2}$ near the critical point.

\begin{figure}
  \includegraphics[width=0.5\textwidth]{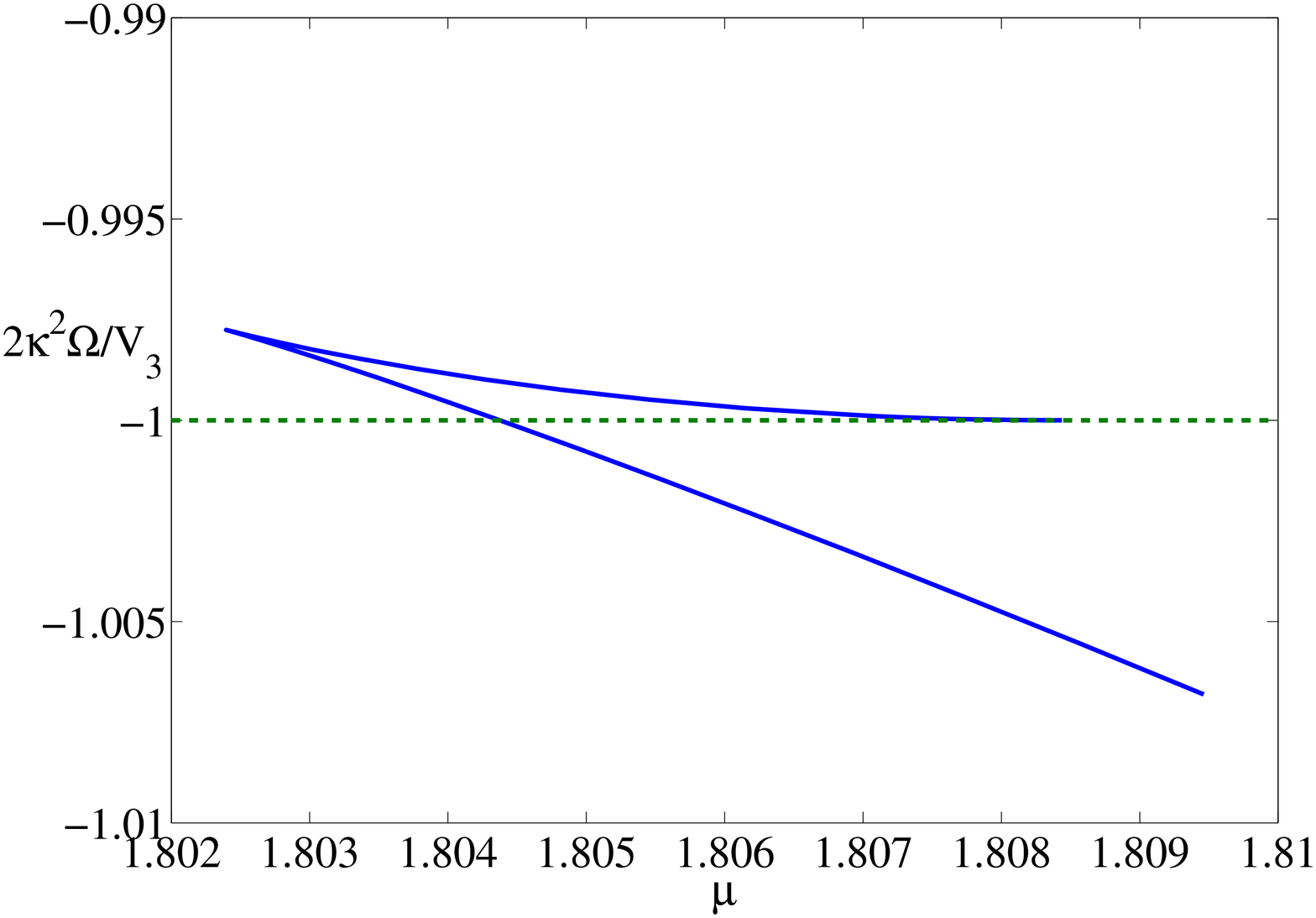}
  \includegraphics[width=0.5\textwidth]{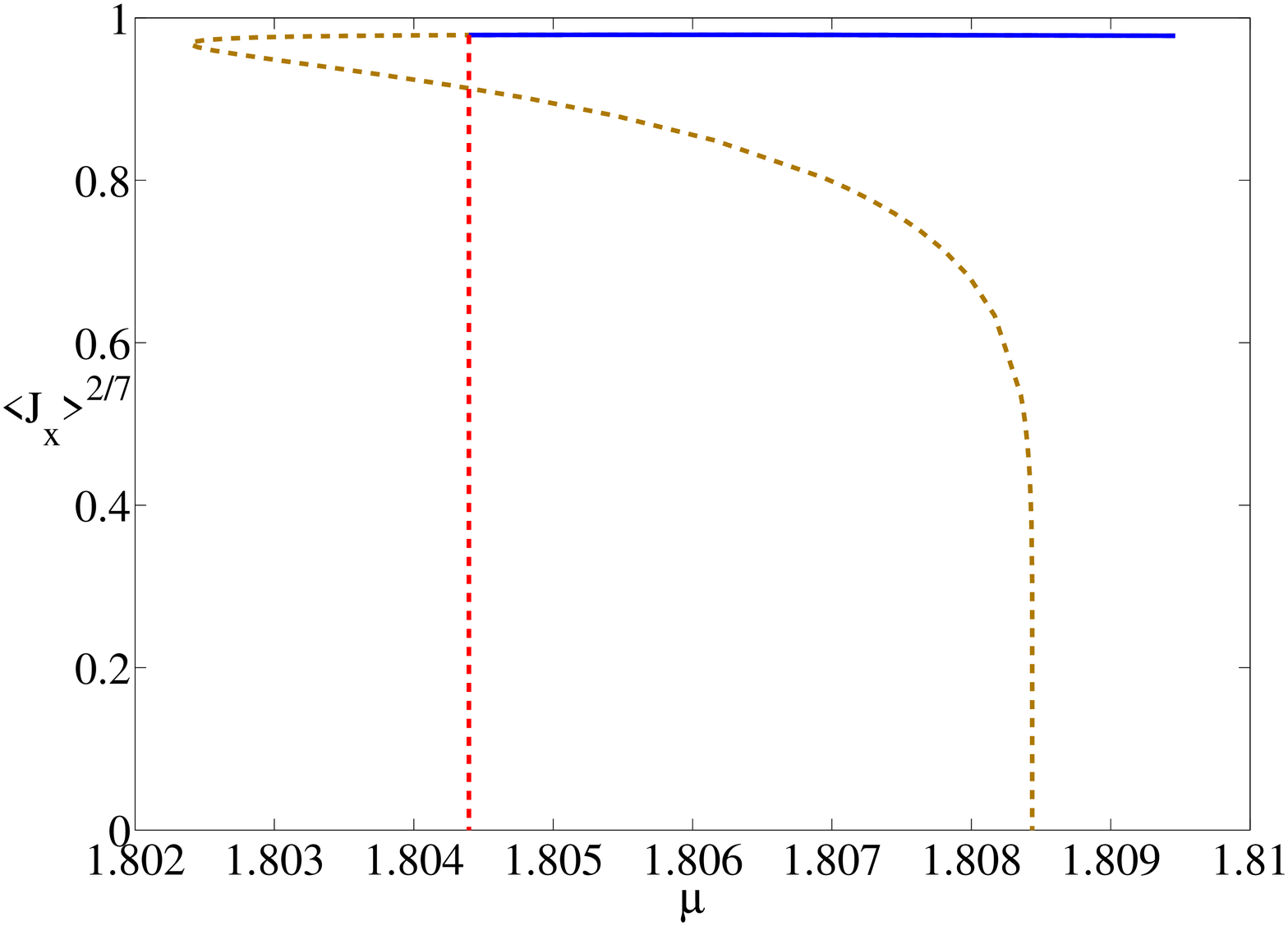}\\
  \caption{The grand potential and condensate with respect to the chemical potential in the case of $m^2=\frac54$, and $q=1.540$. In the left plot, the grand potentials of the pure AdS soliton and hairy soliton are denoted by the green dashed line and blue solid line, respectively. The condensate begins at $\mu_{c1}\simeq 1.8045$. In the right figure, the blue solid line stands for the physical part of the condensate value for the condensed phase, which shows the condensate has a jump from zero to the upper branch when $\mu>\mu_{c1}$. }\label{m2qg4q1.4}
\end{figure}

For the case $q_2\leq q<q_3$, such as $q=1.540$ in figure~\ref{m2qg4q1.4}, the transition from the normal phase to the condensed phase becomes first order. As one can see in figure~\ref{m2qg4q1.4}, the condensate with respect to chemical potential is multi-valued and the free energy develops a characteristic ``swallow tail". The condensate has a jump from zero to the upper branch of the hairy solution at $\mu_{c1}\simeq1.8054$.

\begin{figure}
  \includegraphics[width=0.5\textwidth]{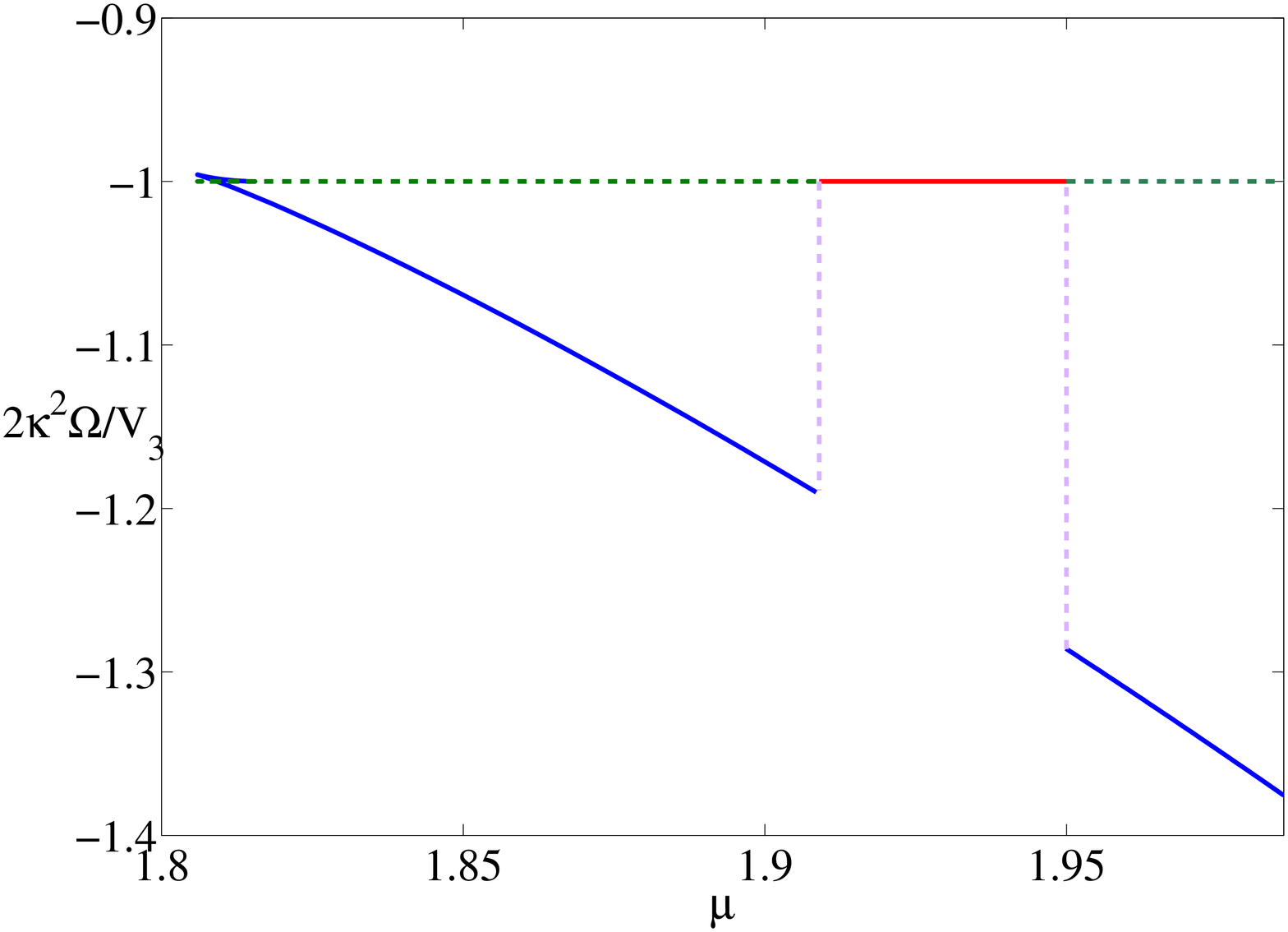}
  \includegraphics[width=0.5\textwidth]{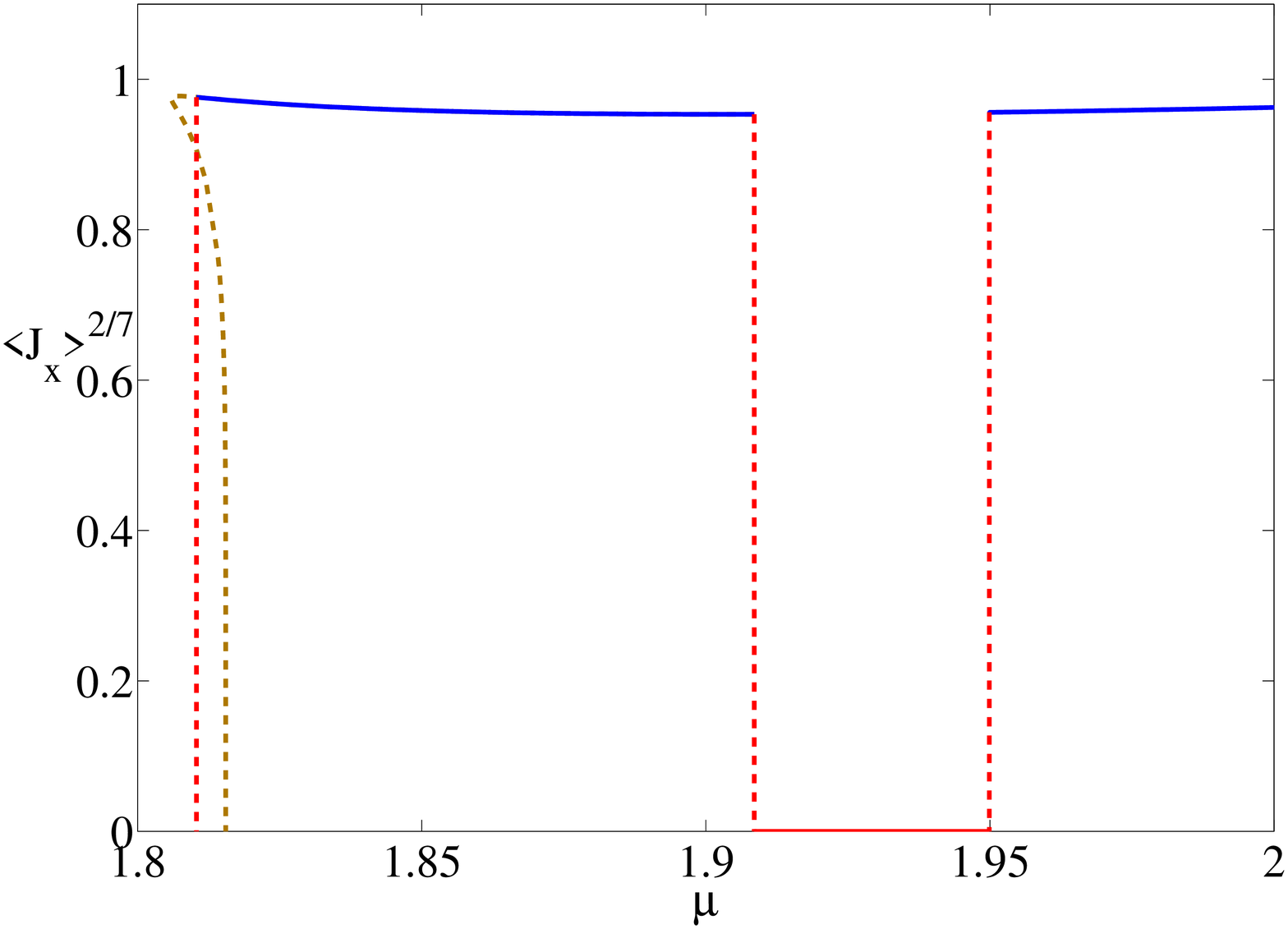}\\
  \caption{The grand potential and condensate with respect to the chemical potential in the case of $m^2=\frac54$ and $q=1.534$. In the left plot, the grand potentials of the pure AdS soliton and hairy soliton are denoted by the green dashed line and blue solid line, respectively. The condensate begins at $\mu_{c1}\simeq1.8094$. However, there does't exist the condensed solution when the chemical potential is in between  $\mu_{c2}\simeq1.9085$ and $\mu_{c3}\simeq1.9499$ (The red solid line region in the figure). The right plot shows the condensate has a jump from zero to the upper branch at $\mu=\mu_{c1}$. When the chemical potential is in the region with the red solid line, the condensate falls into zero again. }\label{m2qg4q1.534}
\end{figure}

There is an additional complication when we decrease the value of $q$ past $q_2$. A concrete example for $q=1.534$ is presented in figure~\ref{m2qg4q1.534}. As one increases the chemical potential, there is a first order transition at $\mu_{c1}\simeq1.8094$. However, if one continues increasing the chemical potential, there will be two discontinuities at $\mu_{c2}\simeq1.9085$ and $\mu_{c3}\simeq1.9499$, between which there does not exist any hairy solution. So there is a kind of zeroth order phase transition from the condensed phase to the normal phase at $\mu_{c2}$ and a kind of zeroth order phase transition from the normal phase to the condensed phase at $\mu_{c3}$. These kinds of zeroth order phase transitions do not come from thermodynamics but from the kinetics of the field equations~\eqref{eoms}. The details will be shown in appendixes~\ref{app1} and~\ref{app1b}.

 Recall that the pure AdS soliton resembles an insulating phase while the hairy soliton mimics a superconducting phase, our result suggests that there is a kind of ``insulator/superconductor/insulator/superconductor" phase transitions by increasing chemical potential. It is interesting to see whether there is any realistic material that can display such a phase structure.

\begin{figure}
  \includegraphics[width=0.5\textwidth]{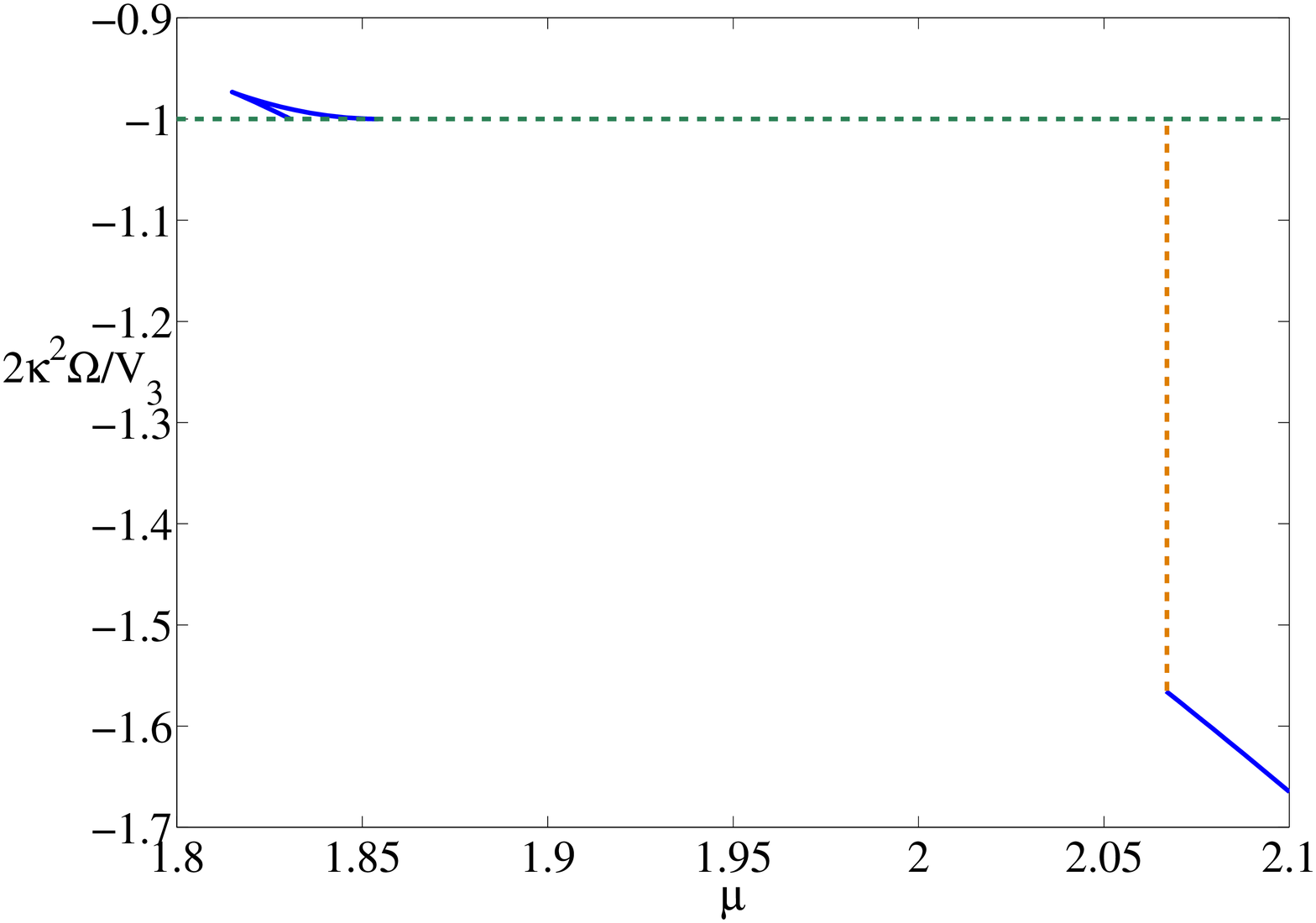}
  \includegraphics[width=0.5\textwidth]{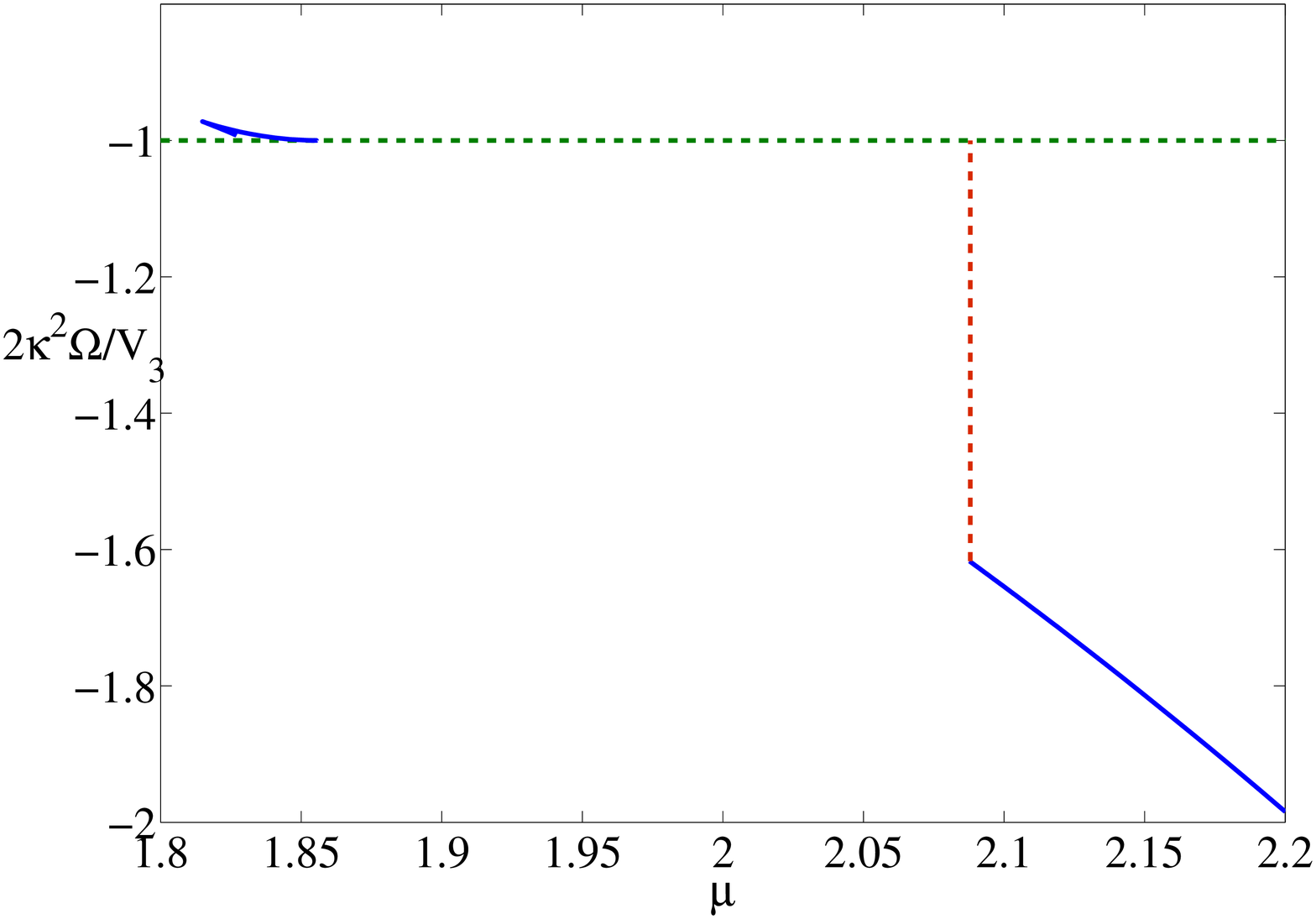}\\
  \caption{The grand potentials with respect to the chemical potential in the case of $m^2=\frac54,~q=q_{1}\simeq1.502$ (left plot) and $q=1.490<q_1$ (right plot). The grand potentials of the pure AdS soliton and hairy soliton are denoted by the green dashed line and blue solid line, respectively. In the case with $q\simeq1.502$, the first point of discontinuity of the grand potential coincides with the first order phase transition point. The right plot shows the case of $q<q_{1}$. Note that only the part which has the lowest free energy is physical. }\label{m2qg4q1.502}
\end{figure}

\begin{figure}
  \includegraphics[width=0.5\textwidth]{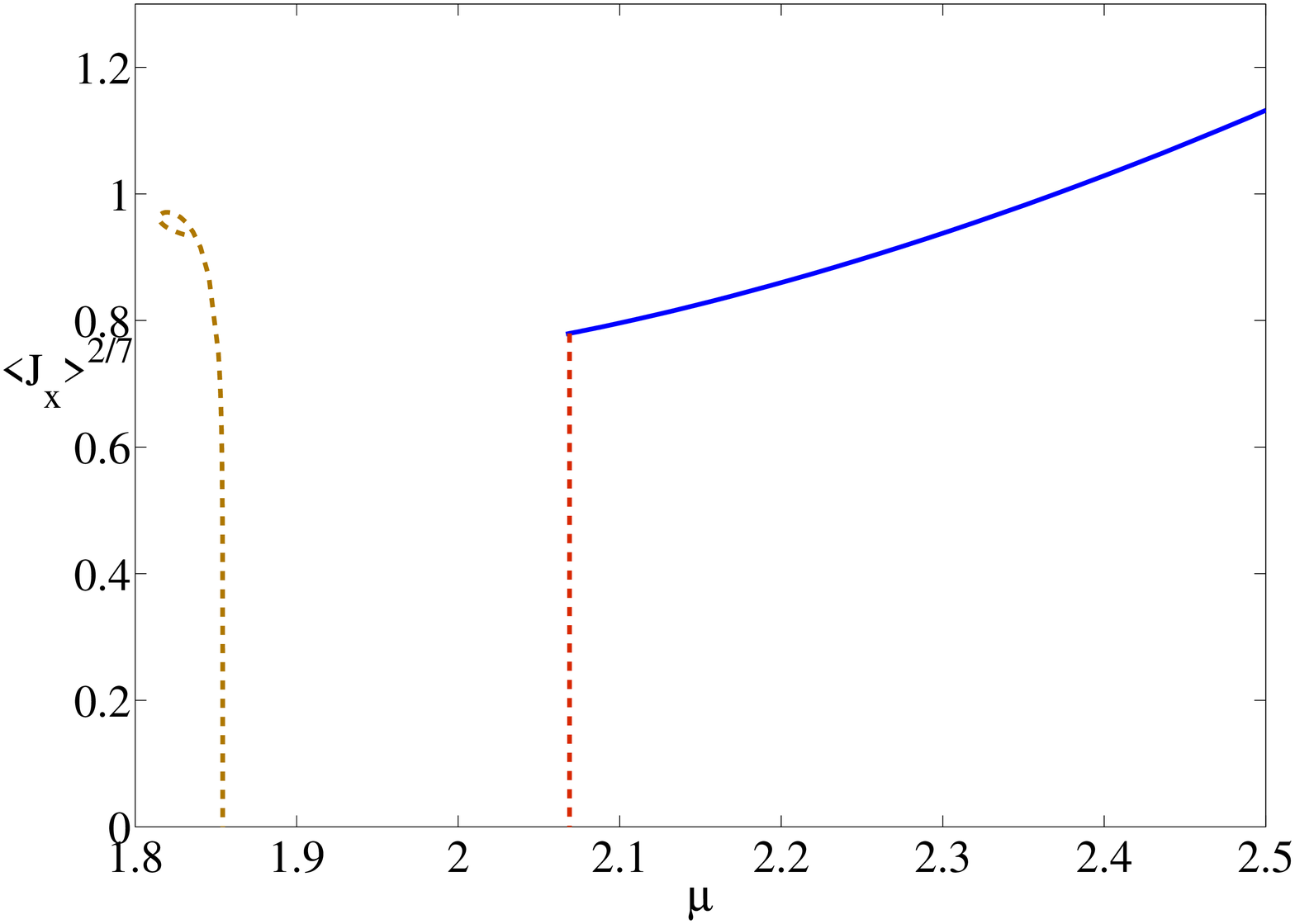}
  \includegraphics[width=0.5\textwidth]{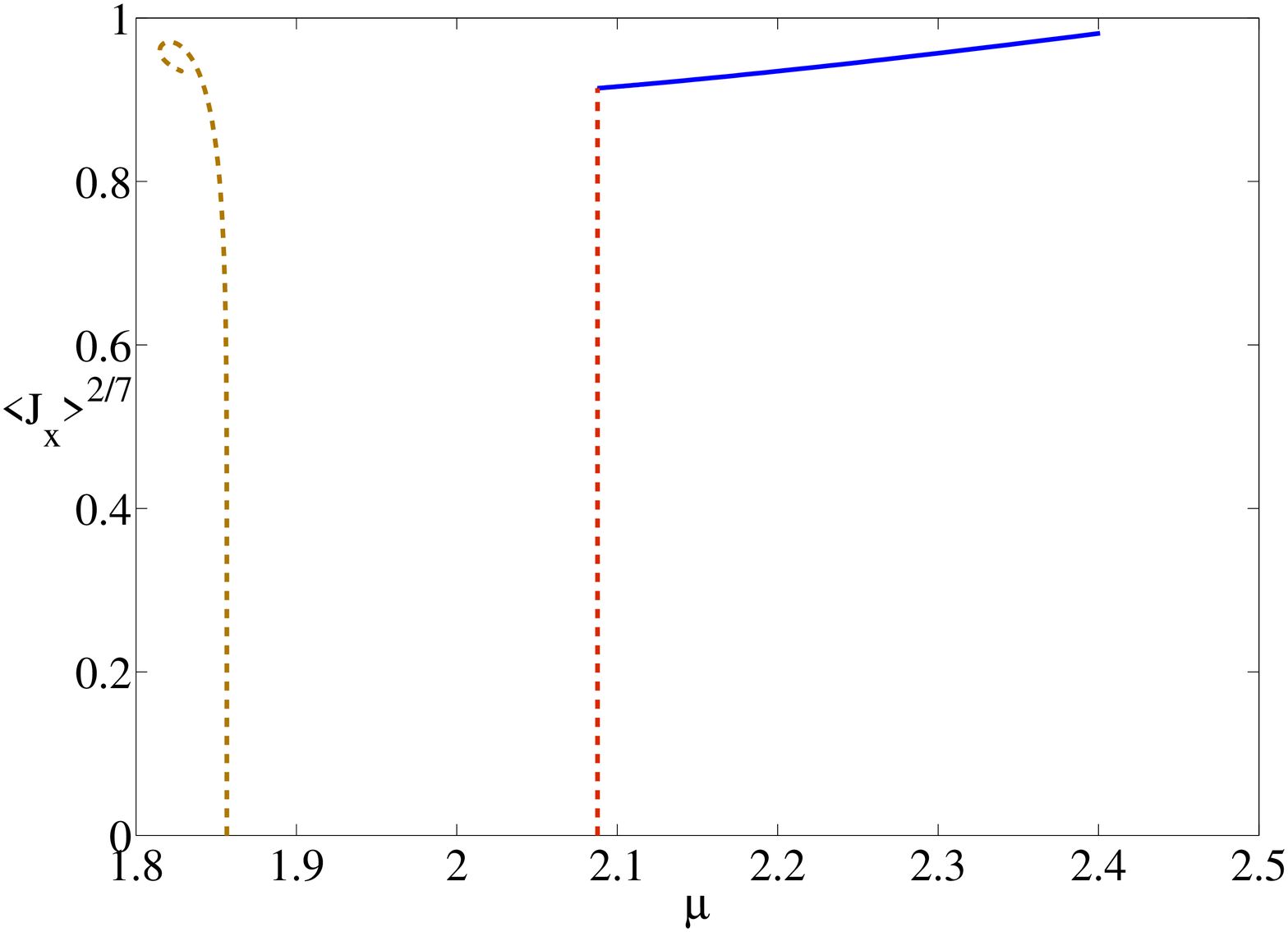}\\
  \caption{The condensate of the vector operator in the case of $m^2=\frac54,~q=q_{1}\simeq1.502$ (left plot) and $q=1.490 <q_1$ (right plot). For the condensed phase, only the blue line is thermodynamically favored, while the dashed orange part is un-physical. Thus, the condensate has a sudden jump form zero to a nonzero value. There exists a zeroth order phase transition from the normal phase to the condensed phase. }\label{m2qJq1.502}
\end{figure}

As we continuously decrease $q$, the value of $\mu_{c1}$ decreases while the value of $\mu_{c2}$ increases. There exists a critical value of charge $q_1\simeq1.5020$ at which the first order phase transition point $\mu_{c1}$ coincides with the first zeroth order point $\mu_{c2}$. In this case,  the first part of condensed phase such as in  figure~\ref{m2qg4q1.534} disappears (see figures~\ref{m2qg4q1.502} and~\ref{m2qJq1.502}). When the value of $q$ is less than $q_1$, the first order transition point disappears and the first discontinuous point $\mu_{c2}$ appears above the line $2\kappa^2\Omega/V_3=-1$, thus is un-physical. This case for $q<q_1$ is shown in figure~\ref{m2qg4q1.502} with $q=1.490$. There is only a zeroth order transition at critical point $\mu_{c1}\simeq1.9499$.

\begin{figure}
\centering
  \includegraphics[width=0.5\textwidth]{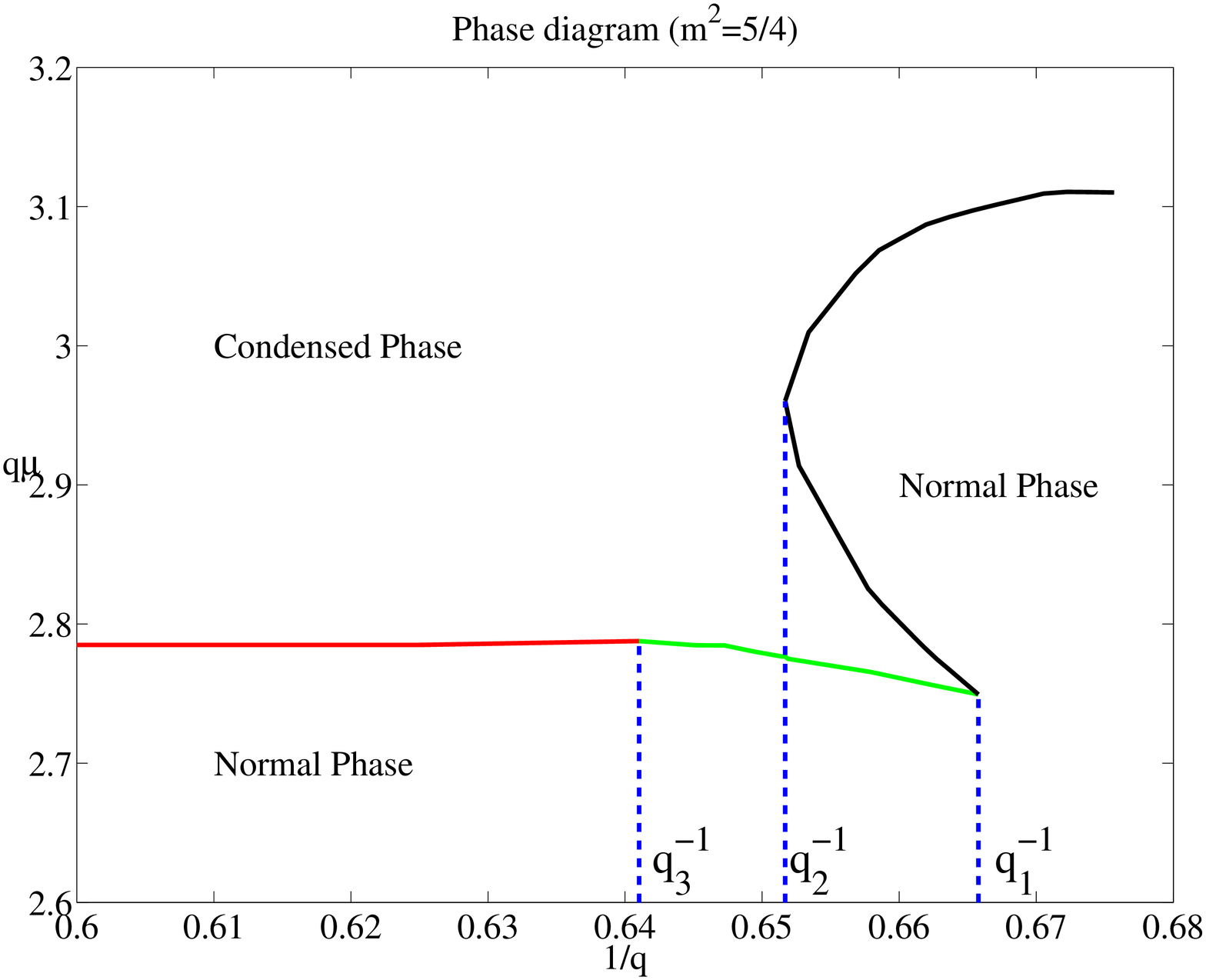}
  \caption{The phase diagram for the case with $m^2=5/4$. The red line stands for the second order phase transition, the green line stands for the first order phase transition, and the black line stands for the zeroth order phase transition. Numerical results show $\mu q\simeq2.7852$ when $1/q$ is close to zero.}\label{PD1}
\end{figure}

The ($\mu$-$q$) phase diagram for the case with $m^2=5/4$ is drawn in figure~\ref{PD1}. One should note that the horizontal axis and vertical axis in the figure are labeled by $1/q$ and $q\mu$,
respectively. It is convenience to compare with the results in the probe limit, where we take the limit $q\rightarrow\infty$ keeping $q\rho_\mu$ and $q A_\mu$ fixed. Therefore, in order to compare our results to the case in the probe analysis, we should make the scaling transformation $\rho_\mu\rightarrow q\rho_\mu$ and $A_\mu\rightarrow qA_\mu$. Under such transformation, the chemical potential $\mu$ becomes $q\mu$ and the temperature $T$ becomes $T/q$. Therefore, they are the quantities $q\mu$ and $T/q$ that are comparable to those in the probe limit. One can see from figure~\ref{PD1} that for sufficiently large $q$, the critical value of $q\mu$ approaches to a constant, which quantitatively agrees with the probe analysis in appendix~\ref{app3}. Furthermore, let us mention that the zeroth order phase transition can only appear in the strong back reaction case. Finally in some range of $q$, one can first see one first order transition and then two zeroth order transitions with the increase of chemical potential.

\subsubsection{Phase transition for $m^2_{c2}\leq m^2\leq m^2_{c1}$}
\label{sect:superconducor3}
In the case of $m^2_{c2}\leq m^2\leq m^2_{c1}$, we find two critical values of $q$ denoted as $q_a$ and $q_b$ with $q_a<q_b$. For small back reaction with $q>q_b$, there is a second order  insulator/superconductor transition at some critical  chemical potential $\mu_{c1}$. As we decrease $q$ to the range $q_a<q<q_b$, the order of transition becomes first order. For larger strength of back reaction with $q<q_a$, the condensed phase appears through a zeroth order phase transition. The main results are shown in table~\ref{Tab3}. The concrete values of $q_a$ and $q_b$ depend on $m^2$. To give more details, we shall consider a typical example with $m^2=0$ in this subsection. The two critical values of $q$ for $m^2=0$ is $q_a\simeq1.172$ and $q_b\simeq1.317$.
\begin{table}
  \centering
  \begin{tabular}{|c|c|c|c|}
    \hline
    Charge &\multicolumn{3}{|c|}{ Phase transition and its order,~~$m^2_{c2}\leq m^2\leq m^2_{c1}$} \\
    \hline
    $q\geq q_b$ & $\mu<\mu_{c1}$, S& $\mu=\mu_{c1},2^{nd}$ & $\mu>\mu_{c1}$, SC \\
    \hline
    $q_a\leq q<q_b$ &$\mu<\mu_{c1}$, S& $\mu=\mu_{c1} ,1^{st}$& $\mu>\mu_{c1}$, SC \\
    \hline
    $q<q_a$ & $\mu<\mu_{c1}$, S& $\mu=\mu_{c1} ,0^{th}$& $\mu>\mu_{c1}$, SC \\
    \hline
  \end{tabular}
  \caption{The phase transition and its order with respect to the charge and chemical potential  in the case of $m^2_{c2}\leq m^2\leq m^2_{c1}$. In the table, S=normal phase/insulator with pure AdS soliton solution, SC=condensed phase/superconductor with hairy soliton
  solution. $0^{th},1^{st}$, and~$2^{nd}$ stand for the zeroth order, first order and second order phase transition, respectively.}\label{Tab3}
\end{table}

\begin{figure}
  \includegraphics[width=0.5\textwidth]{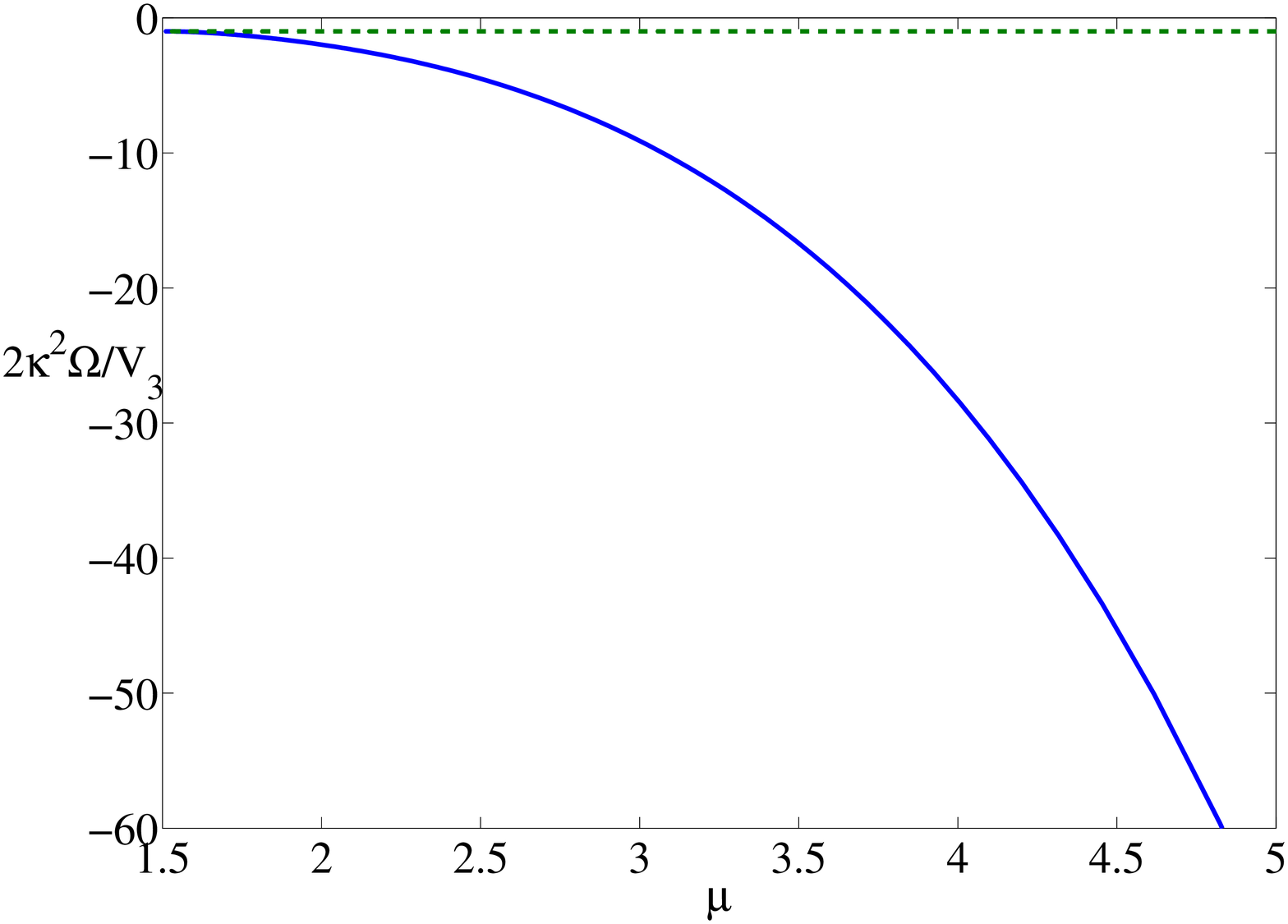}
  \includegraphics[width=0.5\textwidth]{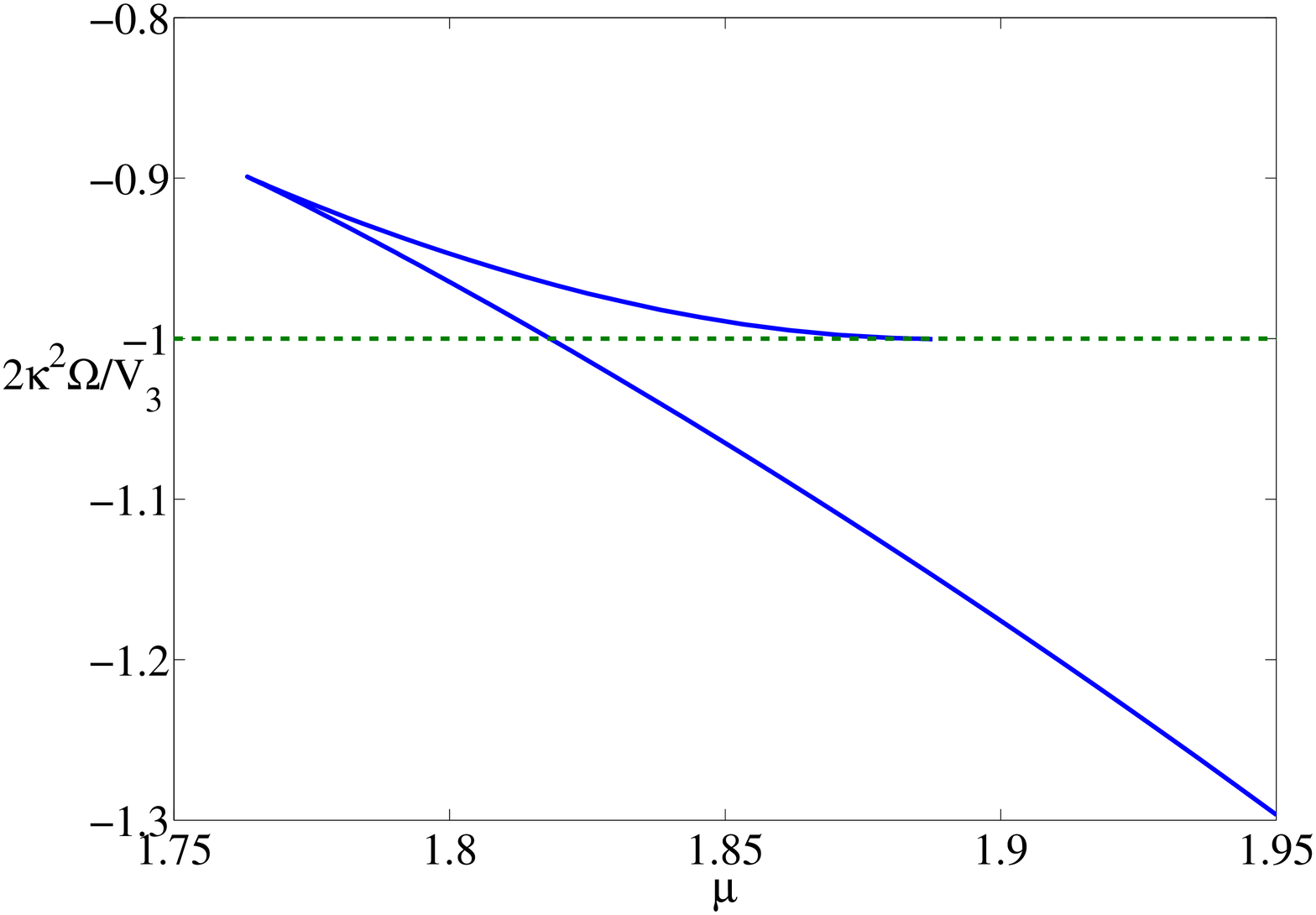}\\
  \caption{The grand potentials with respect to the chemical potential in the case of $m^2=0$, $q=1.500$ (left) and $q=1.200$ (right). In both figures, the grand potentials of the pure AdS soliton and hairy soliton are described by the green dashed line and blue solid line, respectively. In the left plot, the condensate begins at $\mu_{c1}\simeq1.5103$, while in the right plot the condensate begins at $\mu_{c1}\simeq1.8207$. }\label{m2bg4q1.5}
\end{figure}

\begin{figure}
  \includegraphics[width=0.5\textwidth]{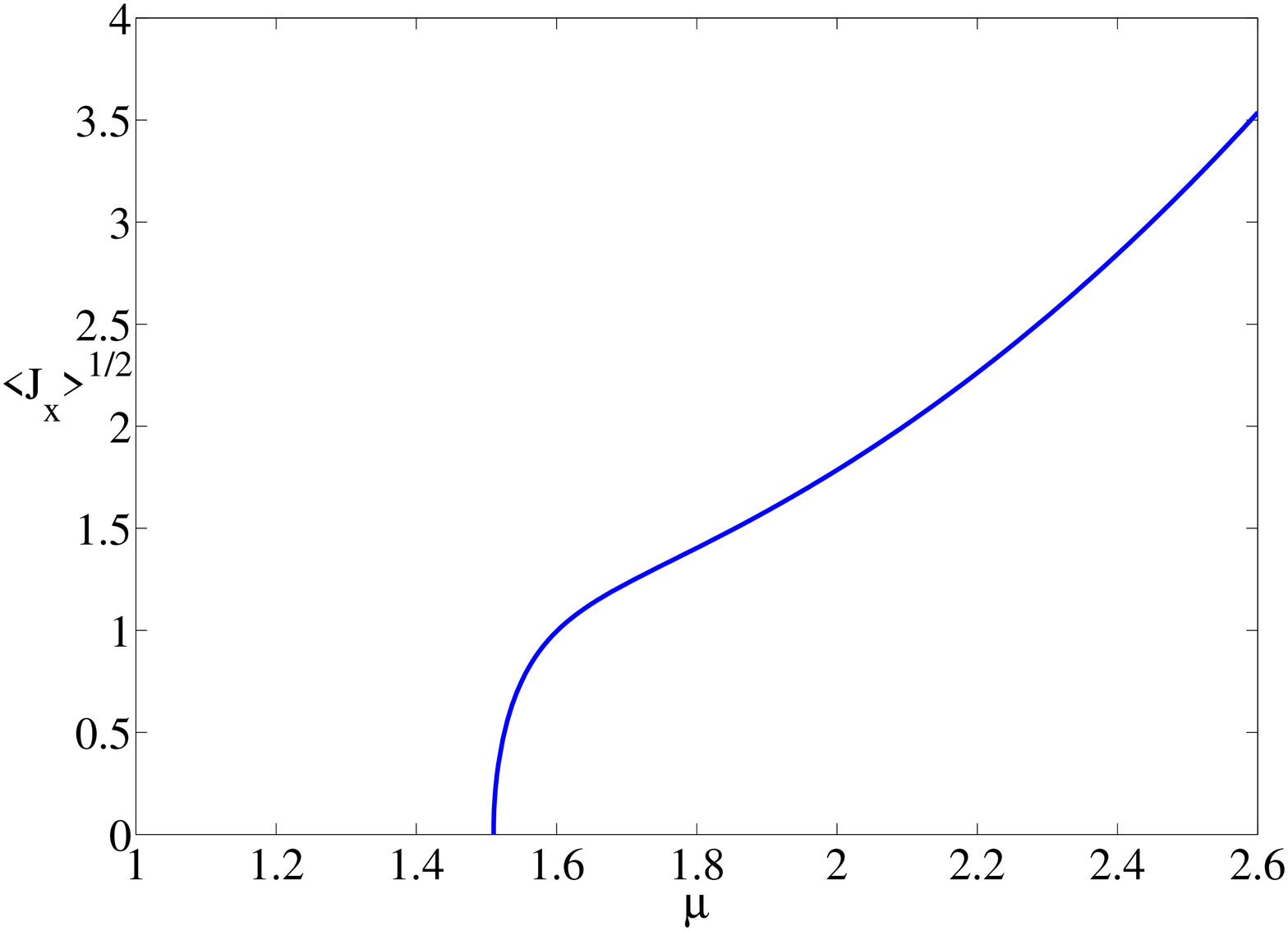}
  \includegraphics[width=0.5\textwidth]{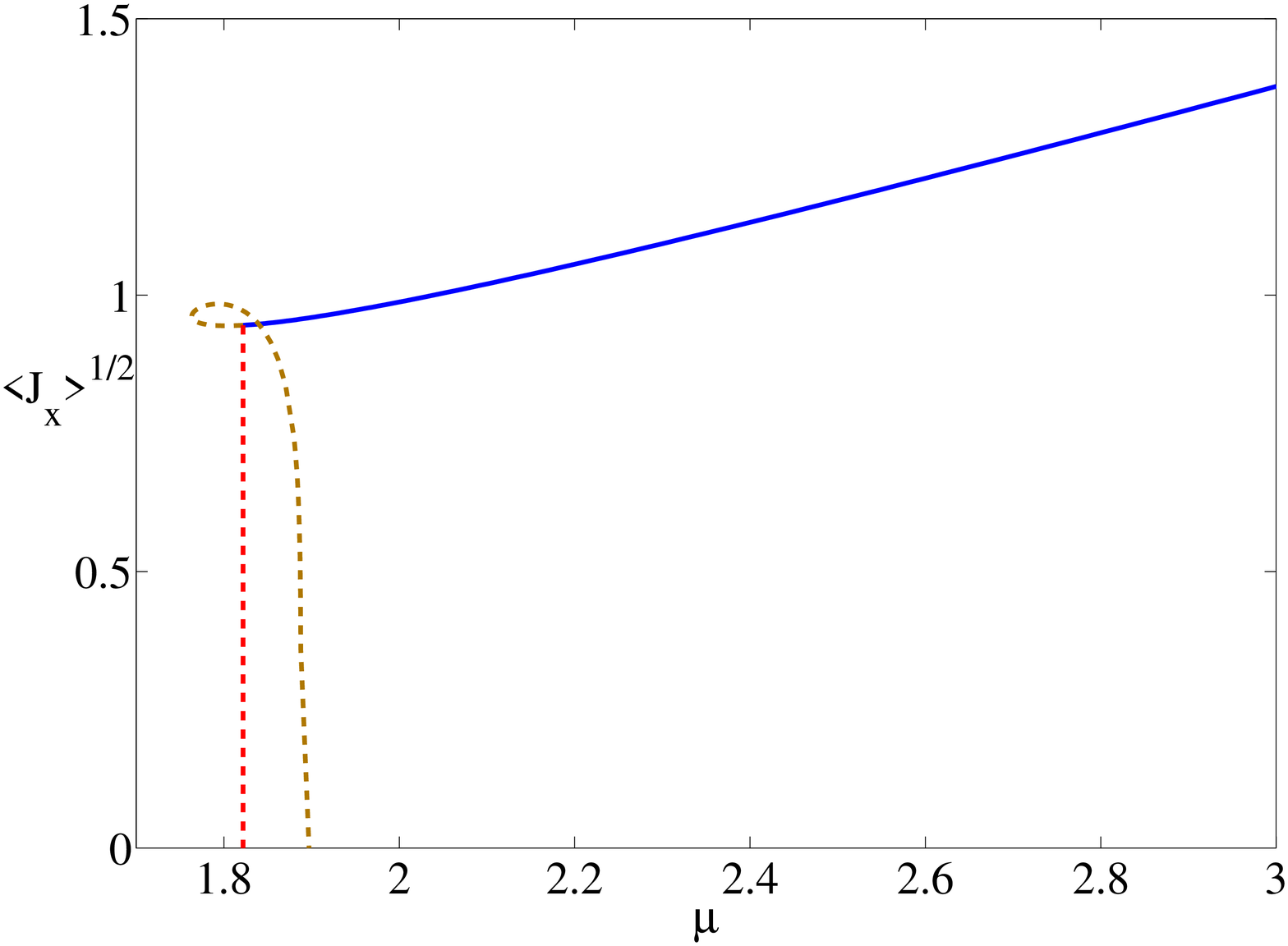}\\
  \caption{The condensate of the vector operator in the case of $m^2=0,~q=1.500$ (left plot) and $q=1.200$ (right plot).
   The left plot shows a second order phase transition, while the right plot shows a first order phase transition from the normal phase to the condensed phase. In the right figure, only the blue part is physical.}\label{m2bJq1.15}
\end{figure}

For $q>q_b$, the situation is very similar to the case with $q>q_3$ in $m^2=5/4$. There exists a critical chemical potential $\mu_{c1}$ at which a second order transition happens. As we go on decreasing $q$ past $q_b$, the transition from the normal phase to the condensed phase is first order. The behaviors of the grand potential and condensate are shown in figures~\ref{m2bg4q1.5} and \ref{m2bJq1.15}, respectively.

\begin{figure}
  \includegraphics[width=0.5\textwidth]{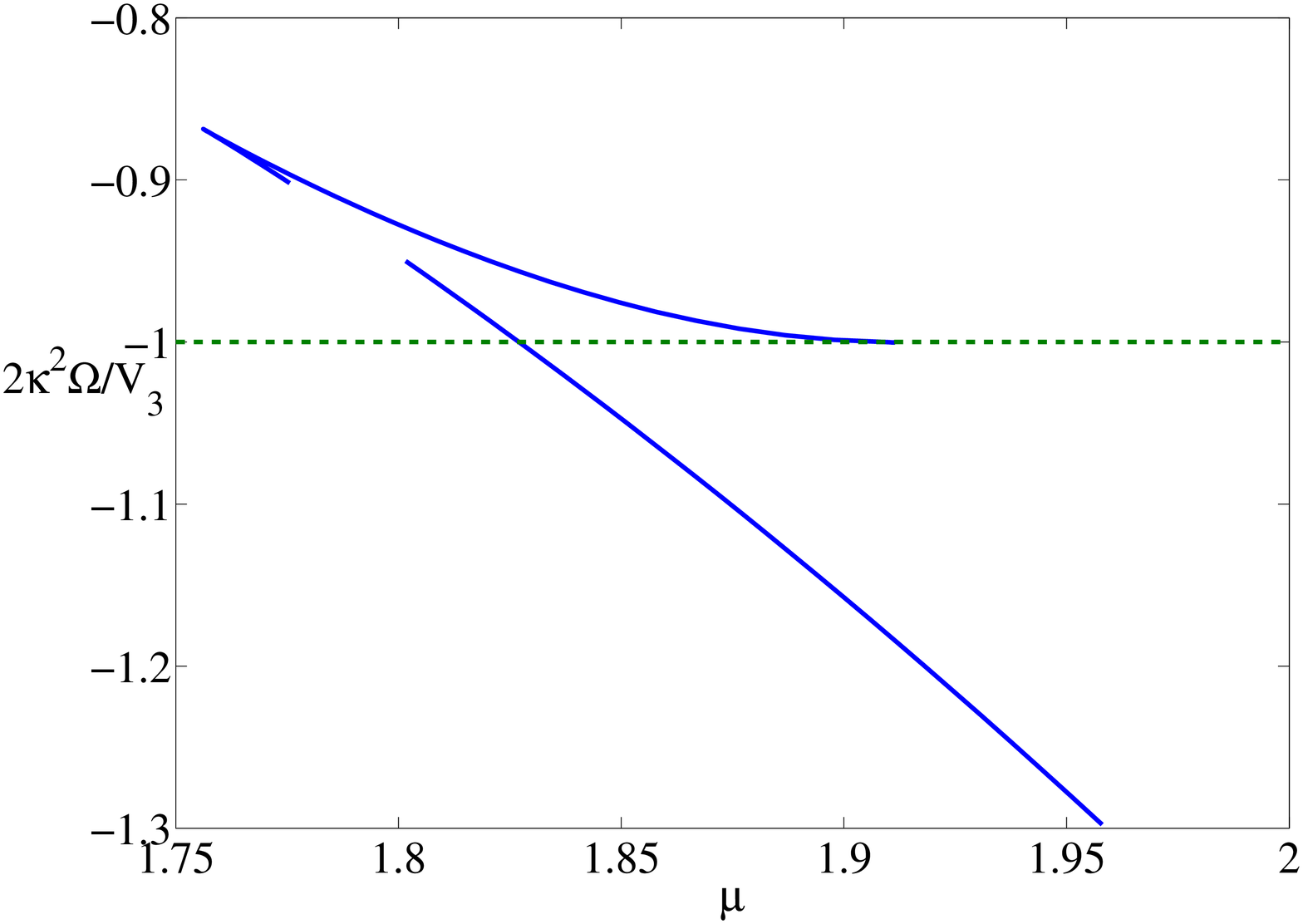}
  \includegraphics[width=0.5\textwidth]{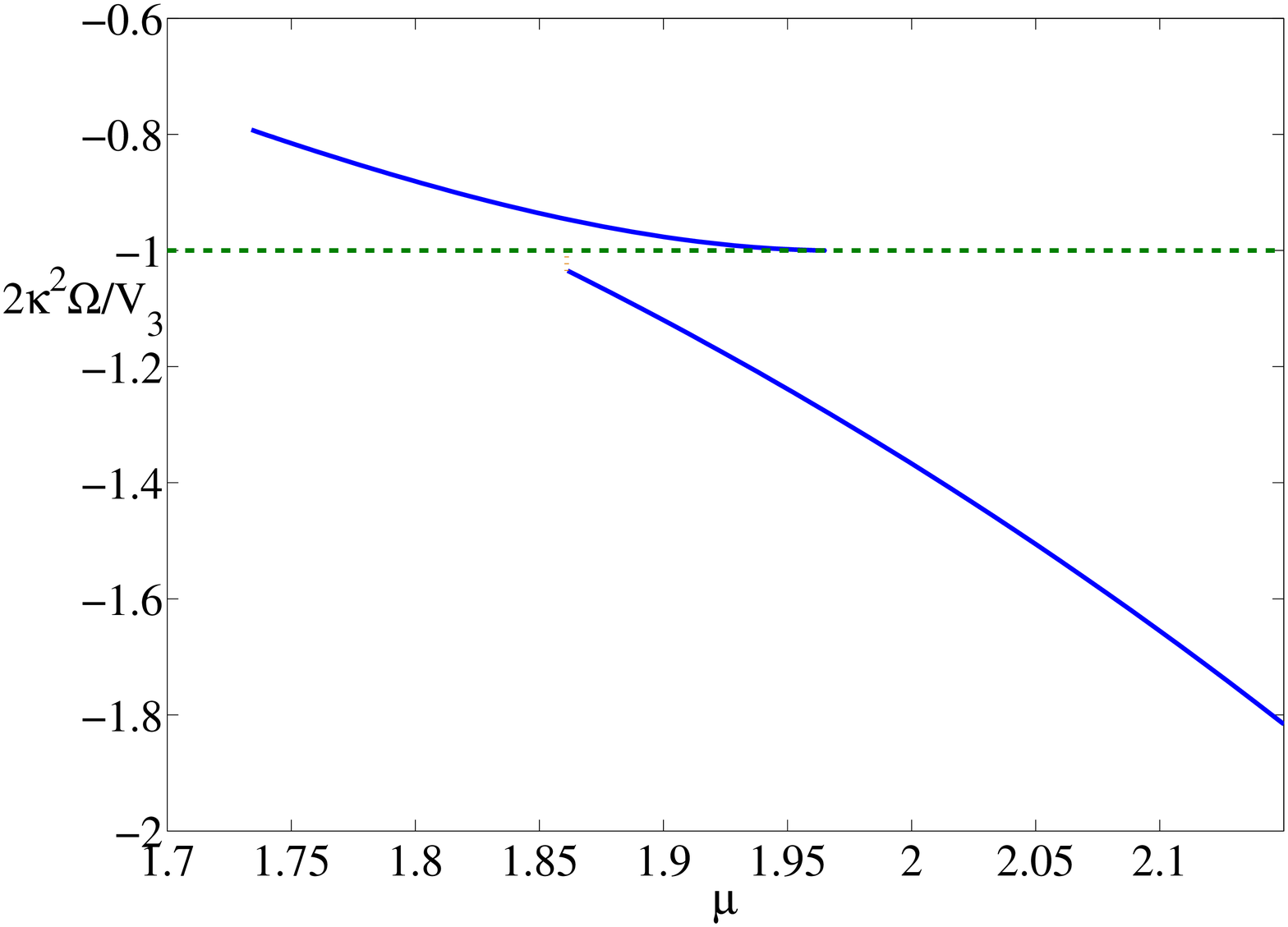}\\
  \caption{The grand potentials with respect to the chemical potential in the case of $m^2=0$, $q=1.185\in(q_a,q_b)$ (left plot) and $q=1.150<q_a$ (right plot). The grand potentials of the pure AdS soliton and hairy soliton are denoted by the green dashed line and blue solid line, respectively. In the case of $q=1.185$, the discontinuity  of the grand potential appears above the normal phase line.  In the right plot for the case of $q=1.150$, one discontinuous point at $\mu_{c1}=1.8597$ is under the normal phase line $2\kappa^2\Omega/V_3=-1$. }\label{m2qg4q1.15}
\end{figure}

 When $q$ is slightly below $q_b$, the curve of condensate (or grand potential) is continuous, while as $q$ is a little larger than $q_a$, the curve becomes discontinuous. In the previous subsection, this discontinuity leads to two zeroth order transitions for $q_1<q<q_2$ (see figure~\ref{m2qg4q1.4}) and one zeroth order transition for $q\leq q_1$ (see figure~\ref{m2qg4q1.502}). In the present case with $q \ge q_a$, although there exist two discontinuous points in the grant potential curve denoted as upper point with larger grand potential and lower point with smaller grand potential, as shown in the left plot of figure~\ref{m2qg4q1.15}, both points appear in the region where the grand potential of the condensed solution is larger than the one in the normal phase, thus are thermodynamically disfavored and will not appear in the physical phase space. So we are left with only a first order transition. The lower point moves down as $q$ decreases and finally its grand potential becomes smaller than the one in the normal phase when $q<q_a$ (see the right plot of figure~\ref{m2qg4q1.15}). Take $q=1.150<q_a$ as an example, the grand potential presented in figure~\ref{m2qg4q1.15} shows clearly that the value of $\Omega$ has a sudden jump from the normal phase to the condensed phase at $\mu_{c1}\simeq 1.8597$, indicating a zeroth order phase transition.


\begin{figure}
\centering
  \includegraphics[width=0.5\textwidth]{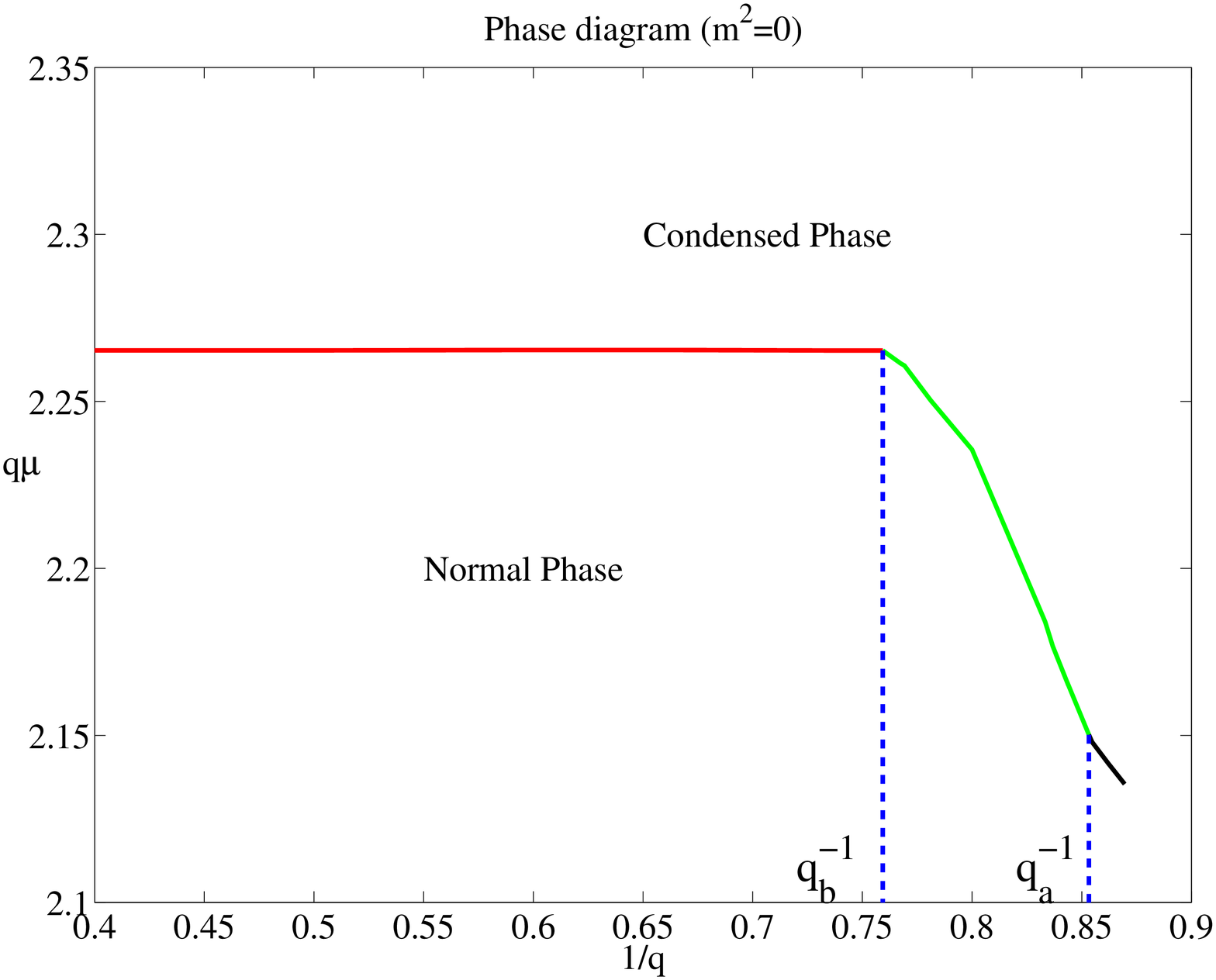}
  \caption{The phase diagram for the case of $m^2=0$. The red, green and black lines stand for the second order, first order and zeroth order phase transitions, respectively.. Numerical results show $\mu q\simeq2.2653$ when $1/q$  approaches to zero.}\label{PD2}
\end{figure}

A phase diagram in terms of $\mu$ and $q$ for the case with $m^2=0$ is shown in figure~\ref{PD2}. Indeed, for large enough $q$, the critical value of $q\mu$ approaches to a constant, which is consistent with the probe limit shown in appendix~\ref{app3}. This phase diagram is reminiscent of the case in the SU(2) holographic p-wave superconductor/insulator model presented in figure~12 in ref.~\cite{Cai:2013oma}.  Comparing our model with the special case $m^2=0$ to the SU(2) model, we find that the equations of motion in our setup look very similar with each other and the asymptotical expansions near the boundary are the same. Actually, we have carefully checked our numerical calculation in ref.~\cite{Cai:2013oma} and found that similar discontinuous points in the grand potential also appear for the back reaction larger than a certain value. More precisely, the discontinuity first appears with grand potential larger than the normal phase, thus can not change the phase structure. However, for very large back reaction, i.e. larger value of $\alpha$ in ref.~\cite{Cai:2013oma}, the lower discontinuous point enters the  region which has a lower grand potential than the one in the insulating phase. Thus we can also find a zeroth order transition from the normal phase to the condensed phase in the SU(2) p-wave model. If we add such a zeroth order transition region to the phase diagram in the SU(2) p-wave model, the figure~12 in ref.~\cite{Cai:2013oma} is quite similar to figure~\ref{PD2} in the present paper. This result is consistent with our probe analysis in the presence of a magnetic field that the Einstein-Maxwell-complex vector field model is a generalization of the SU(2) p-wave model in the sense that the vector field has general $m^2$~\cite{Cai:2013kaa}.

%

\subsubsection{Phase transition for $m^2<m^2_{c2}$}
\label{sect:superconducor2}

When $m^2$ is less than $m^2_{c2}$, there are two $m^2$-dependent critical values of charge $q_c$ and $q_d$ ($q_c<q_d$). The possible phase transitions and their orders are shown in table~\ref{Tab2}. For the above two cases with $m^2\geq m^2_{c2}$, the condensed phase seems to survive even to the limit $\mu\rightarrow\infty$.~\footnote{Due to the precision of numerical control, it becomes much difficult to solve the system numerically for very large chemical potentials. Nevertheless, we have scanned a wide range of chemical potential up to $10^3$ and found no evidence that the curves of grand potential and condensate would turn back to the small chemical potential region.} In contrast, in the case with $m^2<m^2_{c2}$, the hairy solution can not exist above a finite chemical potential. For the weak back reaction with $q>q_d$, we first meet a second order transition at $\mu_{c1}$ from the normal phase to the condensed phase, and then a zeroth order transition from the condensed phase back to the normal phase
 at $\mu_{c2}$. In the range $q_c<q<q_d$, the only difference is that the phase transition happening at $\mu_{c1}$ becomes first order. When we decrease the value of $q$ less than $q_c$, the condensed phase only appears at chemical potentials smaller than a critical one $\mu_{c1}$ and has grand potential larger than the normal phase. The latter means the normal phase is thermodynamical preferred. We take $m^2=-3/4$ as a typical case in the following.
\begin{table}
  \centering
  \begin{tabular}{|c|c|c|c|c|c|}
    \hline
    Charge & \multicolumn{5}{|c|}{Phase transition and its order,~~$m^2<m^2_{c2}$} \\
    \hline
    $q\geq q_d$ & $\mu<\mu_{c1}$, S & $\mu=\mu_{c1}$, $2^{nd}$ & $\mu>\mu_{c1}$, SC & $\mu=\mu_{c2}$,$0^{th}$ & $\mu>\mu_{c2}$,S \\
    \hline
    $q_c\leq q<q_d$ & $\mu<\mu_{c1}$, S & $\mu=\mu_{c1}$, $1^{st}$ & $\mu>\mu_{c1}$, SC & $\mu=\mu_{c2}$,$0^{th}$ & $\mu>\mu_{c2}$,S\\
    \hline
    $q<q_c$ & \multicolumn{5}{|c|}{S} \\
    \hline

  \end{tabular}
  \caption{The phase transition and its order with respect to the charge and  chemical potential in the case of $m^2<m^2_{c2}$. In the table, S=normal phase/insulator with pure AdS soliton solution, SC=condensed phase/superconductor with hairy soliton solution. $0^{th}$, $1^{st}$, and~$2^{nd}$ stand for the zeroth order, first order, and second order phase transition, respectively.}\label{Tab2}
\end{table}

\begin{figure}
  \includegraphics[width=0.5\textwidth]{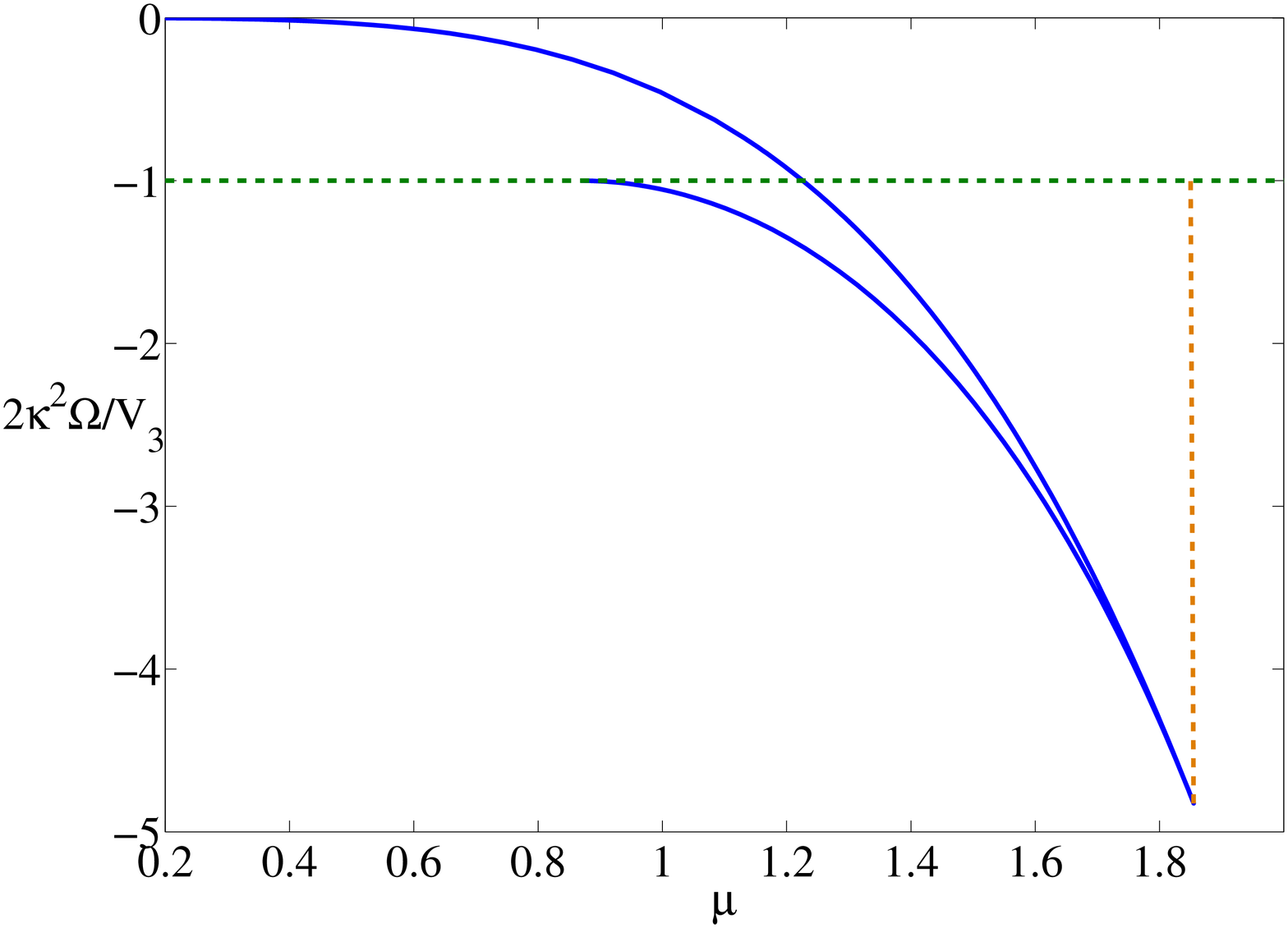}
  \includegraphics[width=0.5\textwidth]{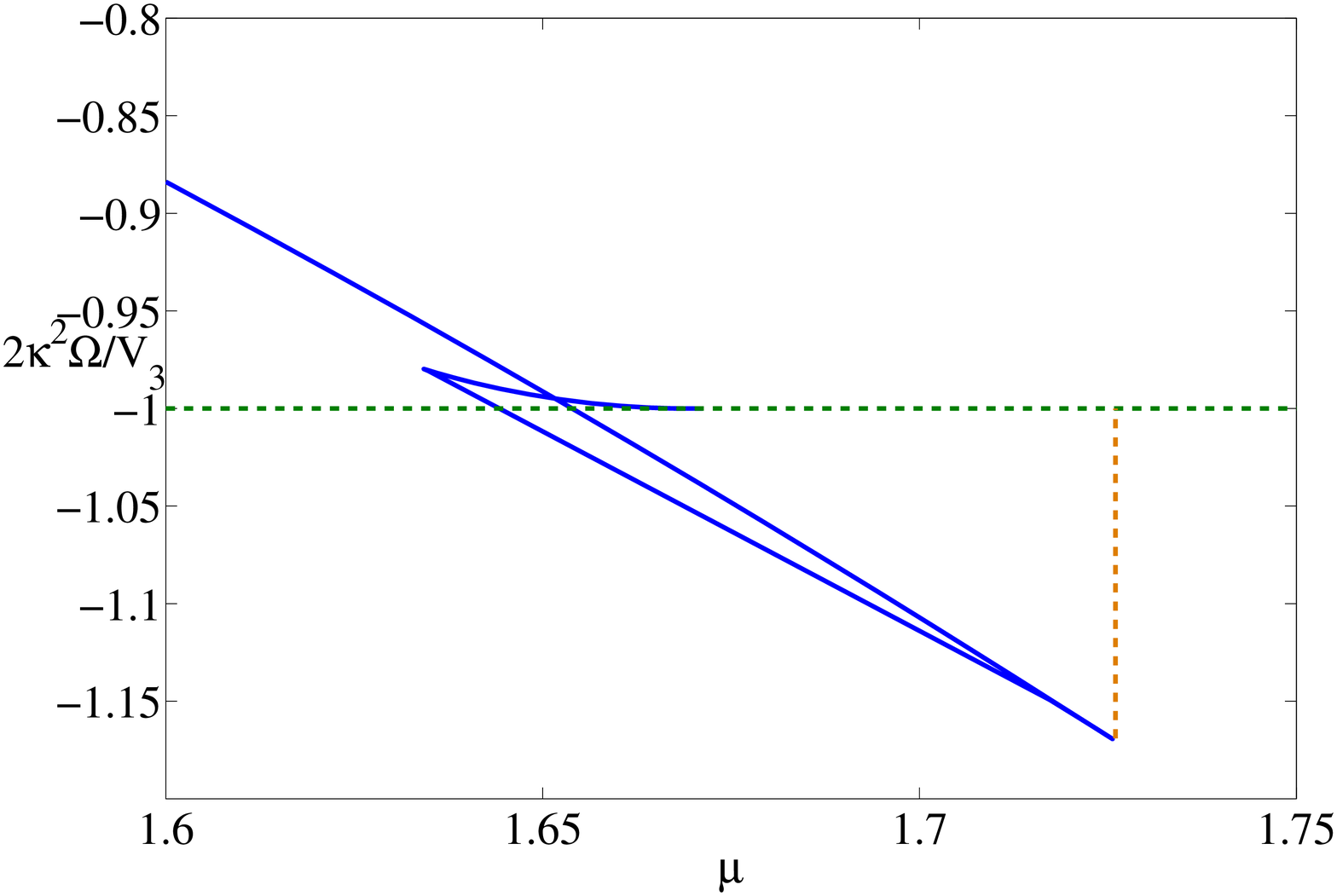}\\
  \caption{The grand potentials with restive to the chemical potential in the case of $m^2=-3/4$, $q=2.000>q_d$ (left plot) and $q=1.040\in(q_c,q_d)$(right plot). The grand potentials of the pure AdS soliton and hairy soliton are denoted by the green dashed line and blue solid line, respectively. In the case of $q=2.000>q_d$, there is a second order phase transition from the normal phase to the condensed phase at $\mu_{c1}\simeq0.8690$ and a zeroth order phase transition from the condensed phase to the normal phase at~$\mu_{c2}\simeq1.8650$ . In the case of $q=1.040\in(q_c,q_d)$, there is a first order phase transition from the normal phase to the condensed phase at $\mu_{c1}\simeq1.6444$ and a zeroth order phase transition from the condensed phase to the normal phase at~$\mu_{c2}\simeq1.7260$ .}\label{m2ag4q2}
\end{figure}

In this case, the two critical $q$ are $q_d\simeq1.0801$ and $q_c\simeq0.9951$. When $q\geq q_d$, there are two critical chemical potentials denoted by $\mu_{c1}\simeq0.8690$ and $\mu_{c2}\simeq1.8650$, respectively. As the chemical potential  is increased to $\mu_{c1}$, a second order phase transition from the normal phase to the condensed phased appears. If one continues scanning the shooting parameters, there is a critical value $\mu_{c2}$ at which the curve of condensate turns back to the small chemical potential  region. We can see from figure~\ref{m2ag4q2} that above $\mu_{c2}$ the hairy solution does not exist, hence, a zeroth order phase transition from the condensed phase to the normal phase happens at $\mu_{c2}$.

\begin{figure}
  \includegraphics[width=0.5\textwidth]{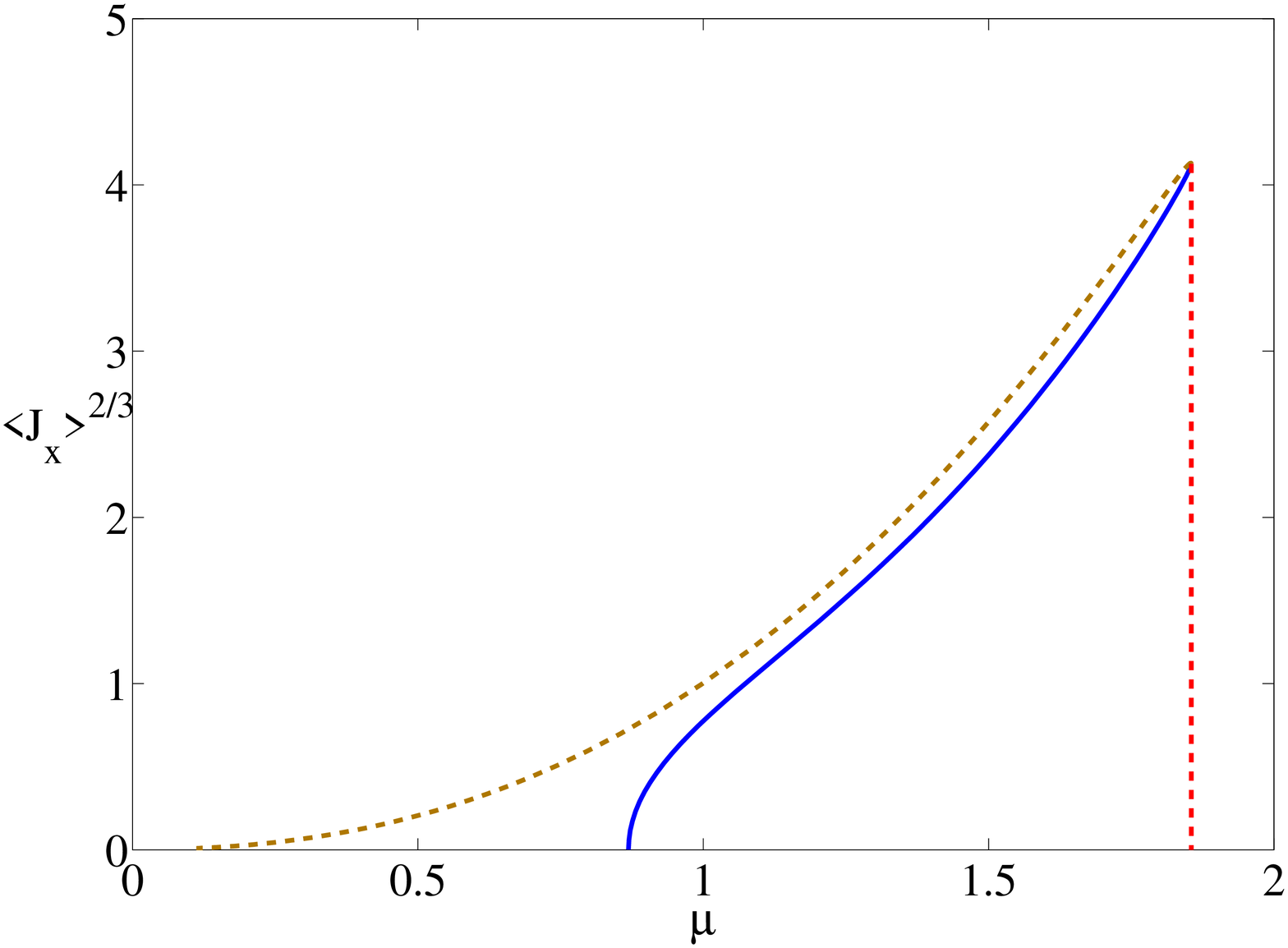}
  \includegraphics[width=0.5\textwidth]{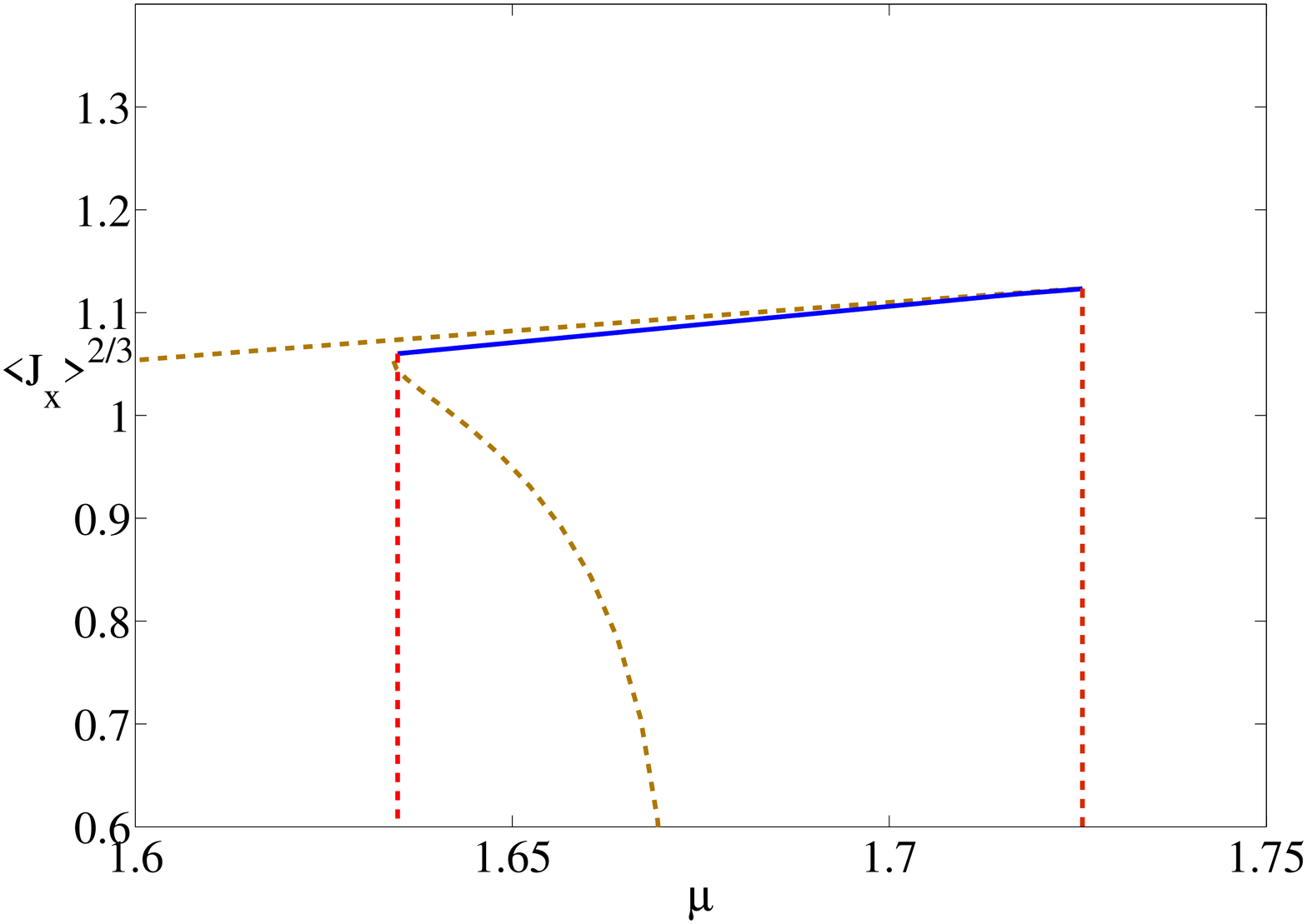}\\
  \caption{The condensation of the vector operator in the case of $m^2=-3/4,~q=2.000$ (left plot) and $q=1.040$ (right plot). In each plot, the thermodynamically preferred hairy phase is drawn by the blue solid line. The left plot shows that the condensate arises from zero as $\mu>0.8690$ and jumps to zero again when $\mu>1.8650$, which represents a second order phase transition and a zeroth order phase transition happening at these two chemical potentials, respectively. In the right plot,  the condensate has a jump at $\mu\simeq1.6444$ and $\mu\simeq1.7260$, respectively, which corresponds to a first order phase transition and a zeroth order phase transition.}\label{m2aJq1.15}
\end{figure}
\begin{figure}
  \includegraphics[width=0.5\textwidth]{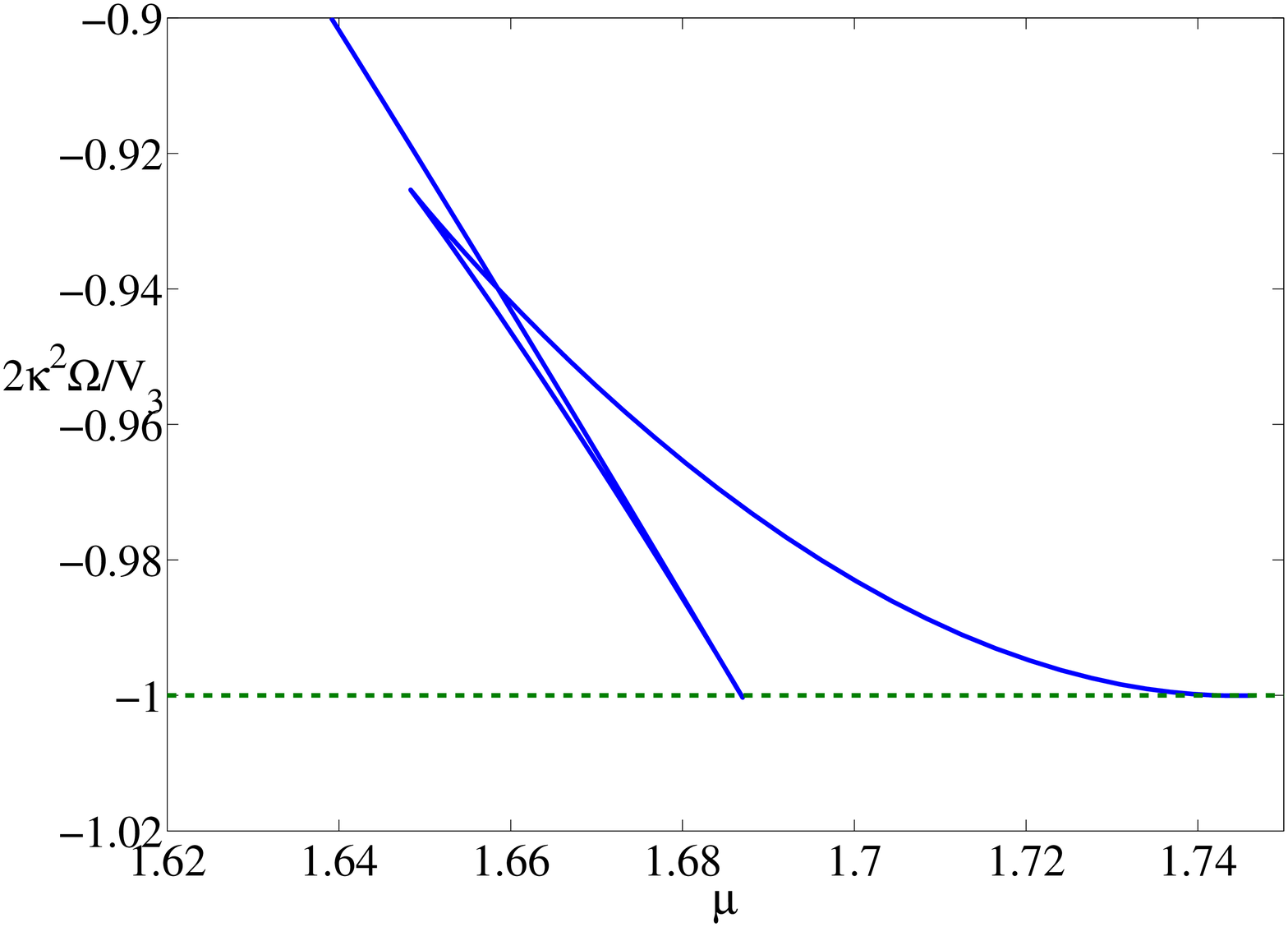}
  \includegraphics[width=0.5\textwidth]{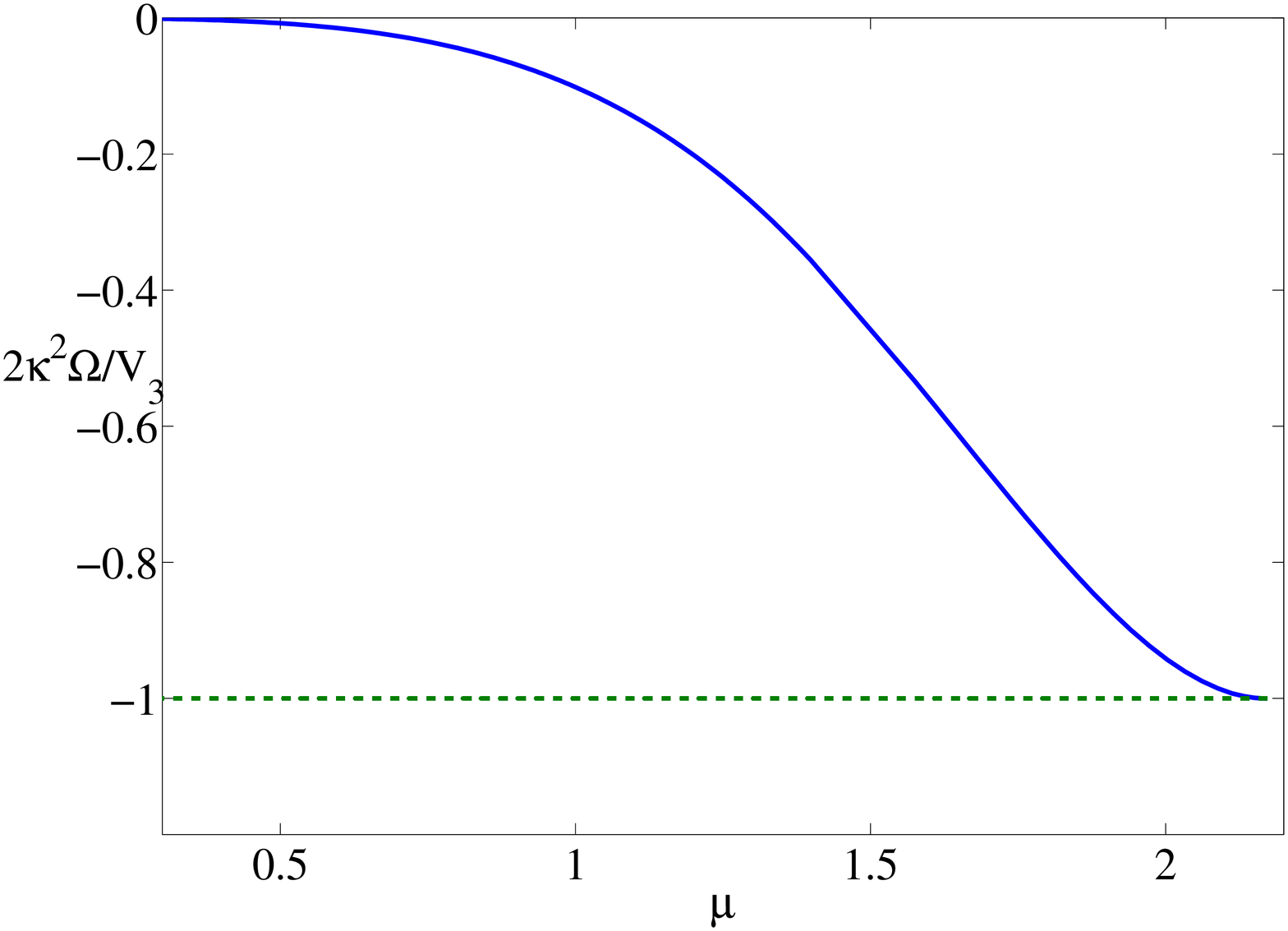}\\
  \caption{The grand potentials with respect to the chemical potential in the case of $m^2=-3/4$, $q=q_c\simeq0.9951$ (left plot) and $q=0.800<q_1$ (right plot). The grand potentials of the pure AdS soliton and hairy soliton are described by the green dashed line and blue solid line, respectively. In the case of $q=q_c$, the two critical chemical potentials for the first order phase and the zeroth order phase transition coincides. In the case of $q=0.800$, the grand potential of the condensed phase in the entire region is larger than the one of the normal phase.}\label{m2ag4q3}
\end{figure}

With the value of $q$ is decreased into the region $q_c\leq q<q_d$, the first phase transition at $\mu_{c1}$ becomes a first order one (see figures~\ref{m2ag4q2} and~\ref{m2aJq1.15}). The zeroth order transition at $\mu_{c2}$ still survives in this case. Our numerical calculation indicates that the difference between $\mu_{c1}$ and $\mu_{c2}$ decreases as $q$ decreases. Therefore, the region of chemical potential where the condensed phase is thermodynamical preferred is reduced by increasing back reaction. At the critical value $q=q_c$, $\mu_{c1}$ is equal to $\mu_{c2}$, thus there is no thermodynamically favored condensed phase.

For sufficiently large back reaction with $q<q_c$, one can see from figure~\ref{m2ag4q3} that hairy solutions only appear below a critical chemical potential and their grand potentials are larger than those of the normal phase. Therefore in this case there is no physical phase transition from the normal phase to the condensed phase when $q<q_c$. This phenomenon is an analog to the ``retrograde condensation" appearing in the case of black hole background in ref.~\cite{Cai:2013pda2}.

\begin{figure}
\centering
  \includegraphics[width=0.5\textwidth]{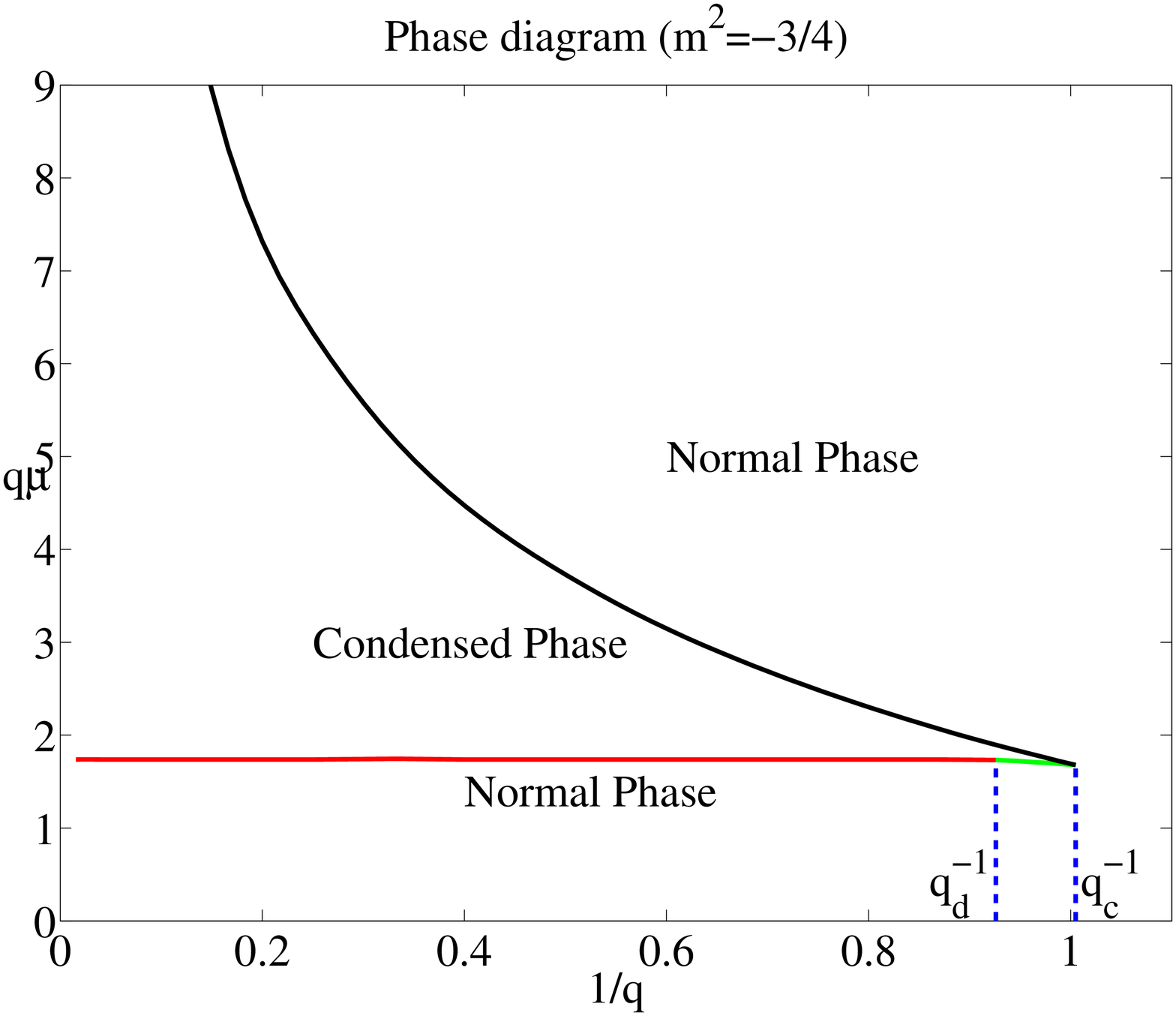}
  \caption{The phase diagram for the case with $m^2=-3/4$. The red, green and black lines stand for the second order, first order and zero order phase transitions, respectively.  Numerical results show $\mu_{c1} q\simeq1.7382$ when $1/q$ approaches to zero. }\label{PD3}
\end{figure}

We construct the ($\mu$,~$q$) phase diagram for the case with $m^2=-3/4$ in figure~\ref{PD3}. We can see from the phase diagram that when $ q >q_c$, there exist the insulator/superconductor/insulator phase transitions as one simply increases the chemical potential.  For very large $q$, the back reaction of matter fields to the geometry is small, one should return to the results obtained from the probe limit. Indeed, we find that the critical chemical potential of the second order transition is very closed to a constant which is just the one by the probe calculation in appendix~\ref{app3}. For a very large $q$, the black curve corresponding to the zeroth order transition from the condensed phase to the normal phase still exists. But we can not see this critical point in the probe limit. Actually, our numerical results for $q\gg1$ suggest the following two scaling relations hold for two critical chemical potentials

\begin{equation}\label{scal1}
\mu_{c1}\propto\frac{1}{q},\  \  \ \mu_{c2}\propto q^\zeta,~~\zeta(m^2)=0.4618m^2+0.0524m^4+{\cal O}(m^6).
\end{equation}
with negative $m^2$. The first relation for $\mu_{c1}$ is consistent with the probe approximation by using general Heun function analytically. The second relation for $\mu_{c2}$ is found by  numerical analysis. For more details please see appendix~\ref{app3}.

This scaling relation on $\mu_{c2}$  shows that the value of $\mu_{c2}$ will be divergent for $q\rightarrow\infty$ once the value of $m^2$ is non-negative. That is to say, there does not exist similar turning point to the normal phase for $m^2\geq0$. This suggests that the value of $m^2_{c2}$ is indeed zero. Furthermore, for the case with $m^2<m_{c2}^2$, one can see that $q\mu_{c2}\propto q^{\zeta+1}$ is divergent as $q\rightarrow\infty$, which means that we can not see the turning point in the probe limit. The ``retrograde condensation" can only appear for strong back reaction case.

\section{AdS Black Hole with Vector Hair}
\label{BHhair}
\subsection{Equations of motion and boundary conditions}
The hairy AdS charged black holes in the Einstein-Maxwell-complex vector field theory have been well studied in four dimensions in the probe limit~\cite{Cai:2013pda} as well as in the case with full back reaction~\cite{Cai:2013pda2}. Especially for the latter case, depending on two model parameters $m^2$ and $q$, the model manifests a rich phase structure. In this section we will construct asymptotically AdS charged black holes with non-trivial vector hair in (4+1) dimensions.

Similar to ref.~\cite{Cai:2013pda2}, we adopt the following metric ansatz
\begin{equation}\label{BHans}
ds^2=-a(r)e^{-b(r)}dt^2+\frac{dr^2}{a(r)}+r^2(c(r)dx^2+dy^2+dz^2).
\end{equation}
The position of horizon is denoted as $r_h$ at which $a(r_h)=0$ and the conformal boundary is located at $r\rightarrow\infty$. For matter fields, we turn on the $x$ component of $\rho_\mu$ and the temporal component of U(1) gauge field, i.e.,
\begin{equation}\label{BHmatter}
\rho_\nu dx^\nu=\rho_x(r)dx,\quad A_\nu dx^\nu=\phi(r)dt.
\end{equation}
Without loss of generality, we can take $\rho_x$ to be real. Then, the independent equations of motion are deduced as follows
\begin{equation}\label{BHeom}
\begin{split}
\phi''+(\frac{c'}{2c}+\frac{b'}{2}+\frac{3}{r})\phi'-\frac{2q^2\rho_x^2}{r^2ac}\phi=0,\\
\rho_x''+(\frac{a'}{a}-\frac{c'}{2c}-\frac{b'}{2}+\frac1r)\rho_x'+\frac{e^{b}q^2\phi^2}{a^2}\rho_x-\frac{m^2}{a}\rho_x=0,\\
b'-\frac{2a'}{a}-\frac{c'}{c}+\frac{2\rho_x'^2}{3rc}-\frac{re^b\phi'^2}{3a}-\frac{2e^b q^2\rho_x^2\phi^2}{3ra^2c}+\frac{8r}{a}-\frac{4}{r}=0,\\
c''+(\frac{a'}{a}-\frac{c'}{2c}-\frac{b'}{2}+\frac{3}{r})c'+\frac{2{\rho_x'}^2}{r^2}-\frac{2e^b q^2\rho_x^2\phi^2}{r^2a^2}+\frac{2m^2\rho_x^2}{r^2a}=0,\\
(\frac{3}{r}-\frac{c'}{2c})\frac{a'}{a}+(\frac{1}{r}+\frac{b'}{2})\frac{c'}{c}-\frac{\rho_x'^2}{r^2c}+\frac{e^b\phi'^2}{2a}+\frac{3e^b q^2\rho_x^2\phi^2}{r^2a^2c}-\frac{m^2\rho_x^2}{r^2ac}-\frac{12}{a}+\frac{6}{r^2}=0,
\end{split}
\end{equation}
where the prime denotes the derivative with respect to $r$.

Those above equations of motion are very similar to those in the (3+1) dimensional spacetime~\cite{Cai:2013pda2}. Therefore, we use the same procedure in ref.~\cite{Cai:2013pda2} to solve the set of equations~\eqref{BHeom} by shooting method. The asymptotical expansion for metric fields and matter fields near the boundary $r\rightarrow\infty$ is as follows
\begin{equation} \label{boundary2}
\begin{split}
\phi=\mu-\frac{\rho}{r^2}+\ldots,\quad \rho_x=\frac{{\rho_x}_-}{r^{{\Delta}_-}}+\frac{{\rho_x}_{+}}{r^{{\Delta}_+}}+\ldots,\\
a=r^2(1+\frac{a_4}{r^4})+\ldots,\quad c=1+\frac{c_4}{r^4}+\ldots,\quad b=0+\frac{b_4}{r^4}+\ldots,
\end{split}
\end{equation}
where the dots stand for the higher order terms of $1/r$. We choose the source free condition ${\rho_x}_-=0$ as before. The coefficients $\mu$, $\rho$, and ${\rho_x}_{+}$ are directly related to the chemical potential, charge density and $x$ component of the vacuum expectation value of the vector operator $\hat{J^\mu}$ in the dual system, respectively.

We focus on black hole configurations that have a regular event horizon located at $r_h$ and require the regularity conditions at the horizon $r=r_h$, which means that all five functions $\{\rho_x(r),\phi(r),a(r),b(r),c(r)\}$ will have finite values at $r_h$ and admit a series expansion in terms of $(r-r_h)$. We have six independent parameters at the horizon $\{r_h,\rho_x(r_h),\phi'_x(r_h),c(r_h),b(r_h)\}$. Other coefficients can be expressed in terms of those parameters.

Two free parameters $b(r_h)$ and $c(r_h)$ can be fixed by AdS boundary conditions that $b(r\rightarrow\infty)=0$ and $c(r\rightarrow\infty)=1$. Without loss of generality, the location of $r_h$ can be fixed to be one in our numerical calculation. We are then left with two independent parameters $\{\rho_x(r_h),\phi'(r_h)\}$. By choosing $\phi'(r_h)$ as the shooting parameter to match the source free condition, i.e., ${\rho_x}_-=0$, we finally have a one-parameter family of solutions labeled by the value of $\rho_x$ at the horizon. After solving the set of equations, we can read off the condensate $\langle \hat{J^x}\rangle$, chemical potential $\mu$ and charge density $\rho$ directly from the asymptotical expansion~\eqref{boundary2}.

\subsection{Thermodynamics}
\label{sect:freeBH}
We will discuss how to extract thermodynamic quantities from our solutions. We will work in the grand canonical ensemble with chemical potential fixed. According to gauge/gravity duality, the Hawking temperature of the black hole is identified with the temperature of boundary thermal state, which is given by
\begin{equation}\label{BHT}
T=\frac{a'(r_h)e^{-b(r_h)/2}}{4\pi},
\end{equation}
and the thermal entropy S is given by the Bekenstein-Hawking entropy of the black hole
\begin{equation}\label{BHS}
S=\frac{4\pi}{\kappa^2}A=\frac{4\pi V_3}{\kappa^2}r^3_h\sqrt{c(r_h)},
\end{equation}
with $A$ denoting the area of the horizon and $V_3=\int dxdydz$.

In order to find thermodynamically favored phase, we should calculate the grand potential $\Omega$ in the grand canonical ensemble. $\Omega$ of the boundary thermal state is identified with temperature $T$ times the on-shell bulk action in Euclidean signature. The Euclidean action can be calculated by
\begin{equation}\label{onshell_1}
-2\kappa^2 S_{Euclidean}=\int d^5x\sqrt{-\bar{g}}(\mathcal{R}+12+\mathcal{L}_m)+\int_{r\rightarrow\infty} d^4x\sqrt{-\bar{h}}(2\mathcal{K}-6),
\end{equation}
where $\bar{h}$ is the determinant of the induced metric on the boundary, and $\mathcal{K}$ is the trace of the extrinsic curvature. We have introduced the Gibbons-Hawking boundary term for a well-defined Dirichlet variational problem. Once again, the potential surface counterterm for the matter fields has been neglected
here since this term does not make any contribution in our setup of the source free condition.

By using the equations of motion~\eqref{BHeom}, the on-shell action reduces to
\begin{equation}\label{onshell_2}
-2\kappa^2 S_{Euclidean}^{on-shell}=2\beta V_3 e^{-b/2}r\sqrt{ac}(\mathcal{K}r-2r-\sqrt{a})|_{r\rightarrow\infty},
\end{equation}
with $\beta=1/T$. Substituting the asymptotical expansion~\eqref{boundary2} into~\eqref{onshell_2}, we obtain
\begin{equation}\label{free}
\Omega=T S_{Euclidean}^{on-shell}=\frac{V_3}{2\kappa^2}(a_4+c_4-b_4).
\end{equation}
In the case with $m^2 >-1$, we find that $c_4-b_4=0$ (see appendix~\ref{app2}). In addition, for the normal phase (conductor) described by the RN-AdS black hole solution in~\eqref{RNmetric}, one has $a_4=-r_h^4-\frac{\mu^2 r_h^2}{3}$ and $c_4=b_4=0$.

In general, different solutions obtained by shooting method with $r_h=1$ will have different chemical potentials. Since we work in the grand canonical ensemble, the chemical potential for each solution must be the same. Let us note that the system admits a useful scaling symmetry
\begin{equation} \label{scalingbh}
r\rightarrow\lambda r,\quad (t,x,y,z)\rightarrow{\lambda^{-1}}(t,x,y,z),\quad(\phi,\rho_x)\rightarrow\lambda(\phi,\rho_x), \quad a\rightarrow\lambda^2a,
\end{equation}
with $\lambda$ a real positive constant. Under this symmetry, the revelent quantities transform as
\begin{equation} \label{transform}
(T,\mu,\Omega)\rightarrow\lambda (T,\mu,\Omega),\quad S\rightarrow S,\quad \rho\rightarrow\lambda^3\rho, \quad {\rho_x}_{+}\rightarrow\lambda^{\Delta_++1}{\rho_x}_{+}.
\end{equation}
Therefore, we can set the chemical potentials of all black hole solutions equal by using above transformation rule directly. Without loss of generality, we shall fix the chemical potential to be $\mu=1$. Finally let us mention that apart from the temperature $T$ and thermal entropy $S$, this scaling transformation (\ref{transform}) is  also true for the soliton case.
\subsection{Phase transition in black hole backgrounds}
Our numerical calculation shows that the phase structure of the model in the (4+1) dimensional black hole geometry looks precisely the same as the case in the (3+1) dimensional black hole spacetime~\cite{Cai:2013pda2}.  Therefore in this section we will just give main results and some relevant details can be found in ref.\cite{Cai:2013pda2}.
The phase transition in the black hole background depends on the values of the mass square $m^2$ and charge $q$. There exists a particular value of $m^2$ denoted by $m_{cb}^2$. For $m^2$ is above or below $m^2_{cb}$, our system exhibits distinguished behaviors. Our numerical analysis suggests that $m_{cb}^2=0$, up to a numerical error,  which is same as the value obtained in the (3+1) dimensional case~\cite{Cai:2013pda2}. For $m^2\geq m_{cb}^2$, the condensed phase exists even for sufficiently low temperatures, while for $m^2<m_{cb}^2$, the condensed phase will be absent when the temperature is lowered to a critical value. All possible phase transitions are summarized in table~\ref{TabBH1} and table~\ref{TabBH2}.

In the case of $m^2\geq m^2_{cb}$, there is a mass dependent critical charge $q_c$. When the charge of the vector field is larger than $q_c$, the condensed phase will become thermodynamically preferred when the temperature is less than a critical temperature at which a second order phase transition occurs. When we decrease $q$ to a certain value smaller than $q_c$, the transition from the normal phase/conductor to the condensed phase/superconductor becomes a first order one. For large $m^2\geq m^2_{cb}$, the
condensed phase seems to survive even if the temperature goes to zero, $T\rightarrow0$.  In contrast, in the case with $m^2 < m^2_{cb}$, the condensed phase cannot exist when the temperature is lower than a certain value. To determine the precise value for $m^2_{cb}$ , we need to solve the coupled equations of motion~\eqref{BHeom} at very low temperatures to see whether the condensate would turn back to the higher temperature region. The precision of numerical calculation prevents the investigation on the limit of $T\rightarrow0$. Nevertheless, our numerical calculation suggests that $m^2_{cb}$ is very closed to zero. We will consider one concrete example for both cases with $m^2 \ge m^2_{cb}$ and $m^2 < m^2_{cb}$, respectively.

\begin{table}
  \centering
  \begin{tabular}{|c|c|c|c|}
    \hline
    Charge &\multicolumn{3}{|c|}{ Phase transition and its order,~~$m^2_{cb}\leq m^2$} \\
    \hline
    $q\geq q_c$ & $T>T_c$, BH& $T=T_c,2^{nd}$ & $T<T_c$, BC \\
    \hline
    $q_c< q$ &$T >T_c$, BH& $T=T_c ,1^{st}$& $T<T_c$, BC \\
    \hline
  \end{tabular}
  \caption{The phase transition and its order with respect to the charge and temperature in the case of $m^2_{cb}\leq m^2$. In the table, BH=normal phase/conductor with the RN-AdS black hole solution, BC=condensed phase/superconductor with hairy black hole solution. $1^{st}$ and~$2^{nd}$ stand for the first order and second order phase transitions, respectively.}\label{TabBH1}
\end{table}
\begin{table}
  \centering
  \begin{tabular}{|c|c|c|c|c|c|}
    \hline
    Charge &\multicolumn{5}{|c|}{ Phase transition and its order,~~$m^2<m^2_{cb}$} \\
    \hline
    $q\geq q_{\alpha}$ & $T>T_{c2}$, BH & $T=T_{c2},2^{nd}$ & $T_{c0}<T<T_{c2}$, BC& $T=T_{c0}$,$0^{th}$ & $T<T_{c0}$,BH\\
    \hline
    $q_{\beta}<q<q_\alpha$& $T>T_{c1}$, BH & $T=T_{c1},1^{nd}$ & $T_{c0}<T<T_{c1}$, BC& $T=T_{c0}$,$0^{th}$ & $T<T_{c0}$,BH\\
    \hline
    $q\leq q_\beta$ & \multicolumn{5}{|c|}{ BH } \\
    \hline
  \end{tabular}
  \caption{The phase transition and its order with respect to the charge and temperature in the case of $m^2_{cb}>m^2$. In the table, BH=normal phase/conductor with the RN-AdS black hole solution, BC=condensed phase/superconductor with hairy black hole solution. $0^{th}$, $1^{st}$ and~$2^{nd}$ stand for the zeroth order, first order and second order phase transitions, respectively.}\label{TabBH2}
\end{table}

\subsubsection{Phase transition for $m^2\geq m_{cb}^2$}
In the case with $m^2\geq m_{cb}^2$, as we lower the temperature, we can always find a phase transition from the normal phase to a condensed phase where a charged black hole develops non-trivial vector hair and is thermodynamically favored. In the dual field theory side, it means that a vector operator acquires a vacuum expectation value $\langle\hat{J_x}\rangle\neq0$ which breaks the U(1) symmetry as well as the rotation symmetry in  $x-y$ plane spontaneously. Furthermore, the order of the transition depends on the strength of back reaction controlled by $q$. For small back reaction, this transition is second order, while for large back reaction, it becomes a first order one. The critical value of $q$ denoted as $q_c$ depends on $m^2$ we choose. We here consider $m^2=5/4$ as a concrete example, where $q_c\simeq2.0972$. One can find similar results for other values of $m^2 \ge m_{cb}^2$.

\begin{figure}
\includegraphics[width=0.5\textwidth]{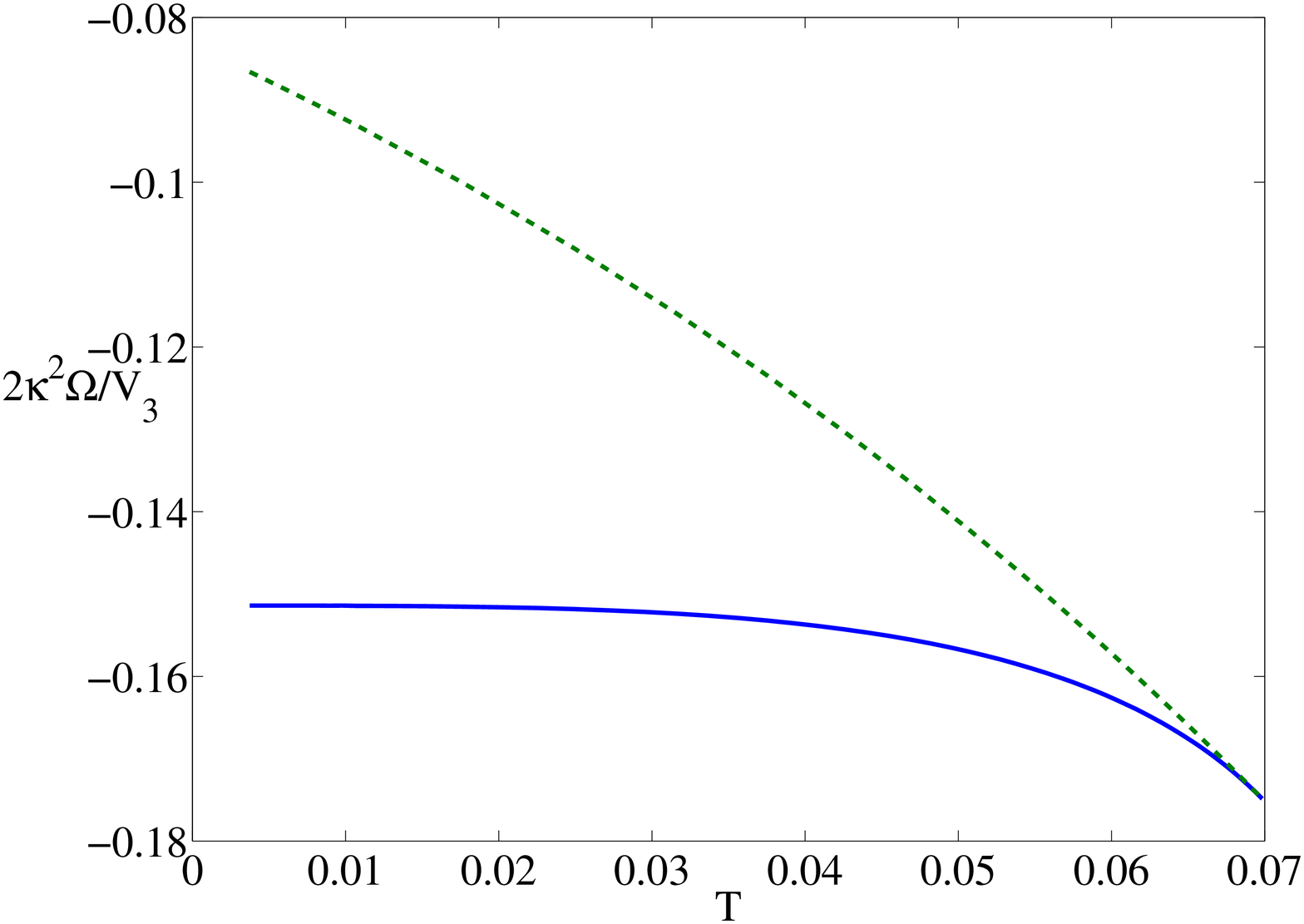}
\includegraphics[width=0.5\textwidth]{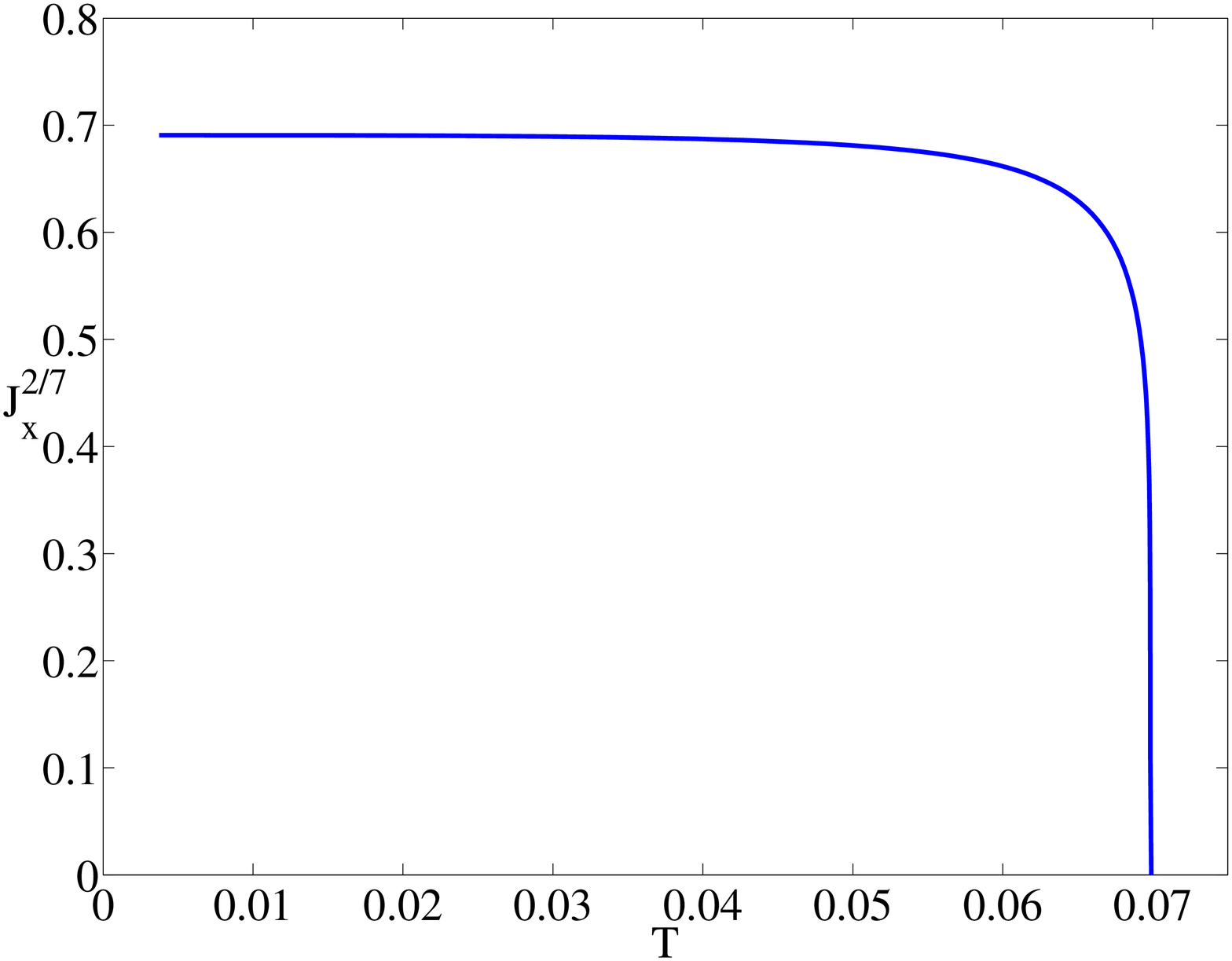}\\
\caption{\label{TomegaA}The grand potential (left plot) and condensate (right plot) as a function of temperature in the case of $m^2=\frac54$, and $q=2.500$. In the left plot, the grand potential of the RN-AdS black hole solution and hairy black hole solution are described by the green dashed line and blue solid line, respectively, and the phase transition happens at $T_c\simeq0.0698$. }
\end{figure}

Let us first consider, for example,  the case with $q=2.500>q_c$. The condensate appears at $T_c\simeq0.0698$ and arises gradually from zero  for lower temperatures. Comparing grand potential between the hairy solution and the RN-AdS solution, one can see from the left plot of figure~\ref{TomegaA} that the condensed phase is thermodynamically favored. It is a second order phase transition with the critical behavior $\langle\hat{J_x}\rangle\sim(1-T/T_c)^{1/2}$ near $T_c$. The critical exponent for the condensate is $1/2$, the same as the one
in mean field theory.

\begin{figure}
\includegraphics[width=0.5\textwidth]{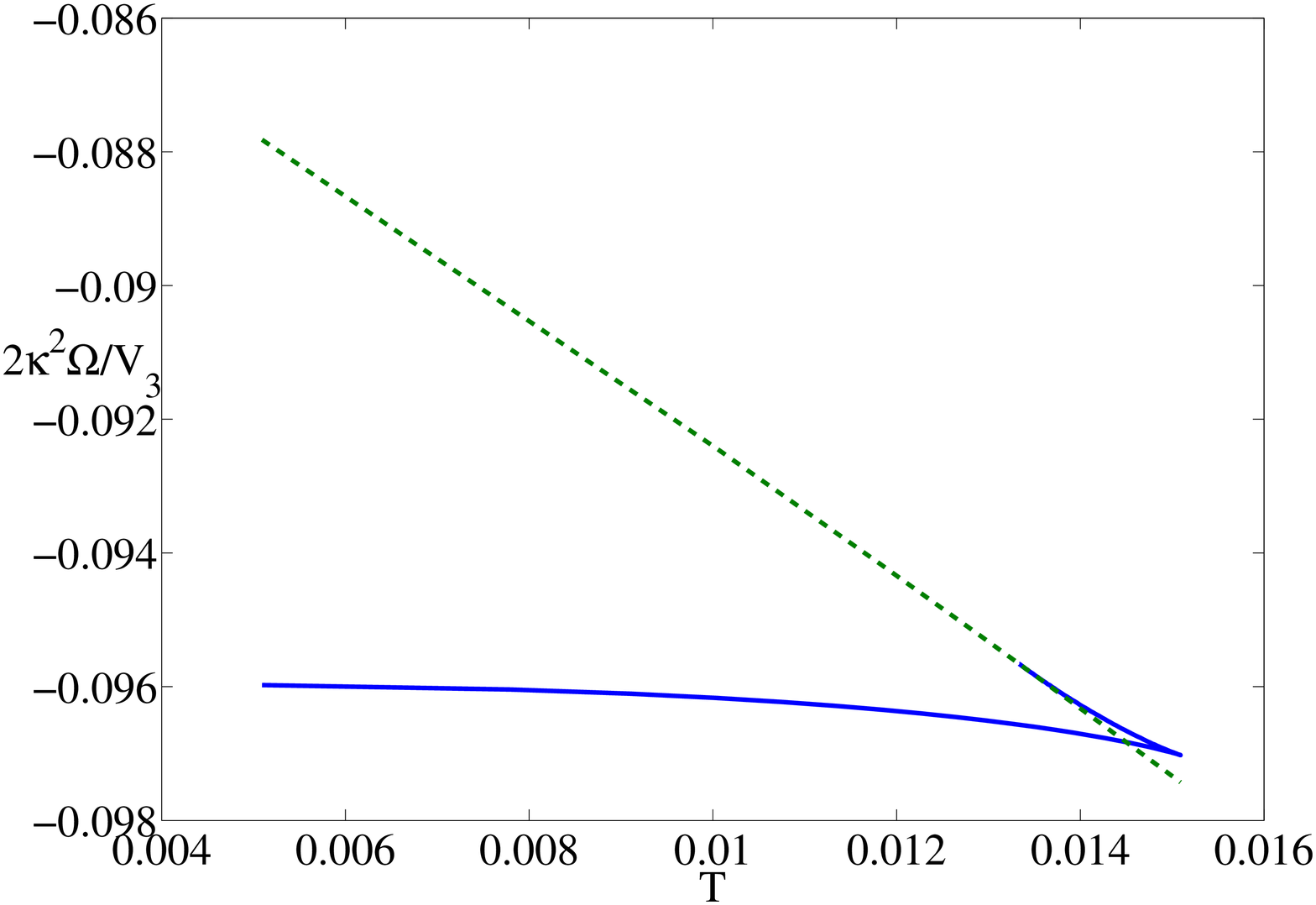}
\includegraphics[width=0.5\textwidth]{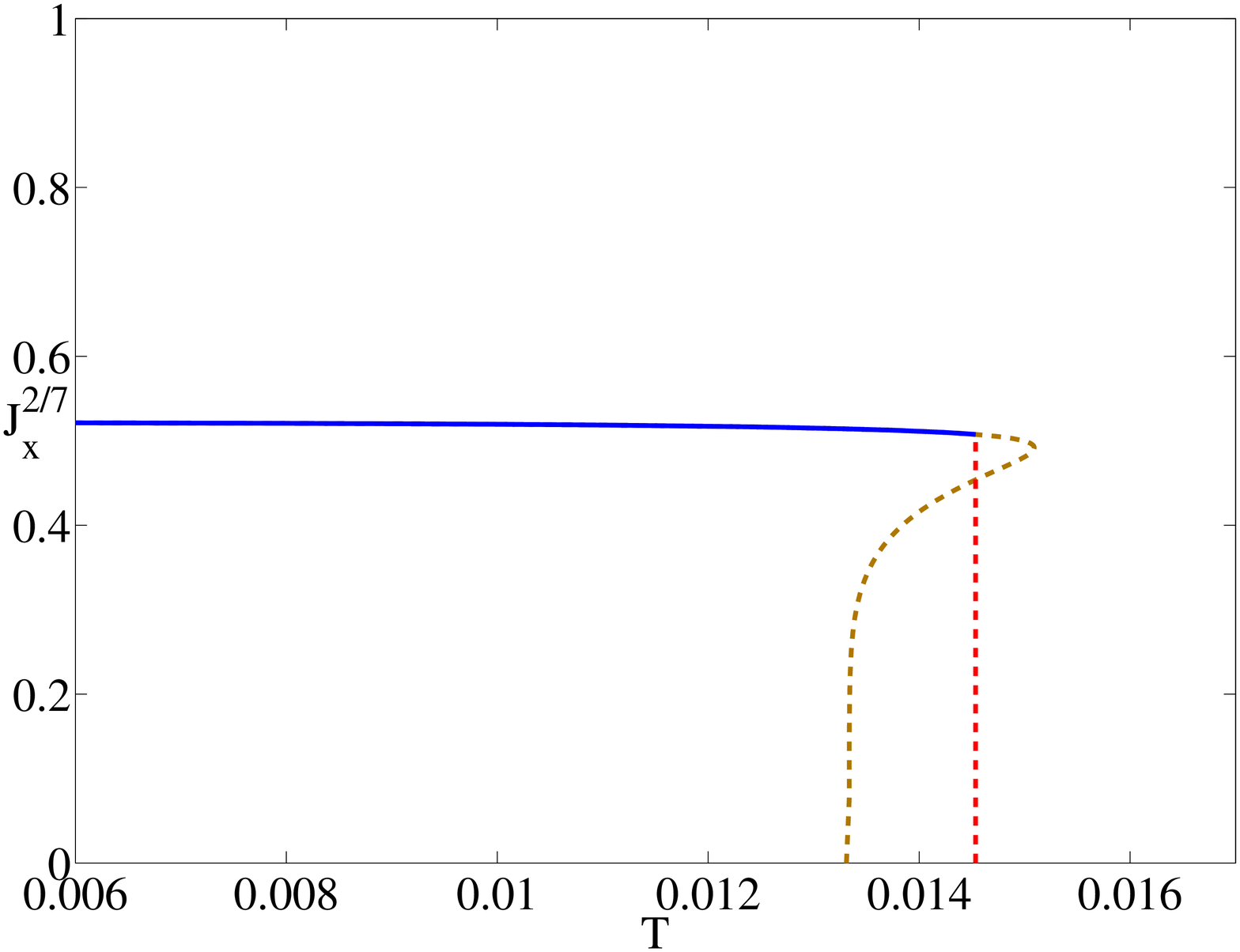}\\
\caption{\label{TomegaA_Sup}The grand potential (left plot) and condensate (right plot) as a function of temperature in the case of $m^2=\frac54$, and $q=1.800$. In the left plot, the grand potential of the RN-AdS black hole solution and hairy black hole solution are described by the green dashed line and blue solid line, respectively. The phase transition happens at $T\simeq0.0145$. The physical branch for the condensate in the right figure is denoted by the blue solid line.}
\end{figure}

A qualitative change happens as $q$ is decreased to below $q_c$. The  grand potential and condensate as a function of temperature for $q=1.800<q_c$ are shown in figure~\ref{TomegaA_Sup}. We can see that a swallowtail appears in the grand potential and in this case, the condensate has a jump from zero to a finite
 value at the critical temperature, which is a typical feature of a first order phase transition. This tells us
that when $q <q_c$, the phase transition becomes first order from the normal phase  to the condensed phase as the temperature is lowered.

\begin{figure}
\centering
  \includegraphics[width=0.5\textwidth]{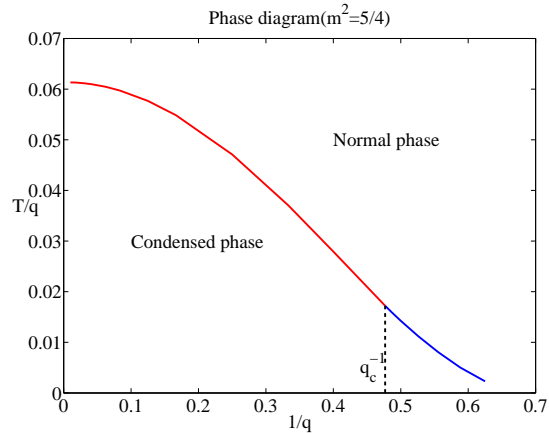}
  \caption{The phase diagram for the case with $m^2=5/4$. The red line and blue line stand for second order and first order phase transitions, respectively. When $1/q\rightarrow0$, we have $T_c/q\simeq0.0613$.}\label{P-D4}
\end{figure}

A phase diagram in terms of temperature and charge is shown in figure~\ref{P-D4} in the case with $m^2=5/4$, where the red and blue curves
stand for the second order and first order phase transitions, respectively.

\subsubsection{Phase transition for $m^2<m_{cb}^2$}

The phase structure for $m^2<m_{cb}^2$ is the same as the one in the (3+1) dimensional black hole background~\cite{Cai:2013pda2}. The parameter space of $q$ is divided into three regions by two special values of $q$, denoted as $q_\alpha$ and $q_\beta$ with $q_\alpha>q_\beta$. The phase behavior changes qualitatively in each region.  To compare our results to the case with one dimension less, we take the same terminology, i.e., the critical temperatures for second order transition, first order transition and zeroth order transition are denoted by $T_{c2}$, $T_{c1}$ and $T_{c0}$, respectively. In this subsection, we consider $m^2=-3/4$ as a concrete example. In this case, $q_{\alpha}\simeq1.766$ and $q_{\beta}\simeq1.700$.

\begin{figure}
  \includegraphics[width=0.5\textwidth]{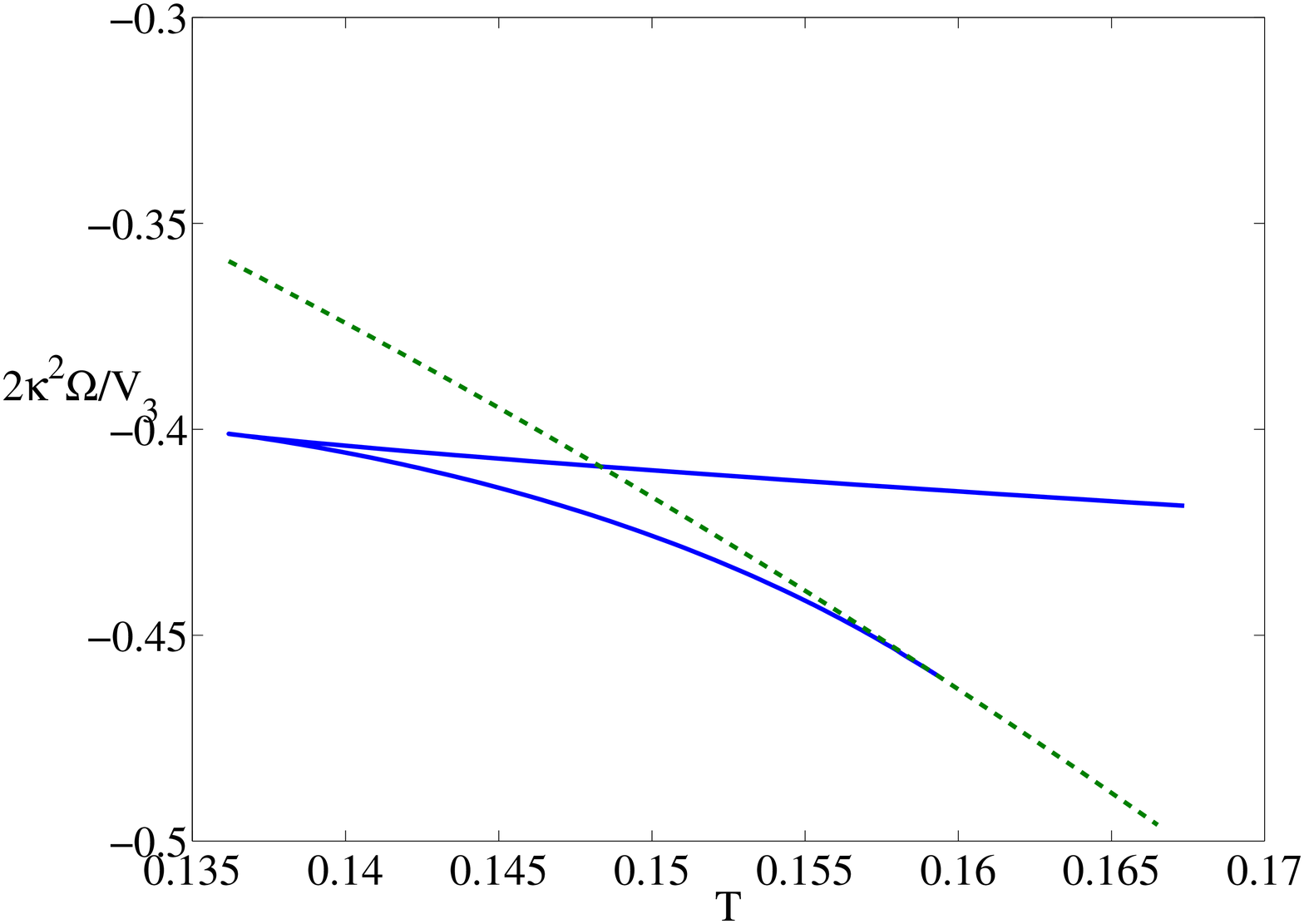}
  \includegraphics[width=0.5\textwidth]{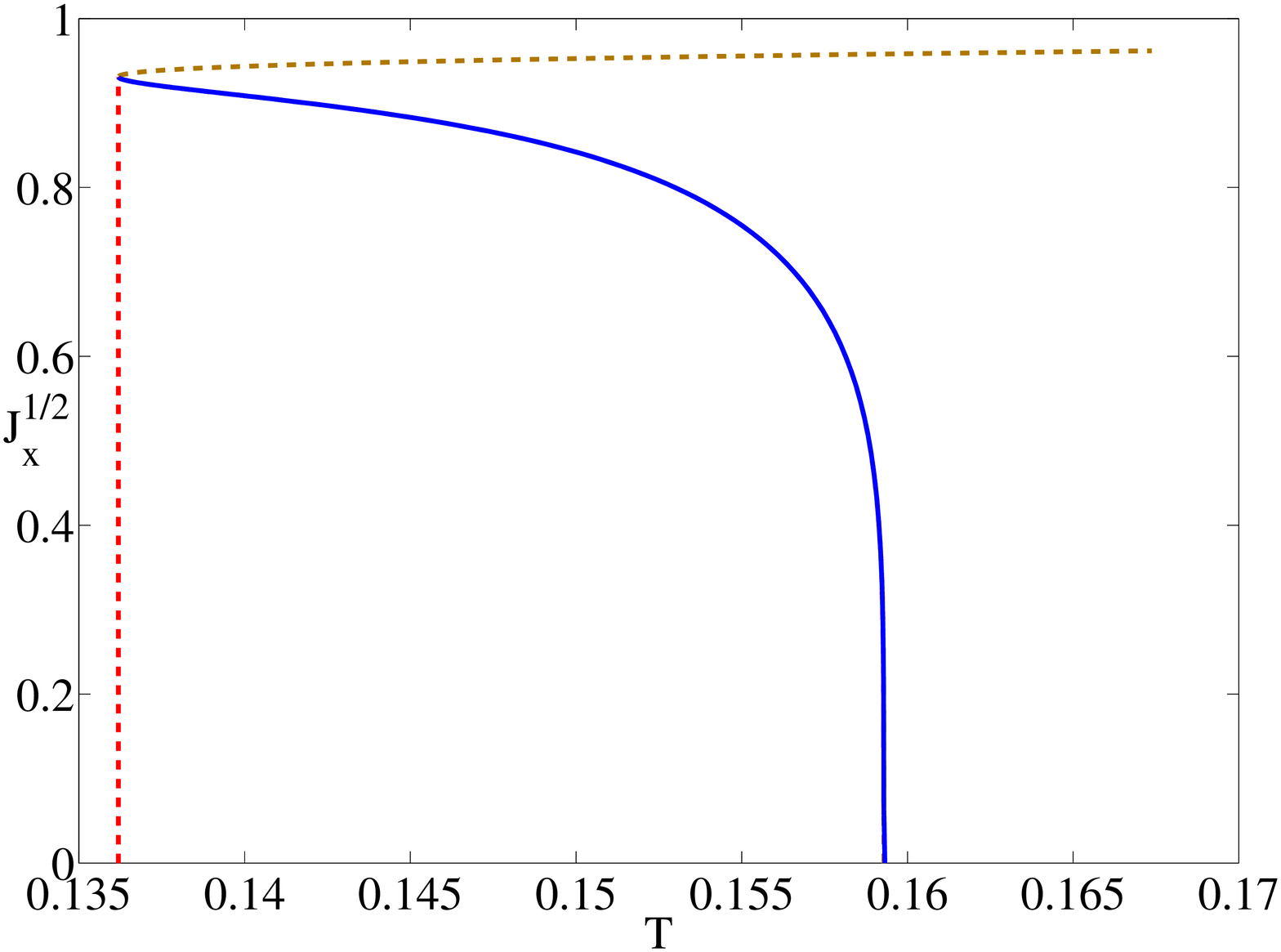}\\
  \caption{The grand potential (left plot) and condensate (right plot) as a function of temperature in the case of $m^2=-\frac34$ and $q=2.000$. In the left plot, the grand potential of the pure RN-AdS  solution and hairy solution are described by the green dashed line and blue solid line, respectively. The condensate begins at $T_{c2}=0.1592$ with a second order phase transition and ends at $T_{c0}=0.1362$ with a zeroth order phase transition. In the right figure, only the blue solid line stands for the physical branch of the condensate.}\label{TomegaB}
\end{figure}

For the weak back reaction $q>q_\alpha$, we show the condensate and grand potential versus temperature in figure~\ref{TomegaB}. We immediately see that the condensate arises below $T_{c2}$ and then turns back to the high temperature region at $T_0$ (see the dashed orange dashed line, which corresponds to the upper branch of the grand potential in the condensed phase). When $T<T_{c0}$, there does not exist hairy black hole solution and thus in the region with $T <T_{c0}$ we have only the RN-AdS black hole solution. By comparing the grand potential for each solution, we find that the thermodynamically favored region of the condensed phase only locates in the region $T_{c0} < T < T_{c2}$. At $T_{c2}$, it is a second order phase transition, while there is a zeroth order transition at $T_{c0}$, since at the critical temperature the grand potential has a sudden jump from the condensed phase to the normal phase.

\begin{figure}
  \includegraphics[width=0.5\textwidth]{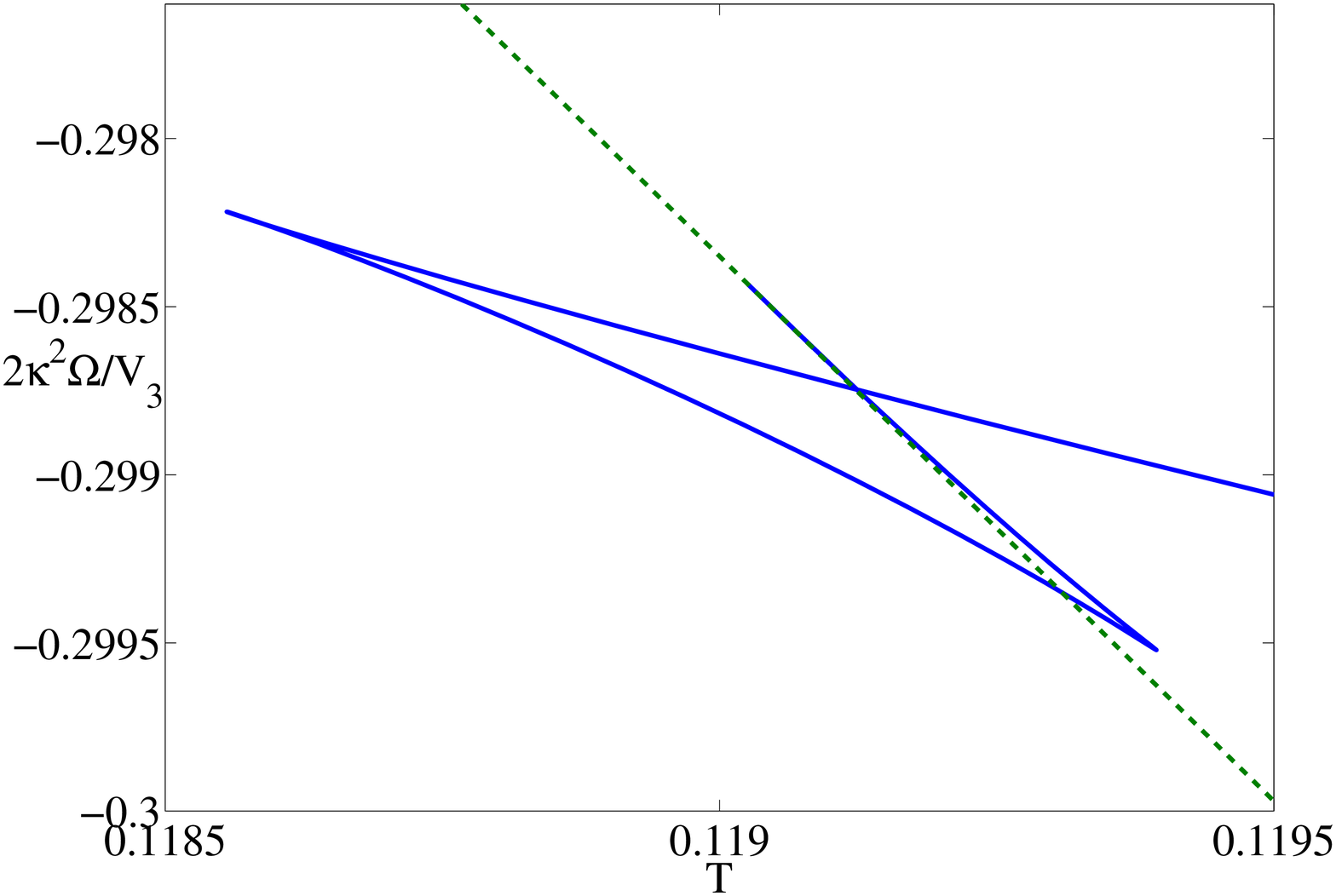}
  \includegraphics[width=0.5\textwidth]{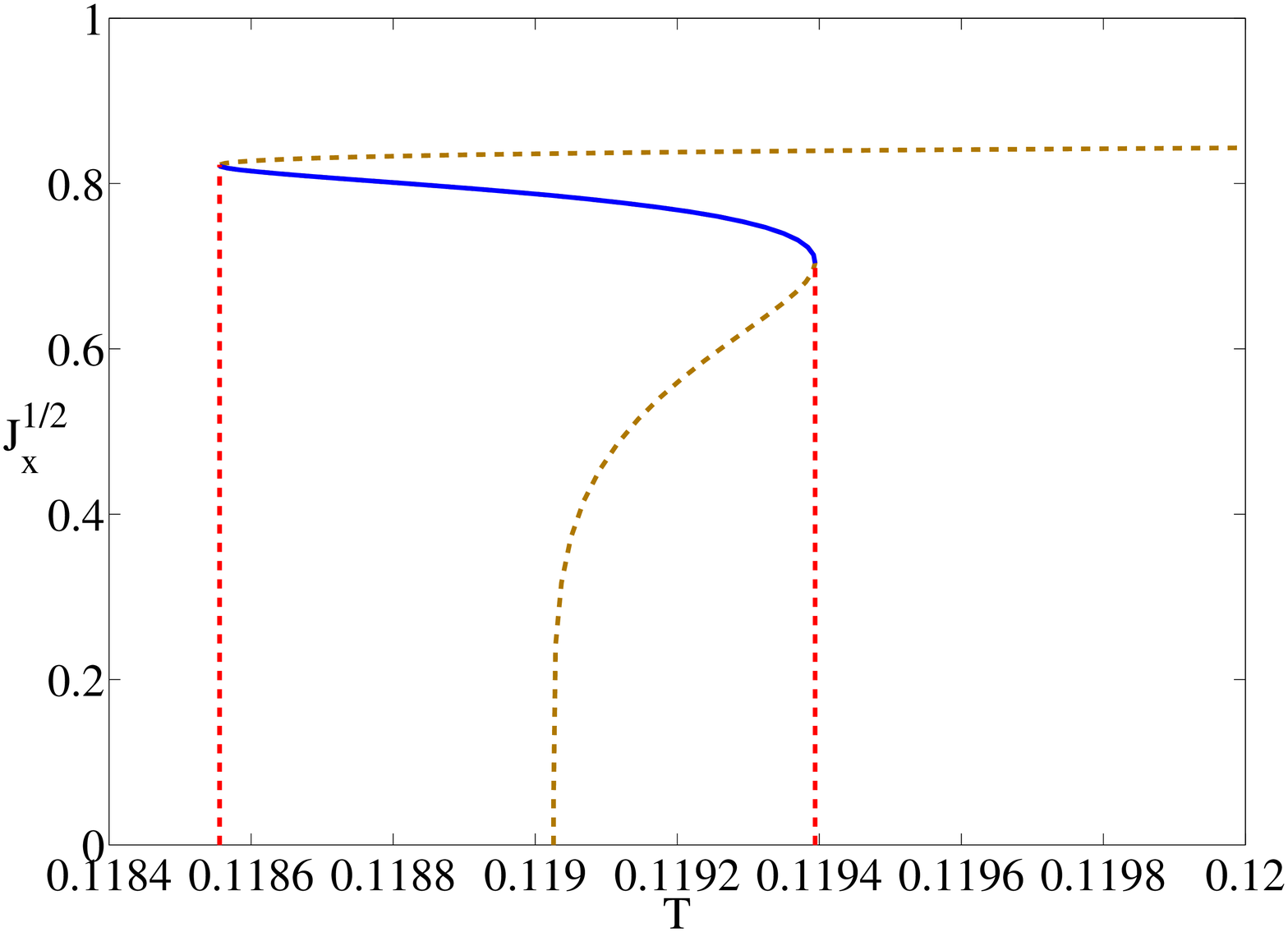}\\
  \caption{The grand potential (left plot) and condensate (right plot) as a function of  temperature in the case of $m^2=-\frac34$ and $q=1.719$. In the left plot, the grand potential of the pure RN-AdS  solution and hairy solution are described by the green dashed line and blue solid line, respectively.  The condensate begins at $T_{c1}=0.1194$ with a first order phase transition and ends at $T_{c0}=0.11856$ with a zeroth order phase transition. In the right plot, the solid blue line is the physical branch of the condensate, while two dashed orange parts are un-physical. }\label{TomegaB_Sup}
\end{figure}

\begin{figure}
  \includegraphics[width=0.5\textwidth]{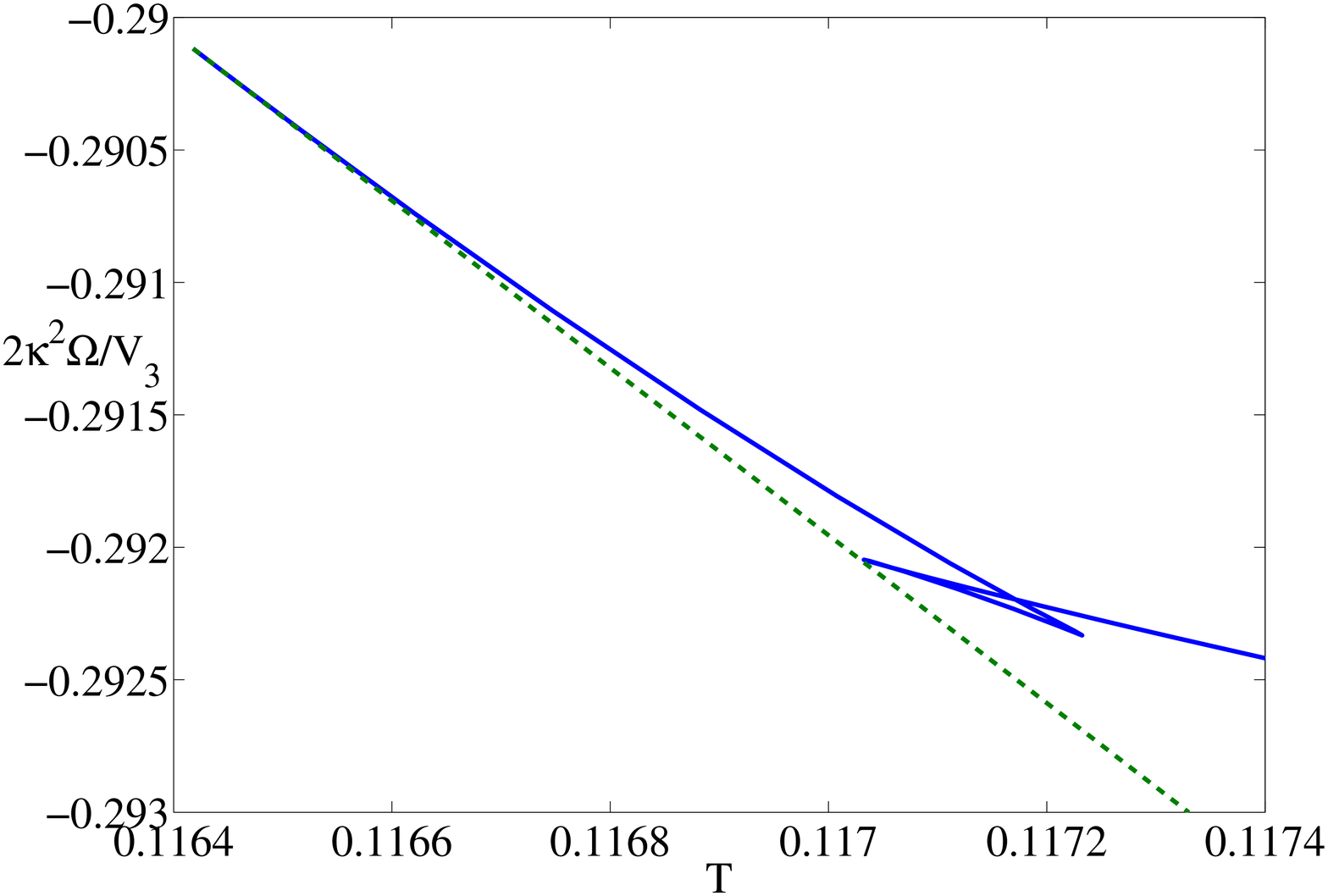}
  \includegraphics[width=0.5\textwidth]{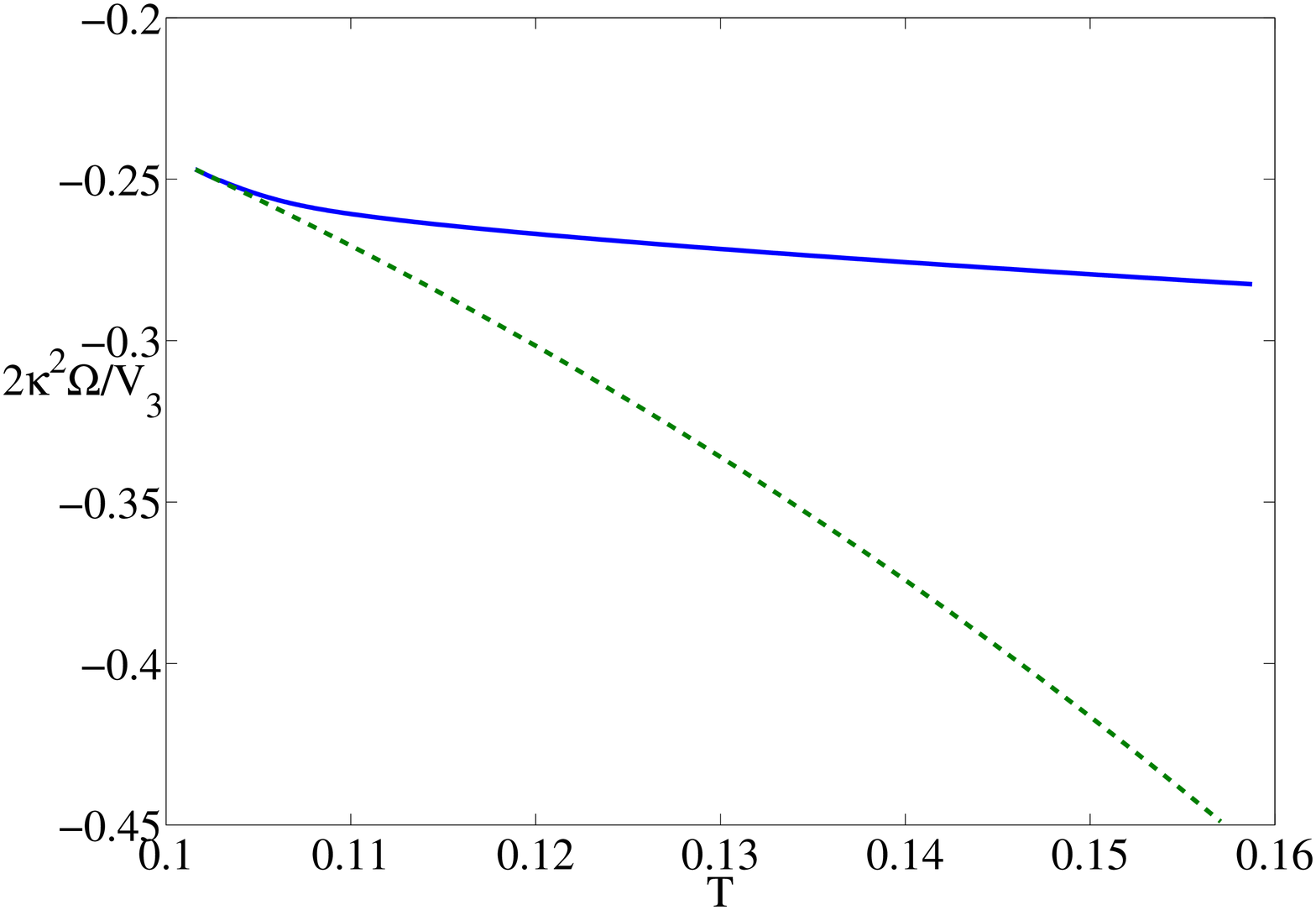}\\
  \caption{The grand potentials as a function of temperature in the case of $m^2=-\frac34$, $q=1.700$(left) and $q=1.600$(right). In both figures, the grand potential of the RN-AdS black hole solution and hairy black hole solution are denoted by the green dashed line and blue solid line, respectively.}\label{TomegaC}
\end{figure}

As we decrease the value of $q$ to the region $q_\beta<q<q_\alpha$, there is a little difference compared to the previous case. As we lower the temperature, we will first see a first order transition from the normal phase to the condensed phase at $T_{c1}$, and then a zeroth order transition back to the normal phase at $T_{c0}$. A typical case with $q=1.719$ is shown in figure~\ref{TomegaB_Sup}.
In the above two cases, as we lower the temperature, there exists a condensed phase which is thermodynamically favored, although only in a narrow temperature range. The story has a dramatic change as $q$ is decreased past $q_\beta$. In this case, the hairy black hole solution only appears in the high temperature region. In figure~\ref{TomegaC} we present the grand potential as a function of temperature  in the cases of $q=1.700$ (left plot) and $q=1.600$ for $m^2=-\frac{3}{4}$. One can clearly see that the grand potential of the hairy black hole solutions is always larger than the one for the RN-AdS black hole at the same temperature.  Thus these hairy black hole solutions are thermodynamically
disfavored. This is just the so called ``retrograde condensation" reported in ref.~\cite{Cai:2013pda2}. Let us notice that similar ``retrograde condensation" also happens in the soliton case in the sense that the hairy soliton solution only appears below a critical chemical potential and is thermodynamically disfavored, as shown in figure~\ref{m2ag4q3}.

\begin{figure}
\centering
  \includegraphics[width=0.5\textwidth]{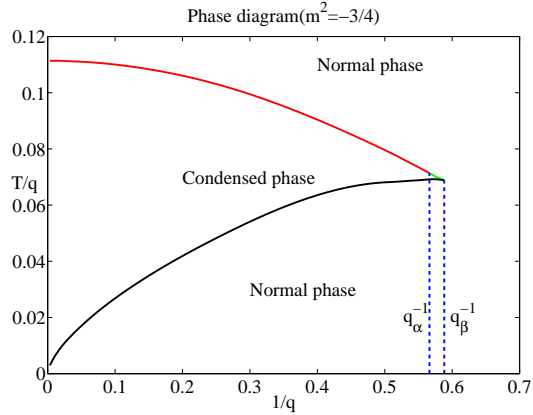}
  \caption{The phase diagram for the case with $m^2=-3/4$. The red and blue lines stand for second order and first order phase transitions, respectively, while the black one for the zeroth phase transition. When $1/q\rightarrow0$, we have $T_{c2}/q\simeq0.1114$ and $T_{c0}/q\rightarrow0$.}\label{P-D5}
\end{figure}

A phase diagram in terms of temperature and charge is shown in figure~\ref{P-D5} in the case with $m^2=-3/4 $. There the red, blue and black
curves represent second order, first order and zeroth order phase transitions, respectively.

\section{The Complete Phase Diagram}
\label{completepd}
The above studies tell us that, depending on two model parameters $m^2$ and $q$, the Einstein-Maxwell-complex vector field theory exhibits rich phase structures. From the perspective of dual field theory, the system admits four phases including an insulating phase described by the pure AdS soliton, a soliton superconducting phase described by the hairy soliton, a conducting phase described by the RN-AdS black hole, and a black hole superconducting phase described by the hairy black hole. For given $m^2$ and $q$, the complete phase diagram can be constructed in terms of temperature and chemical potential. At each point in $T$-$\mu$ plane, one should find the phase which has the lowest grand potential. Since the dual system in spatial directions is homogeneous and infinite, the total grand potential $\Omega$ is divergent. So we consider the grand potential density defined by $\varpi=\Omega/V_3$ for both the soliton and black hole cases. Remember that we fix the size of the compact directions $\eta$ and $z$  to be $\Gamma=\pi$.

As a warmup, let us first show some properties of the grand potential density $\varpi(T,\mu)$. We note that one may associate an arbitrary temperature to the Euclidean soliton for a given chemical potential. So the grand potential density of the soliton case is independent of temperature, i.e., $\varpi_{\text{soliton}}(T,\mu)=\varpi_{\text{soliton}}(\mu)$. Therefore, the phase boundary between insulator and soliton superconductor should be a line parallel to the $T$ axis. According to the scaling law~\eqref{transform}, the grand potential density of black hole for different chemical potential is given by $\varpi_{\text{BH}}(T,\mu)=\varpi_{\text{BH}}(T,\mu=1)\mu^4$. Taking advantage of above properties, one can easily find a first order phase transition between the AdS soliton and RN-AdS black hole,~\footnote{This first order phase transition is the planar analogous of the Hawking-Page transition for black holes in global AdS coordinates~\cite{SWHP}.} and  the phase boundary between the RN-AdS black hole and the AdS soliton at zero temperature is at a chemical potential $\mu=12^{1/4}\simeq1.8612$, while the boundary at zero chemical potential locates at $T=1/\pi$. In addition,  note that since the critical temperature between the RN-AdS black hole and the hairy black hole is proportional to $\mu$, this phase boundary between them is a straight line passing through the original point $(T,\mu)=(0,0)$.

Now we begin to construct the $T$-$\mu$ phase diagram. To find the thermodynamically favored phase among four phases, we adopt a method similar to the one in ref.~\cite{Horowitz:2010jq}. In the first step, we scale all solutions to the ones with the same value of chemical potential by using above scaling rules. Then we scan temperature to find the physical phase boundary at this $\mu$ by comparing grand potential densities for all solutions. Repeating this procedure with different values of $\mu$, we can assemble the full $T$-$\mu$ phase diagram.

\begin{figure}
\includegraphics[width=0.5\textwidth]{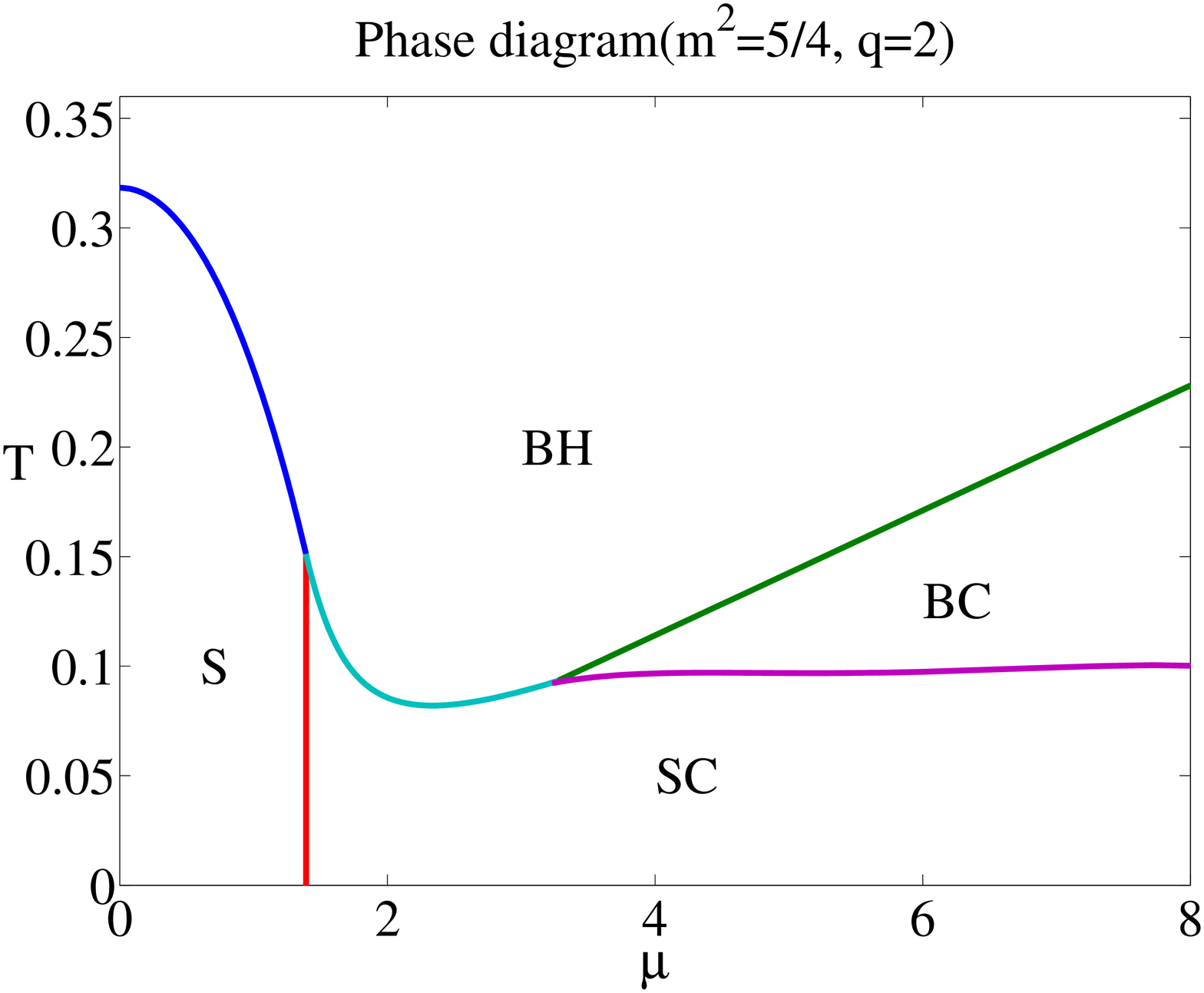}
\includegraphics[width=0.5\textwidth]{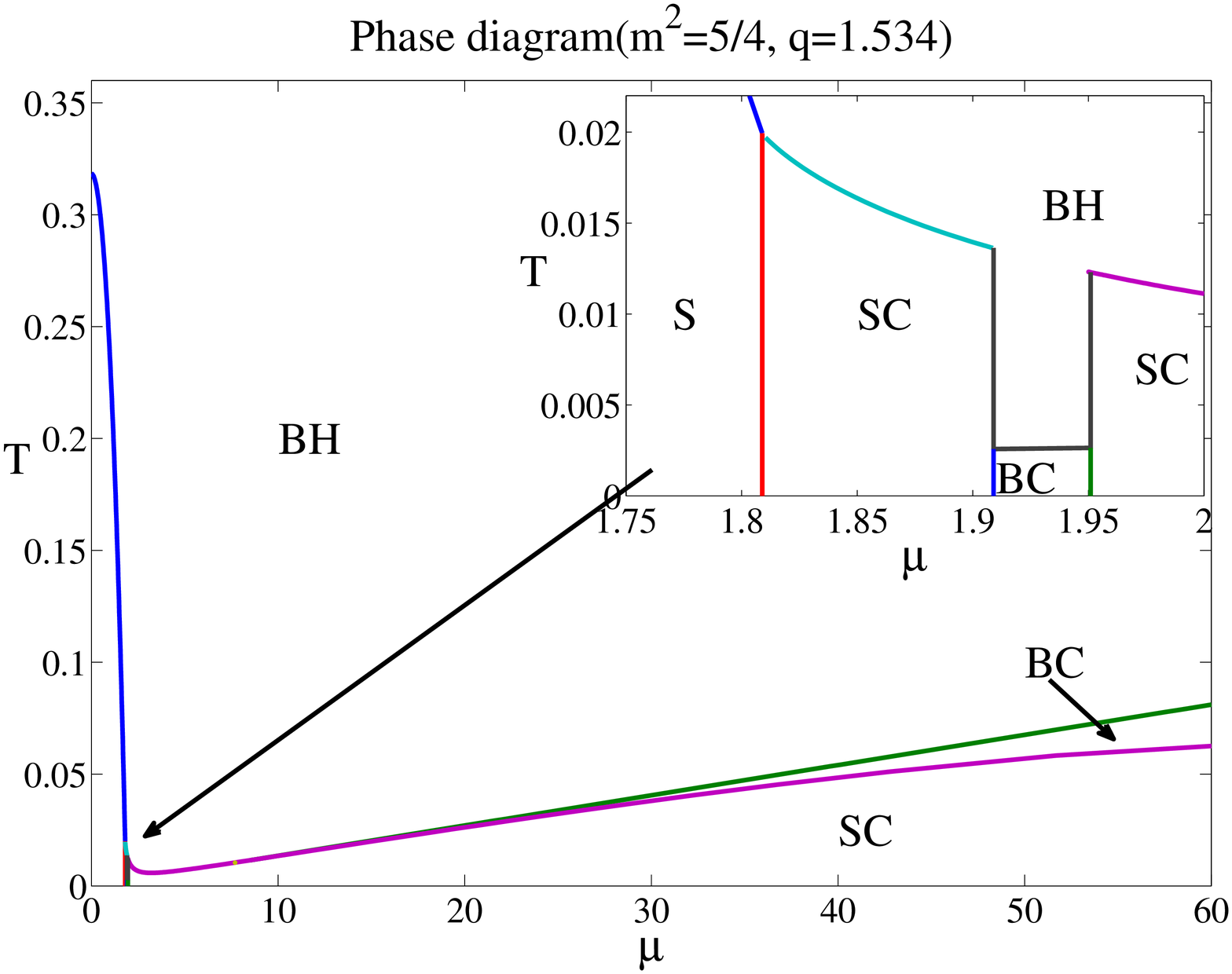}\\
\includegraphics[width=0.5\textwidth]{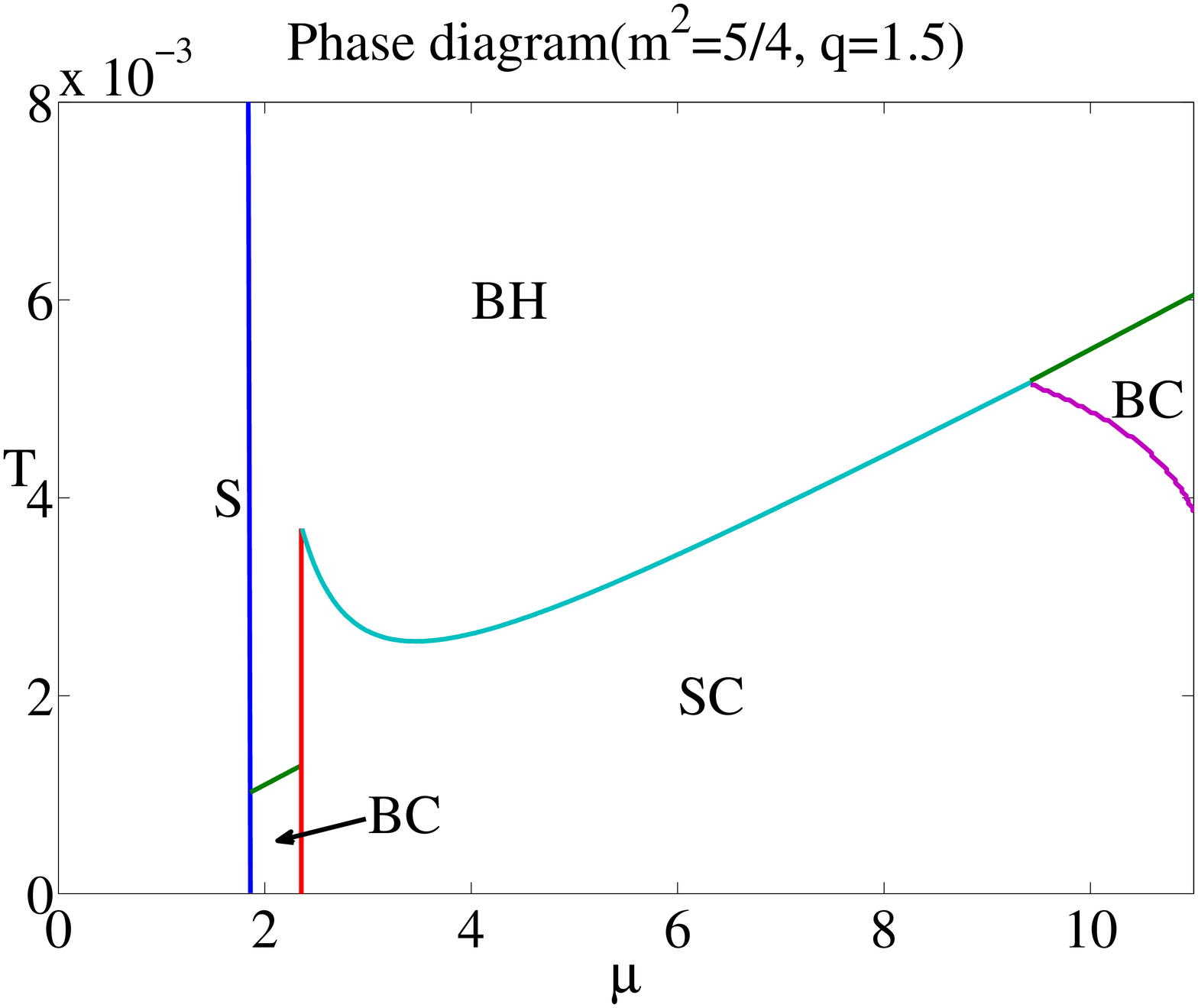}
\includegraphics[width=0.5\textwidth]{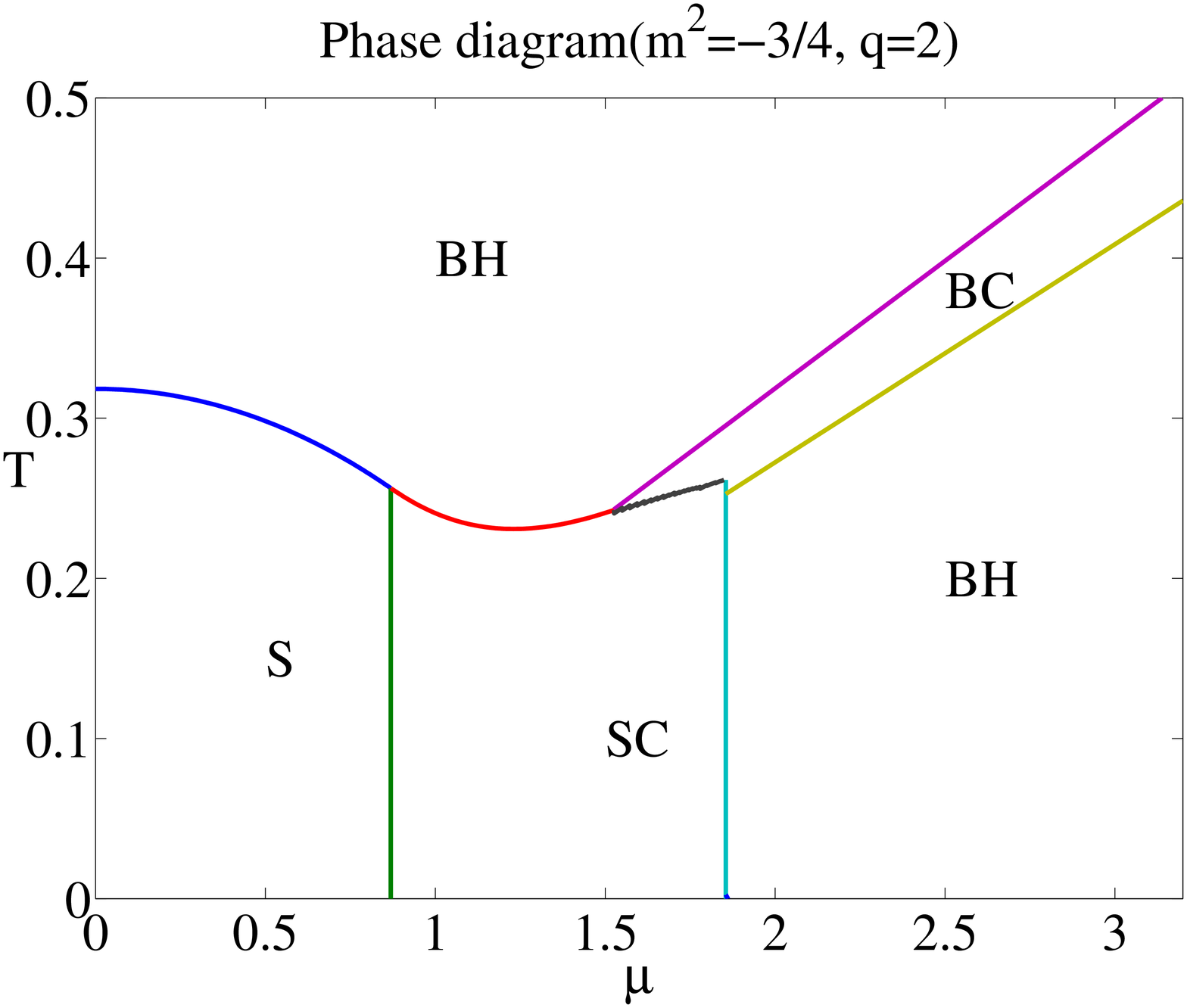}\\
\includegraphics[width=0.5\textwidth]{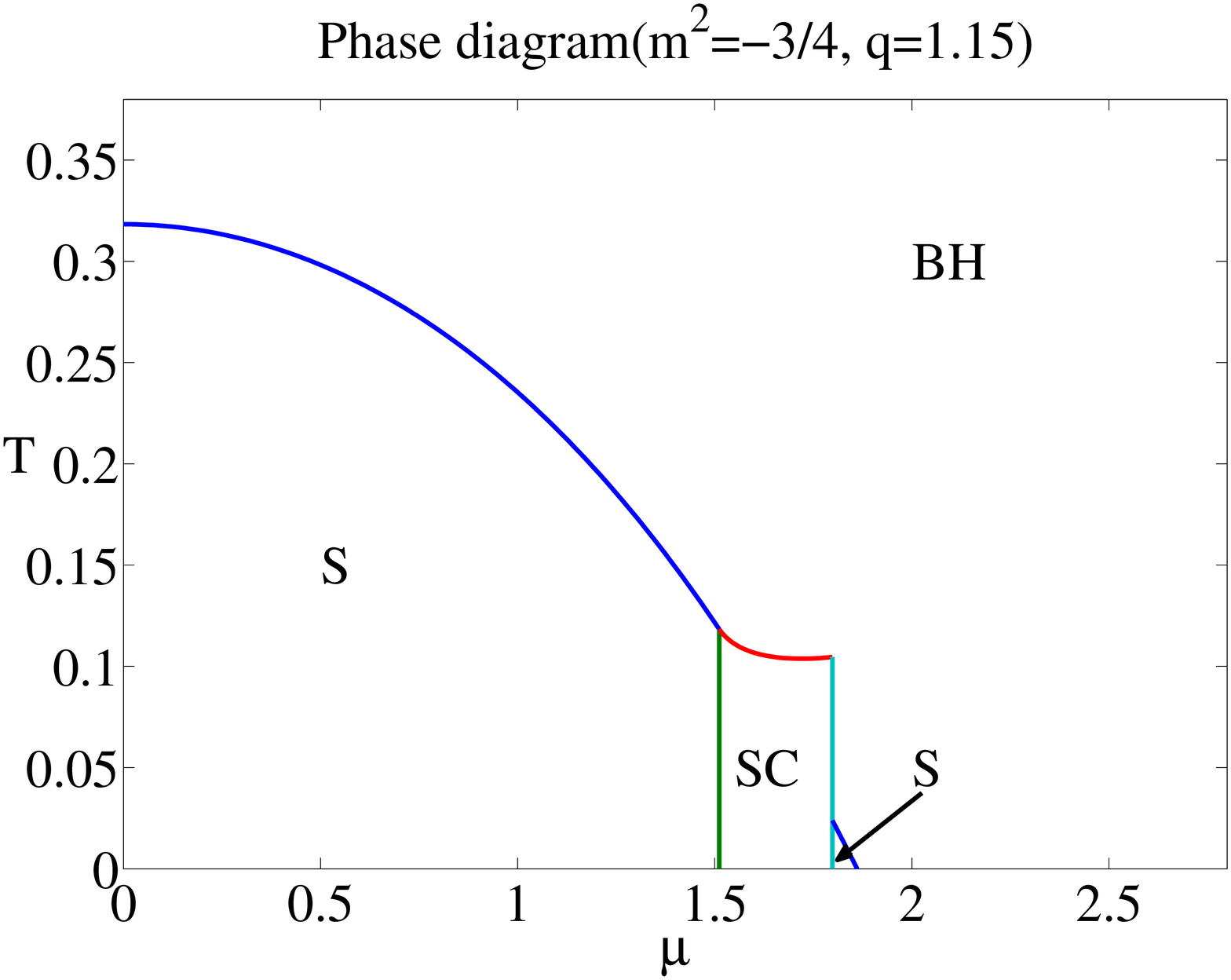}
\includegraphics[width=0.5\textwidth]{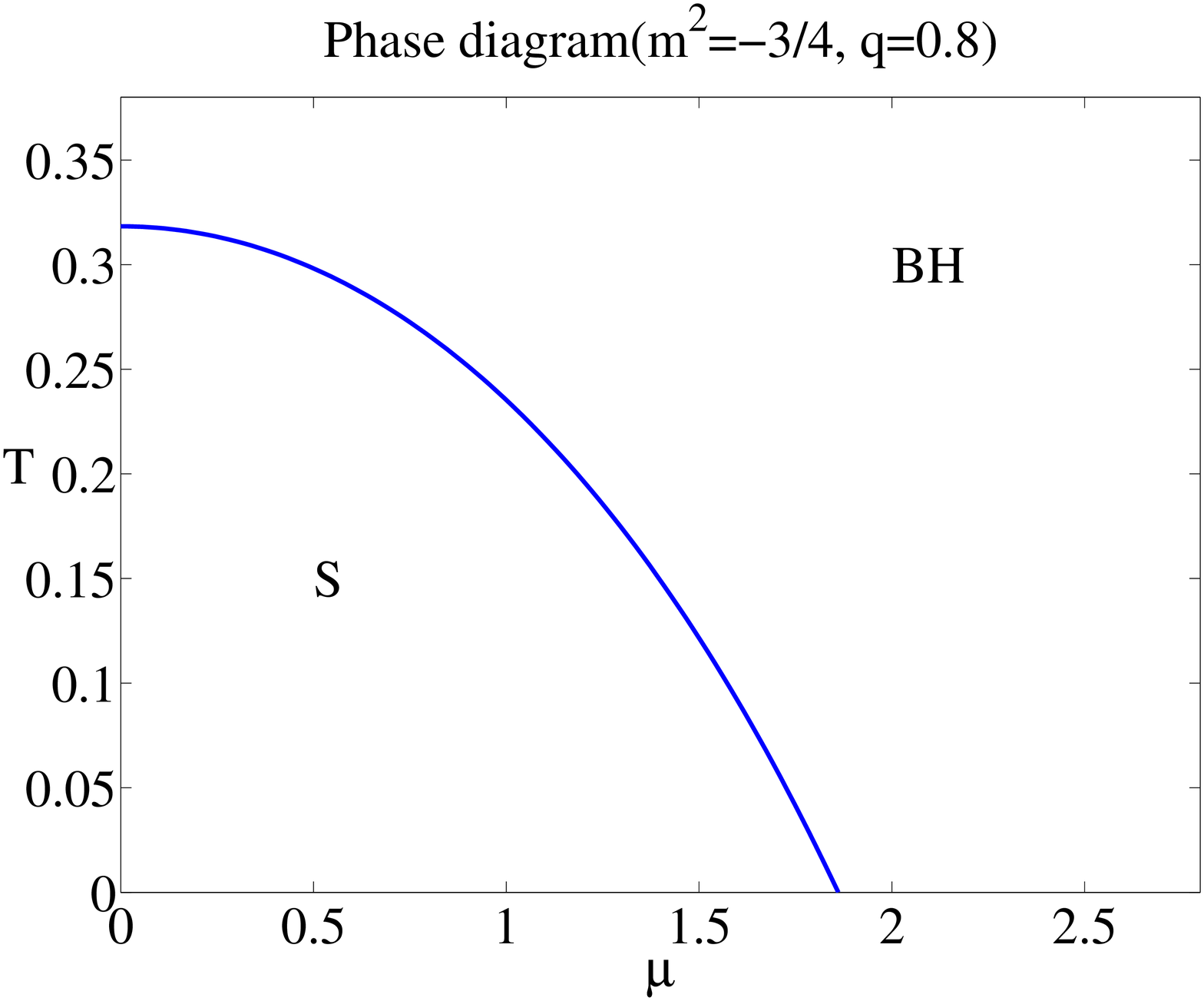}\\
\caption{\label{completephase} The complete phase diagram for the model with different mass square and charge of the vector field. In the figure, S=pure AdS soliton, BH=RN-AdS black hole, SC=hairy soliton, BC=hairy black hole.}
\end{figure}

Remember that for different choice of model parameters $m^2$ and $q$, the phase behaviors can qualitatively change. More precisely, for regions $m^2> m^2_{c1}$, $0\leq m^2\leq m^2_{c1}$ and $m^2<0$, our system exhibits distinguished thermodynamics. It is a very hard work to construct the $T-\mu$ phase diagram for all values of $m^2$. Instead, we consider some typical examples for each parameter range. We hope that the phase diagram will not have a dramatic change for other choice of $m^2$ in each region. We draw some typical examples in figure~\ref{completephase}.

 In the region with $m^2> m^2_{c1}$, let us consider $m^2=5/4$ as an example. The phase structure depends on the strength of back reaction. For weak back reaction with $q>q_2\simeq1.5345$, the phase diagram looks like the top-left plot in figure~\ref{completephase}. In this range of $q$, as we lower $q$, the phase boundary between the insulator and soliton superconductor goes toward larger $\mu$, and the slope of the line separating black hole superconductor from conductor is reduced. This is consistent with the fact that the back reaction makes the phase transition more difficult. Furthermore, the boundary between the two kinds of superconducting phases moves to a lower temperature for smaller $q$. As $q$ is lowered past $q_2$, there will be two discontinuous points in grand potential indicating zeroth order phase transition (see figure~\ref{m2qg4q1.534}). Between two discontinuous points, there is no soliton superconducting phase, thus it changes the phase diagram a lot. As shown in the top-right plot of figure~\ref{completephase}, the soliton superconducting phase is broken into two parts and there is a particular range of chemical potential where the thermodynamically favored phase is replaced by black hole superconductor for lower $T$ and conductor for higher $T$. When the value of $q$ is decreased to the case that the transition from the pure AdS soliton to the hairy soliton becomes zeroth order (see figure~\ref{m2qg4q1.502}), for some suitable values of charge in this region, the insulating phase and soliton superconducting  phase separate from each other, between them there are phases coming from the black hole background. A typical example for this is shown in the middle-left plot in figure~\ref{completephase}.

In the region of $ m_{c2}^2 \le m^2 \leq m_{c1}^2$, we take $m^2=0$ as an example. We find that in this case the $T$-$\mu$ phase diagram is qualitatively similar to the one numerically constructed for the holographic s-wave model~\cite{Horowitz:2010jq} as well as the one schematically plotted in ref.~\cite{Akhavan:2010bf} for the SU(2) p-wave
model. The four phases typically meet in two triple points. As we decrease $q$, the two triple points gradually approach to each other, merge into one quadruple point where four phases can coexist and then pass through each other. There exists a phase boundary in which the black hole superconductor becomes an insulator via a first order phase transition.

In contrast to above two cases, when $m^2$ is negative, the hairy solution can exist only in a certain  chemical potential region for soliton background and temperature region for black hole background. As one can see from figure~\ref{m2ag4q3} and figure~\ref{TomegaC} that the hairy solution becomes thermodynamically irrelevant for sufficiently strong back reaction, even more in the near zero temperature region. Due to this, the $T$-$\mu$ phase diagram has a dramatic change compared to the cases with $m^2\geq0$. Let us consider the case $m^2=-3/4$ as an example. Indeed, the structure of the phase diagram depends on the value of $q$ we choose. A typical diagram for small back reaction is drawn in the middle-right plot of figure~\ref{completephase}. One can see that the lower-right region of the phase diagram is replaced by the conducting phase.  As we lower $q$, the region in which the two superconducting phases dominate is deduced. In the case with $q=1.150$ shown in the bottom-left plot of figure~\ref{completephase}, the soliton superconducting phase shrinks to a smaller region, while the black hole superconducting phase disappears. For sufficiently small $q$, both the black hole superconducting phase and the soliton superconducting phase become thermodynamically disfavored  compared to the uncondensed phases. In this case,  we are left with only the conducting phase and insulating phase at hand. There is a first order phase transition between them shown in the bottom-right plot in figure~\ref{completephase}, which is the well-known confining/deconfining phase transition in gauge field theory~\cite{Witten:1998zw}.

\section{Summary and Discussions}
\label{concdiss}
In this paper we give a detailed study on the holographic p-wave superconductor model based on the Einstein-Maxwell-complex vector field theory with a negative cosmological constant. We work in five dimensional asymptotically AdS backgrounds with full back reaction of matter field on the background geometries. In this model, we find four kinds of bulk solutions given by the pure AdS soliton, RN-AdS black hole and their vector hairy counterparts. According to the AdS/CFT dictionary, the hairy solution is dual to a system with a non-zero vacuum expectation value of the charged vector operator which breaks the U(1) symmetry and the spatial rotation symmetry spontaneously.
 The above four solutions in the bulk correspond to an insulating phase, a conducting phase, a soliton superconducting phase and a black hole superconducting phase, respectively.

In this model there exist two model parameters, i.e., the charge $q$ and the mass square $m^2$ of the charged vector filed $\rho_\mu$. The phase structure of the model
heavily depends on the two parameters. In the black hole background case,  the phase behavior is qualitatively same as the one in the (3+1) dimensional case studied in~\cite{Cai:2013pda2}. One can see second order transition, first order transition, zeroth order transition as well as the ``retrograde condensation". The latter two cases can only appear for the case  $m^2<m_{c}^2$ with $m_{c}^2=0$ suggested by numerical analysis.

For the soliton case, our system exhibits distinguished behavior depending on concrete value of $m^2$.  Qualitatively, the parameter space for $m^2$ is divided
into three regions with $m^2 \ge  m^2_{c1}=0.218 \pm 0.001$, $0 \le m^2 \le m_{c1}^2$ and $m^2 <0$, respectively. In each region, the phase behavior depends on the
strength of back reaction $q$. There also exist second order, first order and zero order phase transitions as well as the ``retrograde condensation". An interesting
observation is that when $m^2 <0$, the hairy black hole solution does not exist as the temperature is lower than a certain value (see figure \ref{TomegaB}), which leads to the zeroth phase transition at the critical point.  We draw the phase
diagrams (see figures~\ref{PD1}, \ref{PD2} and \ref{PD3}) in each region of mass parameter space. Compared to the black hole background case, an additional
 complication appears in the soliton background case.

Recall that our one parameter family of hairy solutions are labeled by the value of $\rho_x$ at the tip of the soliton. For $m^2\geq 0$, as we lower $q$ to a critical value, there exists a region of chemical potential, in which the hairy soliton solution does not exist. As a result, the curve of the condensate (or grand potential) parameterized by $\rho_x(r_0)$ breaks down, which results in two discontinuous points (see figures~\ref{m2qg4q1.534}, \ref{m2qg4q1.502}, \ref{m2qJq1.502}, and \ref{m2qg4q1.15}).  For the existence of this discontinuous point and the hairy soliton solution at this point, we give a detail analysis in appendix~\ref{app1} and appendix~\ref{app1b}. Some physical meanings of this discontinuous point both at the gravity side and the dual field theory should be further studied.
 For the case with $m^2 < 0$, the hairy soliton solution terminates at a finite chemical potential for weak back reaction, while for sufficiently large back reaction, it becomes thermodynamically subdominant compared to the pure AdS soliton. See figures \ref{m2ag4q2} and \ref{m2ag4q3} for details.

With all four phases at hand, we construct some complete phase diagrams in terms of chemical potential and temperature. As one can see,  there are many types of phase transitions in both soliton and black hole backgrounds, which depend on the values of parameters $m^2$ and $q$. The $T$-$\mu$ phase diagrams are expected to be much more complicated than the ones for the holographic s-wave model and SU(2) p-wave model~\cite{Horowitz:2010jq,Akhavan:2010bf}. Some typical examples are shown in figure~\ref{completephase}. It has been suggested that the Einstein-Maxwell-complex vector field model is a generalization of the SU(2) p-wave model in the sense that the vector field has a general mass and gyromagnetic ratio $\gamma$. Comparing our complex vector field model to the SU(2) model with a constant non-Abelian magnetic field, we find that the SU(2) p-wave model can be recovered by the restriction $m^2=0$ and $\gamma=1$ at least in our setup in ref.~\cite{Cai:2013kaa}. Similar statement holds for the black hole case. Indeed, for the special case $m^2=0$, our model exhibits very similar phase structures as the SU(2) model~\cite{Cai:2013oma,Ammon:2009xh}. In particular, the complete phase diagrams in our model with $m^2=0$ are qualitatively same as those drawn schematically in ref.~\cite{Akhavan:2010bf}. We have freedom to choose other values of $m^2$ in our model, thus it can be used to describe much rich phenomena in the dual strong coupled systems. As one can see, depending on the values of $m^2$ and $q$, we have indeed many other types of phase diagrams which do not appear in the SU(2) p-wave model~\cite{Akhavan:2010bf} as well as in holographic s-wave model~\cite{Horowitz:2010jq}. In addition, let us mention that in the soliton background case, we have checked
that there also exists a zeroth order phase transition in the SU(2) p-wave model as the back reaction is strong enough, which is missed in
the previous studies.

Finally some remarks are in order. (1) There exist two critical masses of the vector field in our model, $m_{c1}^2=0.218\pm 0.001$ and $m_{c2}^2=0$, in the soliton background case, a critical mass, $m^2_{cb}=0$, in the black hole background case. Note that the latter also appears in the (3+1) dimensional black hole case~\cite{Cai:2013pda2}, there some possible implications of the critical mass have been discussed and therefore we will not repeat here. While the critical mass $m_{c1}^2=0.218\pm 0.001$ looks spacetime dimension dependent, its appearance is interesting and is worthy to further investigate. Unfortunately at the moment we have
no any physical interpretation for this. (2) In both the soliton and black hole backgrounds, our model exhibits zeroth order phase transition, which has a discontinuity of grand potential (free energy) at the critical point. In ref.~\cite{Maslov:2004} the author argues that a zeroth order
phase transition could appear in the theory of superfluidity and superconductivity and  presents an exactly solvable model for such a phase transition. The authors of ref.~\cite{SYZhao} show that a zeroth order phase transition exists in an exactly solvable pairing model
for superconductors with $p_x+ip_y$-wave symmetry. (3) In the soliton background case, our model shows the insulator/superconductor/insulator/superconductor phase transitions when the chemical potential continuously increases. Indeed such
a series of phase transitions is expected to appear from theoretical perspective in some superconducting materials, for example, see the
phase diagram in the figure 10 of ref.~\cite{Sachdev}.  (4) We can see from the complete phase diagrams in terms of temperature and chemical potential that in some cases, more than one superconducting phase appear in a phase diagram in our model. The phase diagram for some realistic superconducting materials is usually complicated, and indeed, more than one superconducting phases can occur, for example, see ref.~\cite{Kordyuk,Bauer,Wu,Chubukov,Richter,Yuan}. Definitely, it is of great interest to see whether our model is relevant to these superconducting materials.


\section*{Acknowledgements}

This work was supported in part by the National Natural Science Foundation of China (No.10821504, No.11035008, No.11205226, No.11305235 and No.11375247), and in part by the Ministry of Science and Technology of China under Grant No.2010CB833004.

\appendix
\section{The regular boundary conditions for the equations of motion}
\label{app1}
In this appendix, some details for solving the coupled equations of motion~\eqref{eoms} will be given. In particular, we will
show why in the soliton background case, the hairy soliton solution does not exist in some region of chemical potential, shown in subsection~\ref{sect:superconducor1}.  The black hole case will be compared in the end of this appendix.

In order to find the solutions for all the six functions $\mathcal{F}=\{g(r),f(r),h(r),\chi(r),\rho_x(r),\phi(r)\}$, one must impose suitable boundary conditions at boundary~$r\rightarrow\infty$ and the tip $r=r_0$. The asymptotic behaviors for these functions near the infinite boundary can be easily obtained from the equations~\eqref{eoms}, which are shown in equations~\eqref{boundary}. The condition of $\rho_{x_-}=0$ leads the problem to be a boundary value problem. In this paper, the method to solve them is shooting method, i.e, by finding some suitable initial values at the tip for each function in~$\mathcal{F}$ in order to give result of $\rho_{x_-}=0$ near the boundary when we integrate the equations~\eqref{eoms} from the tip to the boundary. To solve the equations~\eqref{eoms}, one needs ten initial conditions, i.e, $\{g(r_0),f(r_0),h(r_0),\chi(r_0),\rho_x(r_0),\phi(r_0),f'(r_0),h'(r_0),\rho'_x(r_0),\phi'(r_0)\}$.

By the scaling symmetries in~\eqref{scaling1}-\eqref{scaling4}, one can first set $\{r_0=1,f(r_0)=1,h(r_0)=1,\chi(r_0)=0\}$ .  The first order derivatives of $f(r),h(r),\rho_x(r)$ and $\phi(r)$ at the tip can be obtained by imposing the requirement that each term in equations~\eqref{eoms}  behaves  regular  at the tip.  By some calculations, we have

\begin{equation}\label{init1}
f'(r_0)=\phi'(r_0)\phi_0=2q^2\rho_{x0}^2\phi^2_0/g'(r_0),
\end{equation}
\begin{equation}\label{init2}
h'(r_0)=-2\rho'_{x0}\rho_0=2\rho_{x0}^2(m^2-q^2\phi^2_0)/g'(r_0),
\end{equation}
\begin{equation}\label{init3}
g'(r_0)=4-q^2\rho_{x0}^2\phi^2_0/3,
\end{equation}
 where $\rho_{x0}=\rho_x(r_0)$ and $\phi_0=\phi(r_0)$, which are two free parameters at hand. We can choose fixing $\rho_{x0}$ using $\phi(r_0)$ as the shooting parameter or vice versa to match the source free condition, i.e., ${\rho_x}_-=0$. After solving the coupled differential equations, we should use the scaling symmetries~\eqref{scaling1}-\eqref{scaling4} again to satisfy the asymptotic conditions $f(\infty)=1$, $h(\infty)=1$ and $\chi(\infty)=0$ and the fixed compacted length $\Gamma=\pi$ of the coordinate $\eta$. Then we can obtain the condensate $\langle \hat{J^x}\rangle$, chemical potential $\mu$,  charge density $\rho$, and grand potential  by reading off the corresponding coefficients in~\eqref{boundary}, respectively.

\begin{figure}
  \includegraphics[width=0.5\textwidth]{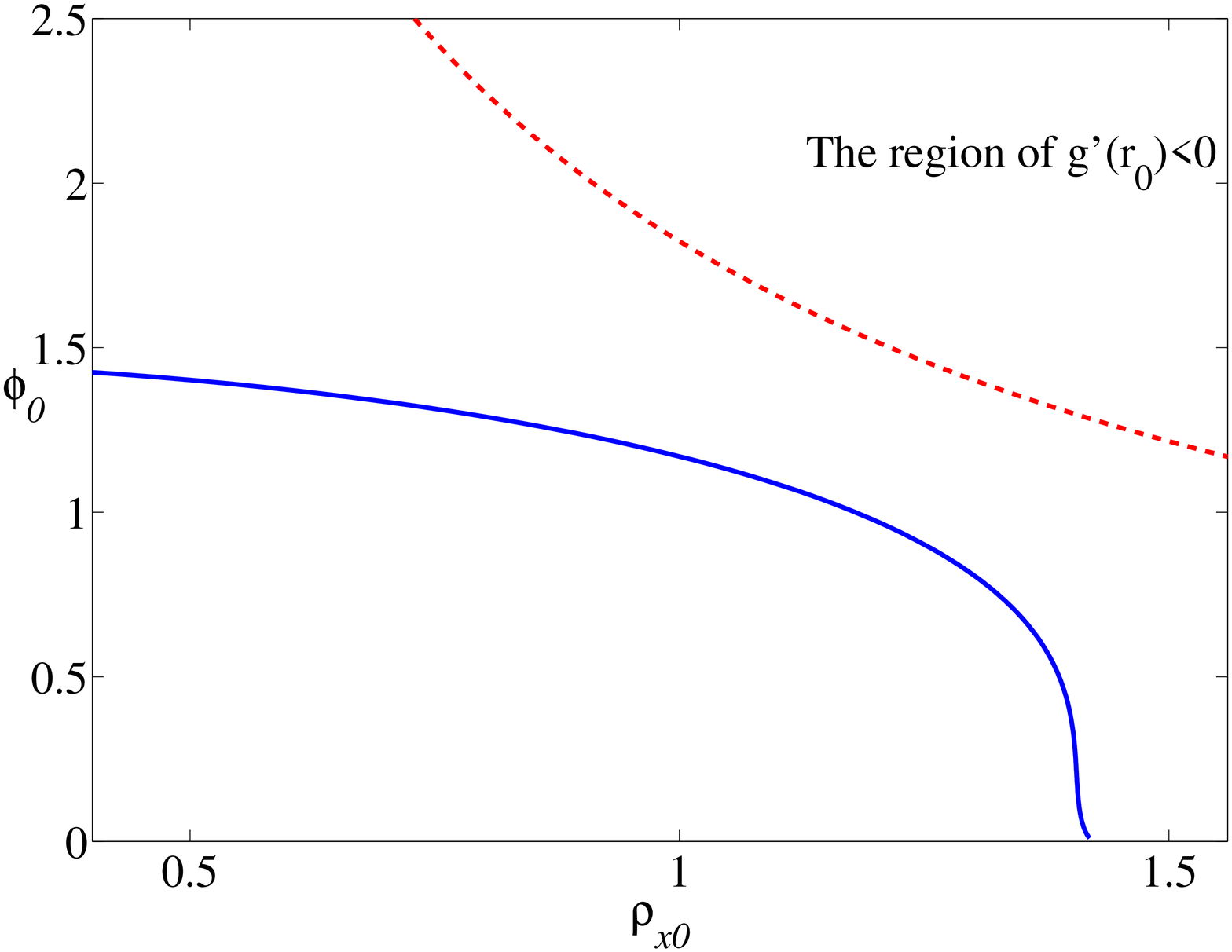}
  \includegraphics[width=0.5\textwidth]{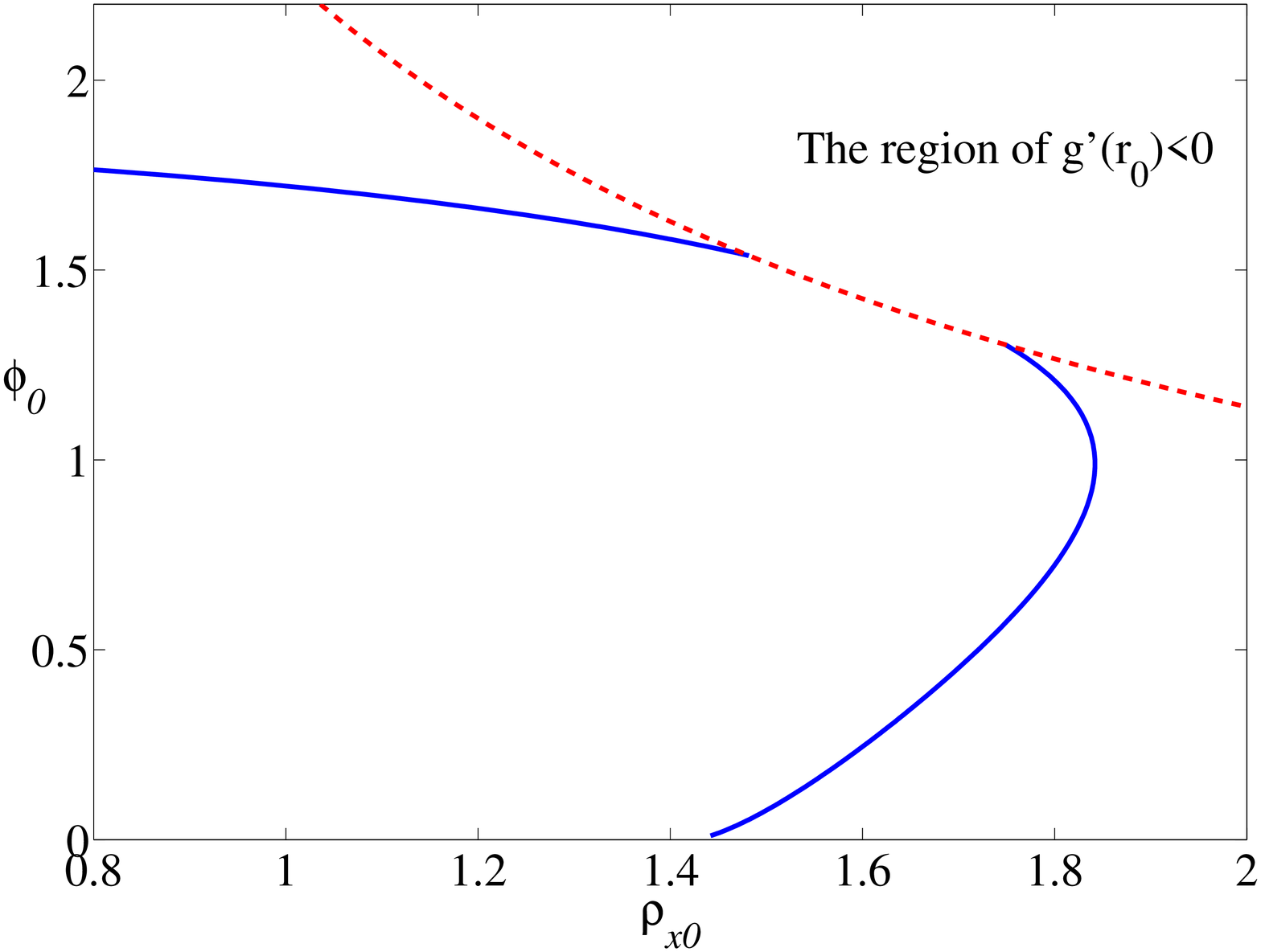}\\
  \caption{The orbit of the initial values $\rho_{x0}$ and~$\phi_0$ for the shooting method in the case of $m^2=5/4$ with $q=1.900$ (left plot) and $q=1.520$ (right plot). The blue solid lines are the orbits for the initial values $\rho_{x0}$ and~$\phi_0$, and the red dashed lines stand for the curves $g'(r_0)=0$.  The regions below the red dashed lines give $g'(r_0)>0$. In the left plot, the blue line is continuous and has no any intersection with the curve $g'(r_0)=0$. In the right plot, the blue lines have two intersection points at~$\phi_0\simeq1.5378$ and~$\phi_0\simeq1.3025$, respectively.}\label{rhox-phi0}
\end{figure}

The equation~\eqref{init3} should be considered  carefully, as it gives a restriction on the solution for equations~\eqref{eoms}. By definition, the tip $r_0$ is the maximal zero point of $g(r)$. Therefore, $g'(r_0)$ must be non negative. For the non-extremal case, we should have a positive value of $g'(r_0)$, which gives a restriction for the initial values of $\rho_{x0}$ and~$\phi_0$,
\begin{equation}\label{restrci1}
4-q^2\rho_{x0}^2\phi^2_0/3>0
\end{equation}
It is precisely this restriction so that we cannot get the hairy soliton solution within some region of chemical potential in the condensed phase, which is shown in figures~\ref{m2qg4q1.534}, \ref{m2qg4q1.502} and \ref{m2qg4q1.15}. In order to show this discontinuous phenomenon
 more visualizable,  in figure~\ref{rhox-phi0} we show two kinds of orbits for the initial values $\rho_{x0}$ and~$\phi_0$ for the case with $m^2=5/4$ and $q=1.900$ and $1.520$, respectively. In the figure, The blue solid line is the orbit for the initial values $\rho_{x0}$ and~$\phi_0$, while the red dashed lines stands for the curve $g'(r_0)=0$. The region below the red line has $g'(r_0)>0$.

From the figure~\ref{rhox-phi0}, one can see that the orbit of $\phi_0$ and $\rho_{x0}$ behaves well and has no any intersection with the dashed red line when $q=1.900$.  In this case, the solution orbit leads to the continuous grand potential and condensate in the condensed phase, as shown, for example, in figure \ref{m2qg4q1.6}. However, when $q=1.520$ as shown in the right plot of figure~\ref{rhox-phi0}, the solution orbit is broken into two parts by the curve $g'(r_0)=0$. It leads to the discontinuity of the grand potential in the condensed phase, just as shown in figures~\ref{m2qg4q1.534}, \ref{m2qg4q1.502} and \ref{m2qg4q1.15}. Between these two intersection points, there do not exist any hairy
soliton solutions and thus in that region we have to replace hairy soliton solutions by the pure AdS soliton because the latter can
associated with any chemical potential.

For the black hole case, we can also construct similar boundary conditions at the black hole horizon and AdS boundary. In this case, we can use the values of $\rho_x(r_h)$ and $\phi'(r_h)$ as the shooting parameters. There is also a restriction at the horizon as
\begin{equation}\label{restrci2}
a'(r_n)=4-\frac16\phi'^2(r_h)>0.
\end{equation}
One can see this restriction only depends on $\phi'(r_h)$, which is very different from the one \eqref{restrci1} in the soliton case. The numerical results in the region where we have scanned show this restriction does not cause any discontinuity in the condensed phase in the black hole case.

\section{The extremal hairy soliton solution}
\label{app1b}
In the main text we have assumed $g'(r_0)>0$ in seeking for the solutions with the regular condition at the tip. In this appendix
we will give the extremal solution with $g'(r_0)=0$. In order to find the extremal solution, it is  helpful to observe the behaviors of these functions in ${\cal F}$ when $g'(r_0)$ approaches to zero.

\begin{figure}
  \centering
  \includegraphics[width=0.5\textwidth]{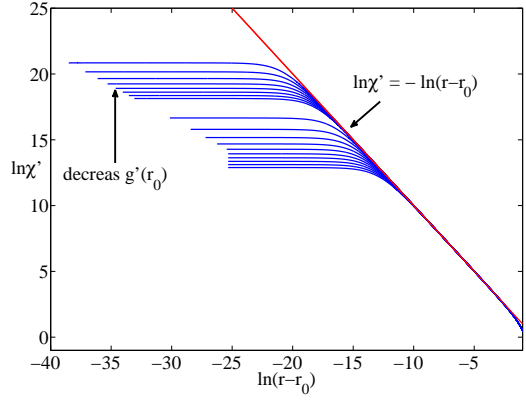}
  \caption{The asymptotic behavior of $\chi'$. From the figure, one can see  that when $g'(r_0)\rightarrow0$, there is an asymptotic behavior $\chi'=\frac1{r-r_0}$. Here the values of $g'(r_0)$ are chosen from $2\times10^{-2}$ to $4\times10^{-4}$.}\label{lnchi}
\end{figure}

 In figure~\ref{lnchi} we plot the behavior of $\ln\chi'$ near the tip when $g'(r_0)$  tends to zero. It shows that when $g'(r_0)$ goes to zero, the behavior of $\chi'$ behaves as  $\chi'=\frac1{r-r_0}$.  Taking this behavior into account, we propose the following expansion near the tip
\begin{equation}\label{expansion}
\begin{split}
&g(r)=g_j(r-r_0)^{1+j}+\cdots,~~\chi(r)=\ln(r-r_0)+\cdots,\\
&\phi(r)=\phi_0+\phi_{10}(r-r_0)^{1-j}+\cdots,~~\rho_x(r)=\rho_{x0}+\rho_{x10}(r-r_0)^{1-j}+\cdots,\\
&f=1+f_{10}(r-r_0)^{1-j}+\cdots, ~~h(r)=1+h_{10}(r-r_0)^{1-j}+\cdots.
\end{split}
\end{equation}
Here the scaling symmetry has been used to scale $f(r)$ and $h(r)$ to be unity at the tip. Substituting these assumptions into the equations~\eqref{eoms}, we find only $j=\frac12$ can give a self-consistent expansion with the coefficients
\begin{equation}\label{expacoeff}
\begin{split}
\phi_{10}=f_{10}/\phi_0=\frac{8q^2\rho_{x0}^2\phi_0}{g_j},~~\rho_{x10}=-h_{10}/2\rho_{x0}=\frac{4\rho_{x0}(m^2-q^2\phi_0^2)}{g_j},\\
\phi_0^2\rho_{x0}^2q^2-12=0,~~g_j=\frac4{q\phi_0}\sqrt{48q^2\phi_0^2-24m^2-4m^2q^2\phi_0^2-72q^2+3q^4\phi_0^4+m^4}.
\end{split}
\end{equation}
In this case with $j=1/2$, a remarkable fact is that although $g'(r_0)=0$, the solution with the expansion (\ref{expansion}) still has a finite
compacted length $\Gamma = 8\pi/(g_j r_0)$ for the coordinate $\eta$.  In addition, the solution looks singular at the tip because $\chi(r)$ is divergent there. In fact, this is a coordinate singularity and the solution is regular at the tip. To see this, one can introduce
 $\xi=(r-r_0)^{1/4}$,  near the tip $r=r_0$,  the metric then has the following form
\begin{equation}\label{formtip}
ds^2=\frac{16}{g_j}d\xi^2+r_0^2(-dt^2+dx^2+dy^2+g_j\xi^2d\eta^2).
\end{equation}

One can use the expansion~\eqref{expansion} with the coefficients \eqref{expacoeff} to solve the equations~\eqref{eoms} with the source free condition by scanning the shooting parameter~$\phi_0$.~\footnote{To avoid the divergence of $\chi',~f',~h',~\phi',~\rho'$ at the tip in the
numerical calculations, it is better to change the coordinate from $r$ to $\xi$ and make a suitable variable substitution on $\chi$ such as $\chi\rightarrow e^{\chi/4}$.} In the case of $m^2=5/4$ and $q=1.520$, we find the first two values of $\phi_0$ are 1.302900 and 1.537516, which are very closed to the corresponding values in figure~\ref{rhox-phi0}, the latter are obtained by setting $g'(r_0)=4\times10^{-4}$.

\section{The stress energy tensor}
\label{app2}
In this appendix, we study  the  stress energy tensor of the dual field theory on the boundary. From the AdS/CFT correspondence, the stress energy  tensor  can be obtained from the Brown-York tensor of the bulk spacetime. In the asymptotic $AdS_5$ spacetime, the Brown-York tensor on the boundary $\partial M$ can be computed as~\cite{VBPK}
\begin{equation}\label{BY1}
T^{\mu\nu}=-\frac1{\kappa^2}(\mathcal{K}^{\mu\nu}-\mathcal{K}\bar{h}^{\mu\nu}+\frac3L\bar{h}^{\mu\nu}+\frac12 G^{\mu\nu}).
\end{equation}
Here $\bar{h}^{\mu\nu}$ is the boundary metric, $G^{\mu\nu}$ is the Einstein tensor for the boundary metric $\bar{h}^{\mu\nu}$.  $\mathcal{K}^{\mu\nu}$ is extrinsic curvature,  defined by the outward pointing normal vector to the boundary $\widehat{n}$ as
\begin{equation}\label{K1}
\mathcal{K}^{\mu\nu}=\frac12(\nabla^{\mu}\widehat{n}^{\nu}+\nabla^{\nu}\widehat{n}^{\mu}).
\end{equation}
In our case, the boundary spacetime is flat, so the Einstein tensor for the boundary metric vanishes. In what follows we discuss the stress energy tensor for the soliton and black hole backgrounds separately.

{\bf The soliton case}

By the metric ansatz~\eqref{ansatz} and the asymptotic expansion for the metric in~\eqref{boundary},  the leading terms of the Brown-York tensor turn out to be
\begin{equation}\label{BY2}
\begin{split}
 2\kappa^2T^{tt}=r^{-6}(4h_4+g_4-4\chi_4),\\
 2\kappa^2T^{xx}=r^{-6}(4\chi_4-g_4-4f_4),\\
 2\kappa^2T^{yy}=r^{-6}(4\chi_4-g_4-4f_4-4h_4),\\
 2\kappa^2T^{\eta\eta}=r^{-6}(3g_4-4h_4-4f_4).
\end{split}
\end{equation}
In the following, we omit the factor $2\kappa^2$ for convenience. The non vanishing components of the stress energy tensor $t^{\mu\nu}$ for the dual field theory on the boundary are
\begin{equation}\label{BY3}
\begin{split}
t^{tt}=h_4+g_4-4\chi_4, \\
t^{xx}=4\chi_4-g_4-4f_4, \\
t^{yy}=4\chi_4-g_4-4f_4-4h_4, \\
t^{\eta\eta}=3g_4-4h_4-4f_4.
\end{split}
\end{equation}
The trace of the boundary stress energy tensor is
\begin{equation}\label{trac1}
Tr(t^{\mu}_{\nu})=12(\chi_4-f_4-h_4).
\end{equation}
 Note that the equation of motion for $\chi(r)$ reads
\begin{equation}\label{chi1}
\chi'-\frac{f'}{f}-\frac{2g'}{g}-\frac{h'}{h}+\frac{2{\rho_x'}^2}{3rh}-\frac{\phi'^2}{3rf}-\frac{2q^2\rho_x^2\phi^2}{3r^5fgh}+\frac{8}{rg}-\frac{8}{r}=0.
\end{equation}
Substituting the expansion (\ref{boundary}) into the above equation, in the limit $r \to \infty$, we have
\begin{equation}\label{chi2}
4(f_4+h_4-\chi_4)=-\frac{2(1+\delta)^2\rho_{x+}^2}{3r^\delta}+{\cal O} (\frac1{r^2}),~~\delta=\sqrt{1+m^2}.
\end{equation}
It is clear from the above equation that the dual field theory is traceless as $\delta >0$. When $\delta =0$, the trace looks non-vanishing
with $Tr(t^{\mu\nu})=2\rho_{x+}^2$. But this is not true. The reason is as follows. Note that the case $\delta =0$ is equivalent to the
lower bound with $m^2=-1$. As mentioned above, in this case, the expansion in (\ref{boundary}) is no longer valid, instead some logarithmic terms
should appear in the asymptotic expansion.  In fact, it can be shown that with these additional terms the dual stress energy tensor is still traceless.

{\bf The black hole case}

In this case, by the metric ansatz~\eqref{BHans} and the asymptotic form of the metric in~\eqref{boundary2}, we have the nontrivial terms of the Brown-York tensor near the boundary
\begin{equation}\label{BY2b}
\begin{split}
 T^{tt}=r^{-6}(4c_4-3a_4),\\
 T^{xx}=r^{-6}(4b_4-a_4),\\
 T^{yy}=r^{-6}[4(b_4-c_4)-a_4],\\
 T^{zz}=r^{-6}[4(b_4-c_4)-a_4],
\end{split}
\end{equation}
where we have omitted the factor $2\kappa^2$ for convenience. The stress energy tensor $t^{\mu\nu}$ for the dual field theory on the boundary reads
\begin{equation}\label{BY3b}
\begin{split}
t^{tt}=4c_4-3a_4, \\
t^{xx}=4b_4-a_4, \\
t^{yy}=[4(b_4-c_4)-a_4], \\
t^{zz}=[4(b_4-c_4)-a_4],
\end{split}
\end{equation}
with its trace
\begin{equation}\label{trac1b}
Tr(t^{\mu}_{\nu})=12(b_4-c_4).
\end{equation}
In this case, let us consider the equation of motion for the metric function $b(r)$
\begin{equation}\label{b1}
b'-\frac{2a'}{a}-\frac{c'}{c}+\frac{2\rho_x'^2}{3rc}-\frac{re^b\phi'^2}{3a}-\frac{2e^b q^2\rho_x^2\phi^2}{3ra^2c}+\frac{8r}{a}-\frac{4}{r}=0.
\end{equation}
Substituting the expansion \eqref{boundary2} into the above equation, and taking the limit $r \to \infty$, we have the following asymptotic form
\begin{equation}\label{b2}
4(c_4-b_4)=-\frac{2(1+\delta)^2\rho_{x+}^2}{3r^\delta}+{\cal O}(\frac1{r^2}),~~\delta=\sqrt{1+m^2}.
\end{equation}
As the soliton case, one can see that the stress energy tensor is also traceless in the black hole background case.

Finally let us mention an interesting fact on the stress energy tensor of the dual field theory. Naively thinking, one may expect
that the stress energy tensor should be anisotropic with $t^{xx} \ne t^{yy}$ in both hairy soliton and black hole cases, for which the metric functions $h(r)$ and $c(r)$ are nontrivial and from the point of view of dual field theory, the x-component of the vacuum expectation value of the vector operator is non zero, while the y-component vanishes. But this anisotropy does
not appear in the stress energy tensor. Our numerical results suggest that $h_4$ and $c_4$ are zero,~\footnote{In the soliton case with $m^2=-3/4,q=2.000$ , the numerical results show that $f_4$, $g_4$ and $\chi_4$ are of order one, but $h_4$ is zero up to a numerical error $10^{-12}$, while in the black hole case with the same model parameters, the numerical results show that $a_4$ is of order one, but $c_4$ and $b_4$ vanish up to a numerical error $10^{-12}$.}  which leads to $t^{xx}=t^{yy}$ in both cases.

A similar phenomenon has been reported in the paper~\cite{Donos:2013cka}, the authors have given an expression for the variation of the free energy of periodic AdS black holes with respect to the periods of transverse coordinates, which gives an explanation for this phenomenon in the black hole and soliton backgrounds.

\section{The scaling relations between the critical chemical potential and charge}
\label{app3}
In this appendix, we will give some discussions on the scaling relation~\eqref{scal1}, which holds for the large $q$ case. We will focus on the soliton case.  The relation between $\mu_{c1}$ and $q$ can be obtained by semi-analytical method through the general Heun function, while the relation between $\mu_{c2}$ and $q$ can only be obtained by numerical method. We first study the former case.

In the case with large $q$, the probe limit is a good description near the phase transition point. In that case, the chemical potential takes the critical value, while the vector field is very small, whose influence on the background and electromagnetic field can be neglected. This means that we can just treat the vector field as a perturbation and take the background solution as
\begin{equation}\label{prob1}
g(r)=1-\frac1{r^4},~~f(r)=h(r)=1,~~\chi(r)=0,~~\phi(r)=\mu_{c1}.
\end{equation}
Then the equation for the vector reads
\begin{equation}\label{probrhox}
\rho_x''+(\frac{g'}{g}+\frac{3}{r})\rho_x'+\frac{q^2\mu_{c1}^2}{r^4fg}\rho_x-\frac{m^2}{r^2g}\rho_x=0.
\end{equation}
By introducing
\begin{equation}\label{probF}
r=\frac1z,~~\rho_x(r)=F(z)z^{1+\delta},~~\delta=\sqrt{1+m^2},
\end{equation}
the equation~\eqref{probrhox} can be rewritten as
\begin{equation}\label{probF2}
F''(z)+\frac{z^4(5+2\delta)-(2\delta+1)}{z(z^4-1)}F'(z)+ \frac{z^2(\delta+1)(\delta+3)-q^2\mu_{c1}^2}{z^4-1}F(z)=0.
\end{equation}
This equation has a general solution as
\begin{equation}\label{solF}
\begin{split}
  F(z)= & C_1\text{HeunG}(-1,-\frac14q^2\mu_{c1}^2,\frac{\delta+3}{2},\frac{\delta+1}2,\delta+1,1;-z^2)\\
    & +C_2z^{-2\delta}\text{HeunG}(-1,-\frac14q^2\mu_{c1}^2,\frac{3-\delta}{2},\frac{1-\delta}2,1-\delta,1;-z^2).
\end{split}
\end{equation}
Here $C_1$ and $C_2$ are two constants and HeunG stands for the general Heun function with the feature $\text{HeunG}(~,~,~,~,~,~;0)=1$ . To satisfy the source free condition at the boundary, it is easy to see that one has to take $C_2=0$. On the other hand, in order to have a regularity for the vector field $\rho_x$ at the tip, there exists a constraint condition at $z=1$ as
\begin{equation}\label{FBC}
[z^4(5+2\delta)-(2\delta+1)]F'(z)-[z^2(\delta+1)(\delta+3)-q^2\mu_{c1}^2]F(z)|_{z=1}=0.
\end{equation}
Substituting the solution~\eqref{solF} with $C_2=0$ into \eqref{FBC}, we can obtain
a nonlinear equation for $\mu_{c1}^2q^2$. The equation can be easily solved by numerical method. The solution of course depends on  $\delta$ or $m^2$. As a result we can write the solution as
\begin{equation}\label{muq1}
    \mu_{c1}q=\alpha(m^2),~~~\text{or}~~~~~~\mu_{c1}=\alpha(m^2)/q.
\end{equation}
This is just the first scaling relation in \eqref{scal1}. For the cases with $m^2=5/4,0$ and~$-3/4$, we have $ \mu_{c1}q\simeq2.784531829,2.265193164$ and~$1.737772548$, respectively. It is easy to see that they are very close to corresponding ones obtained by using the shooting method in section~\ref{Soliton}. This agreement also provides a piece of evidence to support the results we obtained by shooting method to solve the equations~\eqref{eoms}.

\begin{figure}
  \includegraphics[width=0.5\textwidth]{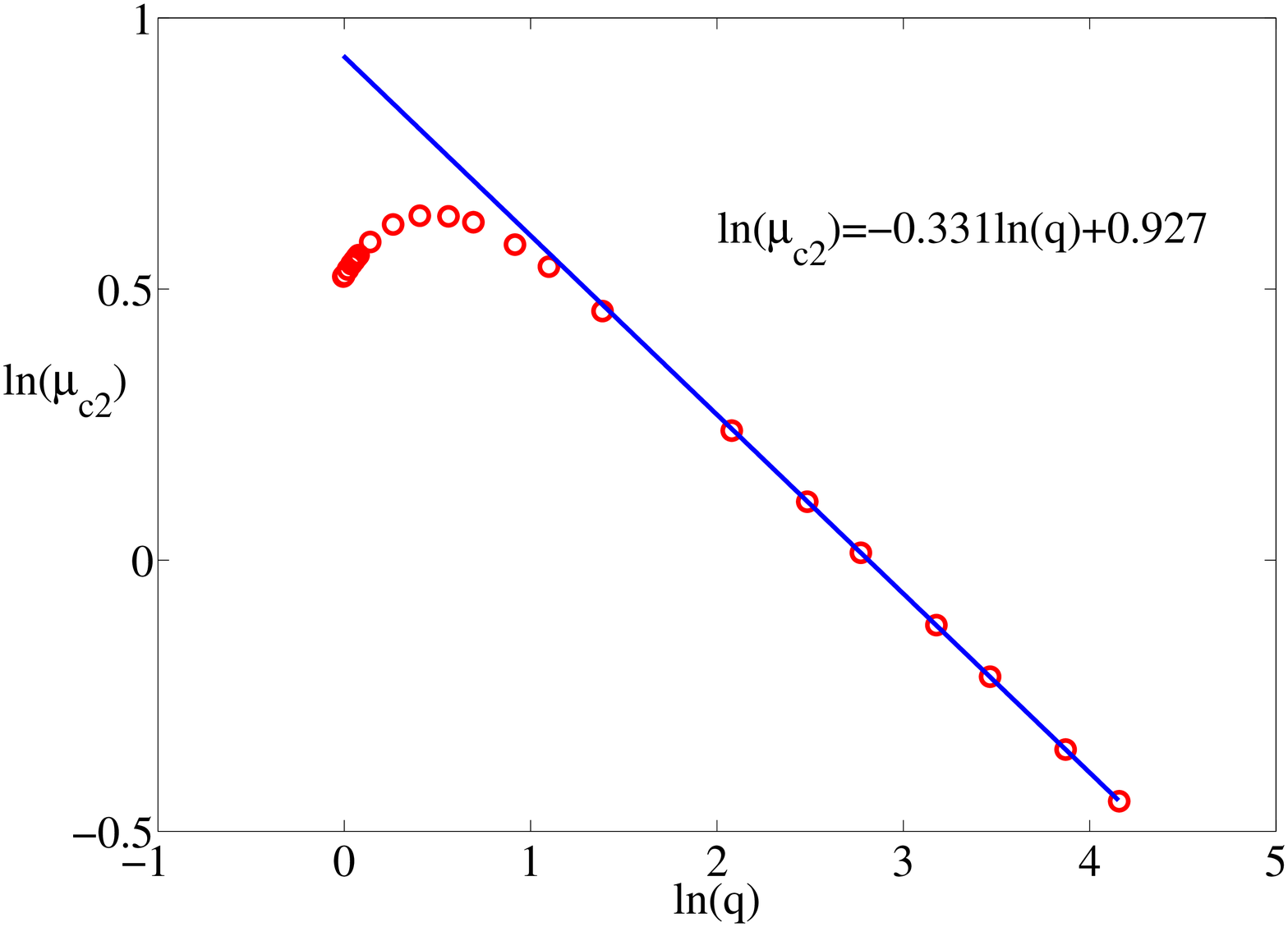}
  \includegraphics[width=0.5\textwidth]{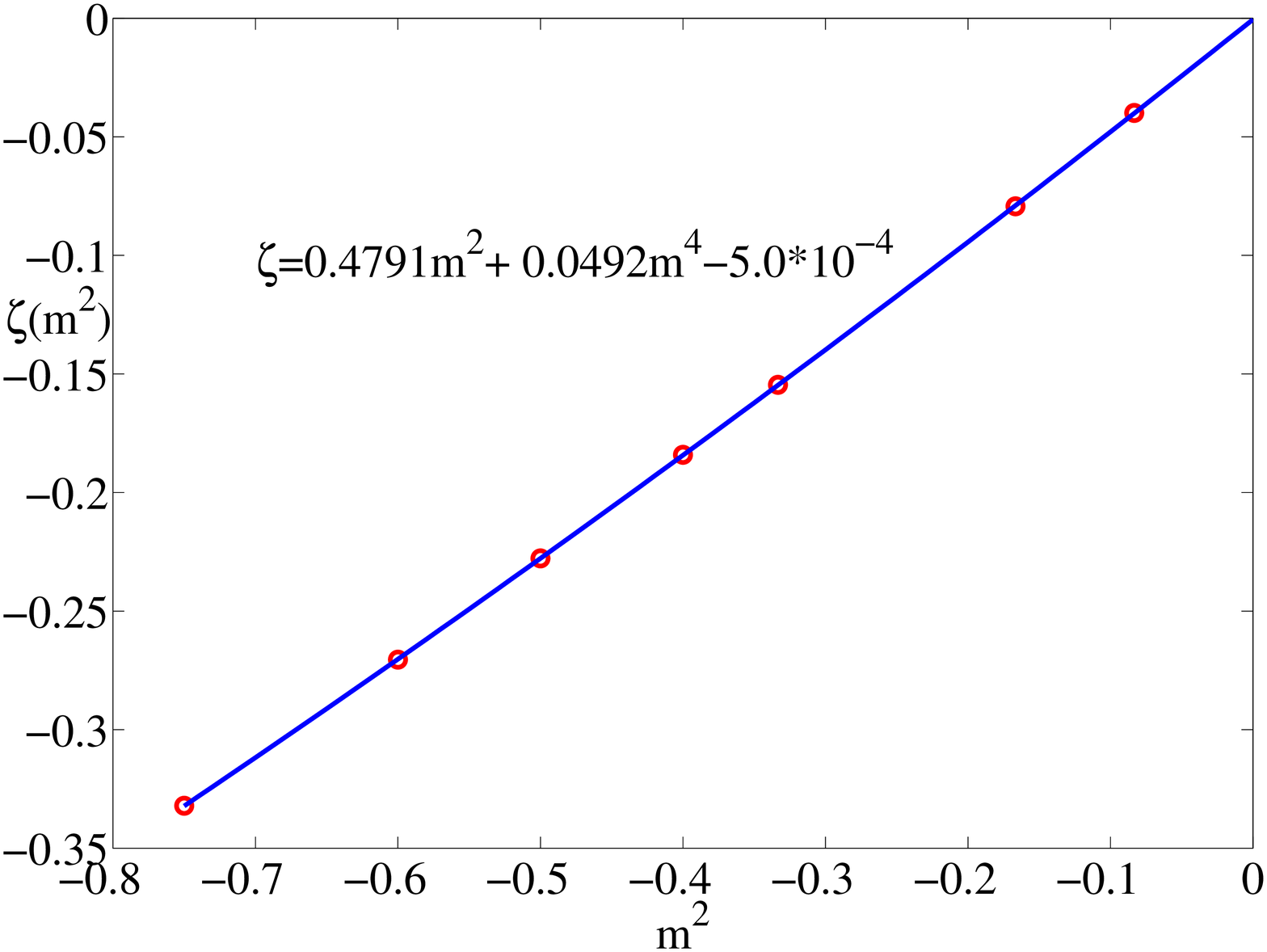}
  \caption{The fitting curve for the second relation in~\eqref{scal1}. In both figures, the red circle and blue line are the numerical results and fitting curves, respectively. The left plot shows the scaling relation between $\ln({\mu_{c2}})$ and $\ln(q)$ in the case of~$m^2=-3/4$. It shows that there is a well-behaved linear dependence when $\ln(q)$ is larger than 2.  In the right figure, we plot $\zeta$ with respect to different values of mass square. The numerical results are fitted well by a second order polynomial with a small constant term.} \label{lnmulnq1}
\end{figure}

It seems impossible to get the second scaling relation in \eqref{scal1} by an analytical method since it happens in the condensed phase with
large condensate. In this case, the back reaction of the vector field can not be neglected, and we have to solve the equations~\eqref{eoms} without any approximation. By  numerical analysis, we find that when $q$ is very large, the critical chemical potential for the phase transition from the condensed phase to the normal phase has a power law form as
\begin{equation}\label{muq2}
\mu_{c2}\propto q^{\zeta(m^2)}.
\end{equation}
A typical case with $m^2 =-3/4$ is shown in the left plot of figure~\ref{lnmulnq1}. The fitting  of $\zeta$ for different $m^2$ is shown in the right plot of figure~\ref{lnmulnq1}. One can see that the relation between $\zeta$ and $m^2$ is fitted well by a second order polynomial with a very small constant term. This constant term in fact is less than our numerical error. Thus if neglect this constant term, we can obtain the following fitting relation
\begin{equation}\label{gamma2}
    \zeta(m^2)=0.4791m^2+0.0492m^4.
\end{equation}

\end{document}